\documentclass[apj]{emulateapj}
\usepackage{natbib}
\usepackage{graphicx}
\usepackage{color}

\usepackage{hyperref}
\usepackage{amsmath}
\usepackage{enumitem}

\newcommand{\ev}[1]{\left < {#1} \right >}
\newcommand{\brac}[1]{\left ( {#1} \right )}
\newcommand{\cl}[1]{\color{black}{#1}\color{black}}

\begin{document}
\title{Testing LSST Dither Strategies for Survey Uniformity and Large-Scale Structure Systematics}

\author{Humna Awan\altaffilmark{1}, Eric Gawiser\altaffilmark{1}, Peter Kurczynski\altaffilmark{1},  \cl{R. } Lynne Jones\altaffilmark{2}, Hu Zhan\altaffilmark{3}, Nelson \cl{D. } Padilla\altaffilmark{4}, Alejandra M. Mu\~noz Arancibia\altaffilmark{4,5}, Alvaro Orsi\altaffilmark{6}, \cl{Sof\'ia A. Cora\altaffilmark{7,8,9}}, and Peter Yoachim\altaffilmark{2}}

\altaffiltext{1}{Department of Physics $\&$ Astronomy, Rutgers University, 136 Frelinghuysen Rd., Piscataway, NJ 08554 }
\altaffiltext{2}{Department of Astronomy, University of Washington, 3910 15th Ave NE, Seattle, WA 98195}
\altaffiltext{3}{Key Laboratory of Space Astronomy and Technology, National Astronomical Observatories, Chinese Academy of Sciences, Beijing 100012, China}
\altaffiltext{4}{Instituto de Astrof\'isica, Pontificia Universidad Cat\'olica de Chile, Av. Vicu\~na Mackenna 4860, Santiago, Chile}
\altaffiltext{5}{Instituto de F\'isica y Astronom\'ia, Universidad de Valpara\'iso, Av. Gran Breta\~na 1111, Valpara\'iso, Chile}
\altaffiltext{6}{Centro de Estudios de Fisica del Cosmos de Aragon, Plaza de San Juan 1, Teruel, 44001, Spain}

\altaffiltext{7}{Instituto de Astrof\'isica de La Plata (CCT La Plata, CONICET, UNLP), Paseo del Bosque s/n, B1900FWA, La Plata, Argentina}
\altaffiltext{8}{Facultad de Ciencias Astron\'omicas y Geof\'isicas, Universidad Nacional de La Plata, Paseo del Bosque s/n, B1900FWA, La Plata, Argentina}
\altaffiltext{9}{Consejo Nacional de Investigaciones Cient\'ificas y T\'ecnicas, Av. Rivadavia 1917, C1033AAJ, CABA, Argentina}
\email{awan@physics.rutgers.edu}

\shorttitle{Testing LSST Dither Strategies}
\shortauthors{Awan et al.}

\slugcomment{Accepted for publication in ApJ}

\begin{abstract}
The Large Synoptic Survey Telescope (LSST) will survey the southern sky from 2022--2032 with unprecedented detail. \cl{Since the observing strategy can lead to artifacts in the data}, we investigate the effects of telescope-pointing offsets (called dithers) on the $r$-band coadded 5$\sigma$ depth yielded after the 10-year survey. We analyze this survey depth for several geometric patterns of dithers (e.g., random, hexagonal lattice, spiral) with amplitude as large as the radius of the LSST field-of-view, implemented on different timescales (per season, per night, per visit). Our results illustrate that per night and per visit dither assignments are more effective than per season. Also, we find that some dither geometries (\cl{e.g., } hexagonal lattice) are particularly sensitive to the timescale on which the dithers are implemented, while others like random dithers perform well on all timescales. We then model the propagation of depth variations to artificial fluctuations in galaxy counts, which are a systematic for large-scale structure studies. We calculate the bias in galaxy counts \cl{caused by } the observing strategy, accounting for photometric calibration uncertainties, dust extinction, and magnitude cuts;  uncertainties in this bias limit our ability to account for structure induced by the \cl{observing } strategy. We find that after 10 years of the LSST survey, the best \cl{dither } strategies lead to uncertainties in \cl{this } bias smaller than the minimum statistical floor for a galaxy catalog as deep as $r$$<$27.5\cl{. A  few of these strategies } bring the uncertainties close to the \cl{statistical } floor for $r$$<$25.7 after only one year of survey.
\end{abstract}

\section{Introduction}
The Large Synoptic Survey Telescope (LSST) is an upcoming wide-field deep survey, designed to make detailed observations of the southern sky. A telescope with an effective aperture of 6.7m and a 3.2 Gigapixel camera, LSST will survey about 20,000 deg$\mathrm{^2}$ of the sky in \textit{ugrizy} bands, over the course of ten years with $\sim$150 visits in each band to each part of the survey area \citep{LSST2009}. While the survey has various goals, from studying near-Earth objects to transient phenomena, its imaging capabilities are particularly promising for studying dark energy. With its wide-deep observation mode, LSST will probe 1) the shear field from weak gravitational lensing, 2) Baryonic Acoustic Oscillations (BAO) in the galaxy power spectrum and correlation functions, 3) evolution of the galaxy cluster mass function, 4) Type Ia supernovae and their distance-redshift relationship, and 5) time delays from strong gravitational lenses, providing an opportunity to study dark energy from one dataset.  The nature of these cosmic probes leads to requirements on the survey observing strategy, understood in terms of cadence, i.e. frequency of visits in a particular filter, and uniformity, i.e. survey depth across various regions of the sky. For goals dependent on spatial correlations, such as BAO and additional large-scale structure (LSS) studies, survey uniformity is of critical importance, while time domain science often depends on high cadence. \

The baseline LSST observing strategy tiles the sky with hexagons, each of which inscribes an LSST field-of-view (FOV) \citep{LSST2009}. Given that the FOV is approximately circular, the hexagonal tiling leads to regions between the FOV and the inscribed hexagon that overlap when adjacent fields are observed. Therefore, observations at fixed telescope pointings lead to deeper data in these overlapping regions, decreasing survey uniformity and inducing artificial structure specifically at scales corresponding to the expected BAO signal at $z$ $\sim$ 1 \citep{Carroll2014}. While the double-coverage data could be discarded to make the survey uniform, the loss would comprise nearly 17$\%$ of LSST data \citep{Carroll2014}, equivalent to 1.5 years of survey time. On the other hand, correction methods have been developed for other surveys \citep[\cl{e.g.,}][]{Ross2012, Leistedt2015} to post-process and correct for the systematics in the observed data -- such an approach could also work for LSST survey uniformity. Here, however, we address the approach of minimizing eventual survey systematics by designing an optimal observing strategy. \

Dithers, i.e. telescope pointing offsets, are helpful in reducing systematics. While LSST plans to implement small dithers to compensate for the finite gaps between the CCDs \citep[\cl{e.g.,}][]{McLean2008}, implementing large dithers on the scale of the FOV appears to offer a solution for LSST survey uniformity, reducing the artificial structure by a factor of 10 as compared to the undithered survey \citep{Carroll2014}. In this paper, we analyze various \cl{dither } strategies, varying in both the geometric pattern and the timescale on which the pattern is implemented. We develop a methodology for a quantitative comparison of these strategies and explore their effects on survey depth and BAO systematic uncertainty.  We introduce the LSST Operations Simulator and the Metrics Analysis Framework in Section~\ref{OpSim_MAF}. Then, in Section~\ref{DitheringStrategies}, we describe the variants of the dithers implemented, followed by a discussion of the impacts of the \cl{dither } strategies on the coadded depth as well as artificial fluctuations in galaxy counts in Section~\ref{Results and Analysis}. We conclude in Section~\ref{Conclusion}, highlighting that our work illustrates the capability to assess the effectiveness of various \cl{dither } strategies for LSST science goals.\\

\section{The LSST Operations Simulator and Metrics Analysis Framework}{\label{OpSim_MAF}}
The LSST Operations Simulator (OpSim) simulates 10-year surveys, accounting for realistic factors that affect the final data; these considerations include scheduling of observations, telescope pointing, slewing and downtime, site conditions, etc. \citep{Delgado2014}. More specifically, OpSim output contains realizations of LSST metadata, stamped with sky position, time, and filter \citep{LSST2009}, allowing post-processing of the output to simulate different \cl{dither } strategies. \

As mentioned earlier, LSST OpSim tiles the sky with hexagonal tiles. In order to effectively account for the overlapping regions between the hexagons, we utilize the Hierarchical Equal Area isoLatitude Pixelization (HEALPix) package to uniformly tile the sky with equal area pixels \citep{Gorski2005}. HEALPix uses nearly-square pixels to tile the sky with a resolution parameter N$\mathrm{_{side}}$, leading to a total number of pixels N$\mathrm{_{pixels}}$= 12N$\mathrm{_{side}^2}$. In our analysis, we use N$\mathrm{_{side}}$= 256, giving a total of 786,432 pixels, and effectively tiling each 3.5$^\circ$ FOV with about 190 HEALPix pixels. Here we note that our resolution is four-fold higher than that used in \citet{Carroll2014}; this improvement ensures that we do not encounter signal aliasing in the angular scale range we study here. \

We carry out our analysis within the Metrics Analysis Framework (MAF), designed for the analysis of OpSim output in a manner that facilitates hierarchical building of the analysis tools.  MAF consists of various classes, of which most relevant here are ${Metrics}$ that contain the algorithm to analyze each HEALPix pixel and ${Stackers}$ that provide the functionality of adding columns to the OpSim database; for details, see \citet{Jones2014}. Some of our code has already been incorporated into the MAF pipeline\footnote{\url{https://github.com/lsst/sims\_maf}}, and the rest can be found in the LSST GitHub repository\footnote{\url{https://github.com/LSST-nonproject/sims_maf_contrib/tree/master/mafContrib}}. 

\section{\cl{Dither } Strategies}{\label{DitheringStrategies}}
We consider \cl{dither } strategies with three different timescales: by season, by night, and by visit. A single visit is a set of two 15 second exposures \citep{Ivezic2008}. Since OpSim output does not have a season assignment for the simulated data, we define seasons separately for each field, starting from zero and incrementing the season number when the field's RA is overhead in the middle of the day. This leads to 11 seasons for the 10-year data, and we assign the 0th and the 10th seasons the same dither position. \

Since fields are scheduled to be visited at least two times in a given night, followed by a typical revisit time of three days \citep{Ivezic2008}, we implement two approaches for the by-night timescale: 1) FieldPerNight, where a new dither position is assigned to each field independently, and 2) PerNight, where a new position is assigned to all fields. The first approach tracks each field and assigns it a new dither position only if it is observed on a new night, while the second approach assigns a dither position to all the fields every night (regardless of whether a particular field is observed or not). For the by-visit timescale, we only consider FieldPerVisit, and for by-season strategies, we consider PerSeason only.  \

For the dithers, we implement a few geometrical patterns to probe the effects of dither positions themselves. Since the sky is tiled with hexagons inscribed within the 3.5$^\circ$ FOV, we restrict all dither positions to lie within these hexagons. For by-season strategies, given that there are only 10 seasons throughout the LSST run, we pick a geometry that allows choosing 10 dither positions uniformly across the FOV:
\begin{itemize}[noitemsep]
	\item[--]{Pentagons: points alongs two pentagons, one inside an inverted, bigger pentagon. }
\end{itemize}
For by-night and by-visit timescales, we consider four different geometries:
\begin{itemize}[noitemsep]
	\item[--]{Hexagonal lattice dithers: 217 points arranged on a hexagonal lattice \cl{\citep{Krughoff2009}}.}
	\item[--]{Random dithers: random points chosen within the hexagon such that every dither position is a new random point.}
	\item[--]{Repulsive Random dithers: after creating a grid of squares inside the hexagon, squares are randomly chosen without replacement. Every dither position is a random point within a chosen square.}
	\item[--]{Fermat Spiral dithers: 60 points are chosen from the spiral defined by $r \propto \sqrt{\theta}$, where $\theta$ is a multiple of the golden angle 137.508$^{\circ}$ (geometry appears in nature; see \citet{MunozR2014}). }
\end{itemize}

\begin{figure*}
	\vspace*{-4em}
	\hspace*{12em}
	\begin{minipage}{0.25\paperwidth}
		\includegraphics[trim={{105 105 105 105}},clip=true,width=.27\paperwidth]{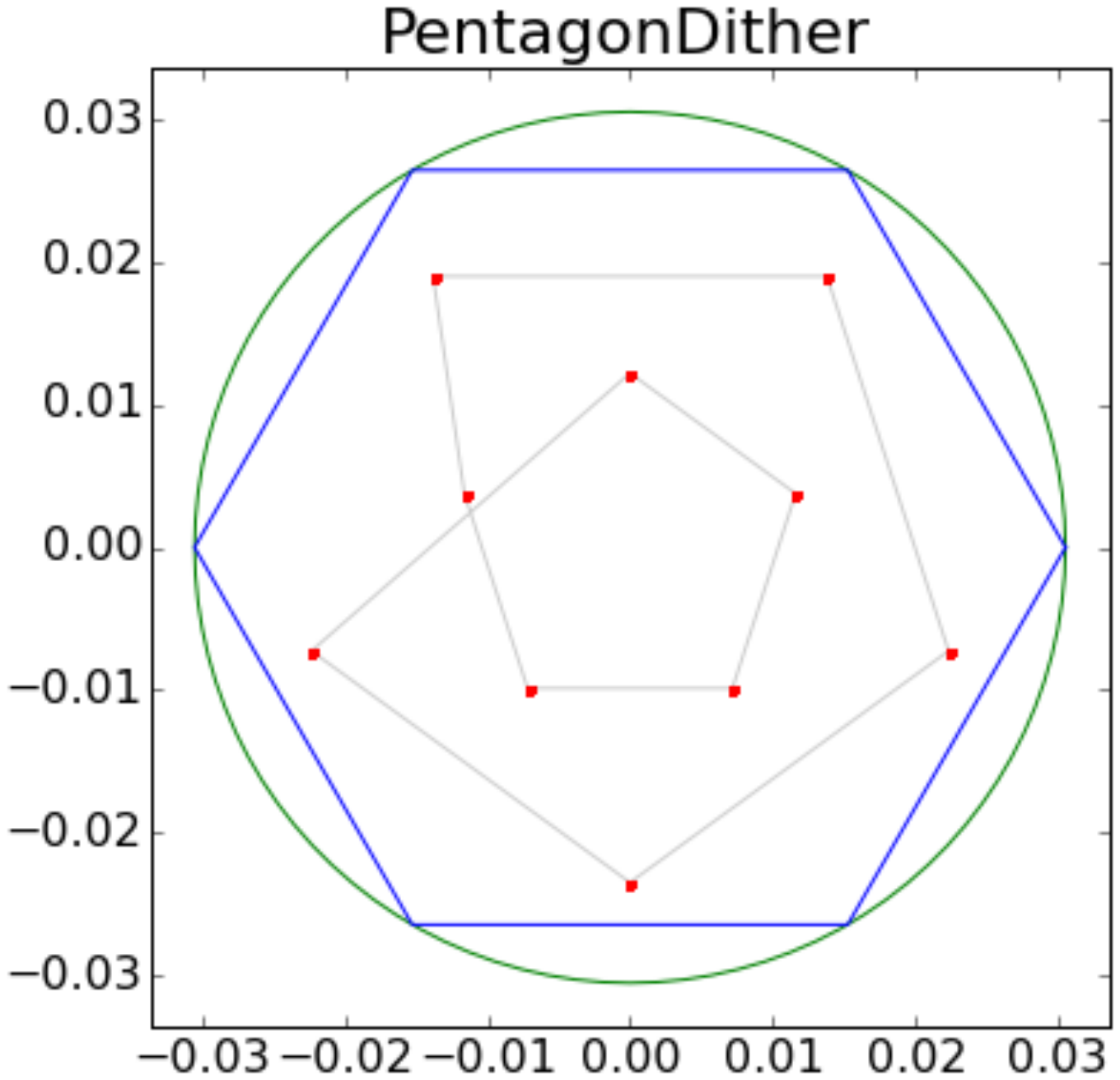}
	\end{minipage}\
	\hspace*{0em}
	\begin{minipage}{0.25\paperwidth}
		\includegraphics[trim={{105 105 105 105}},clip=true,width=.27\paperwidth]{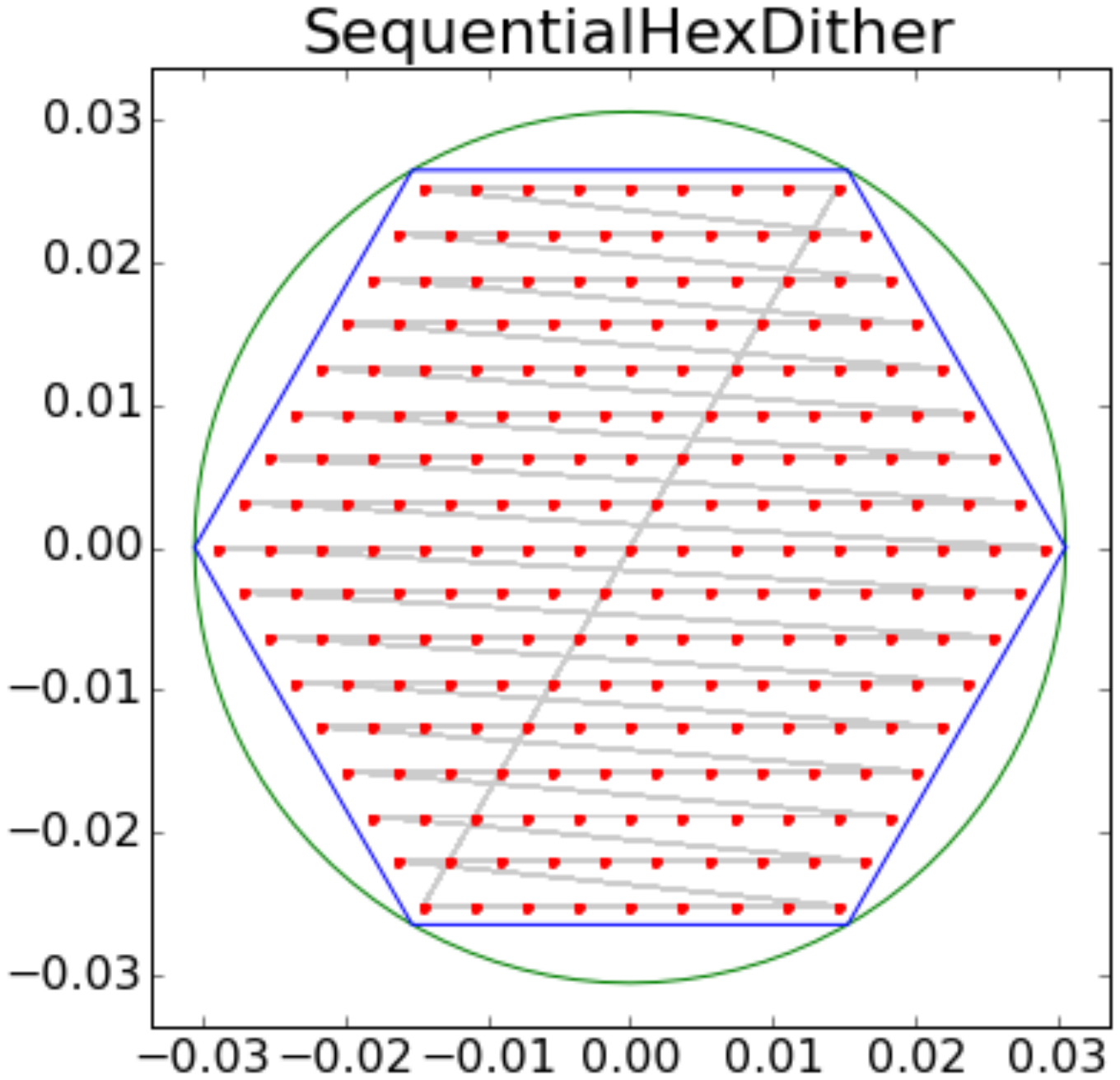}
	\end{minipage}\
	\hspace*{2em}
	\vspace*{0.5em}
	\hfill
	
	\vspace*{-7em}
	\hspace*{1em}
	\begin{minipage}{0.25\paperwidth}
		\includegraphics[trim={{105 105 105 105}},clip=true,width=.27\paperwidth]{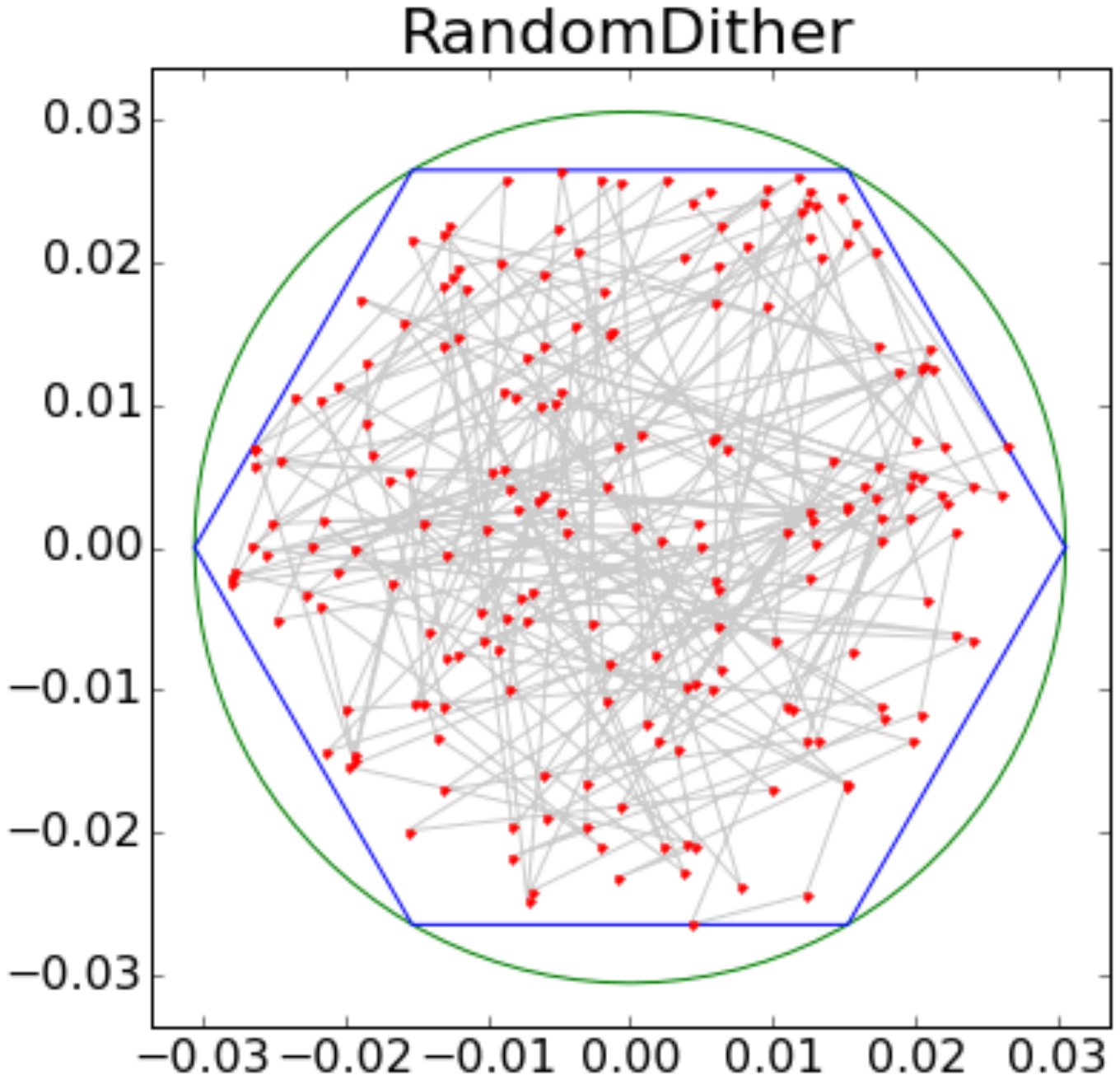}
	\end{minipage}\
	\hspace*{0em}
	\begin{minipage}{0.25\paperwidth}
		\includegraphics[trim={{105 105 105 105}},clip=true,width=.27\paperwidth]{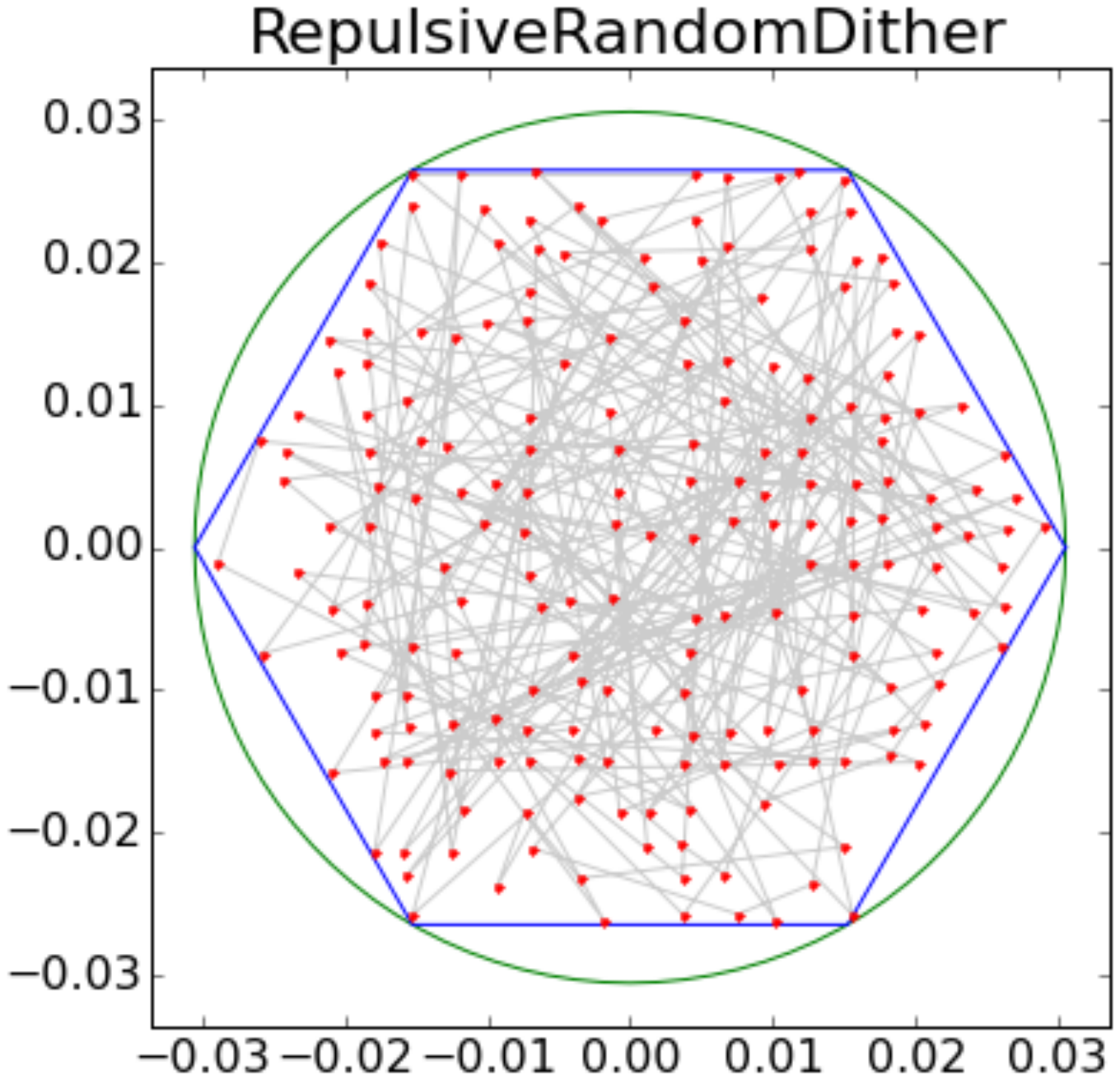}
	\end{minipage}\
	\hspace*{0em}
	\begin{minipage}{0.25\paperwidth}
		\includegraphics[trim={{105 105 105 105}},clip=true,width=.27\paperwidth]{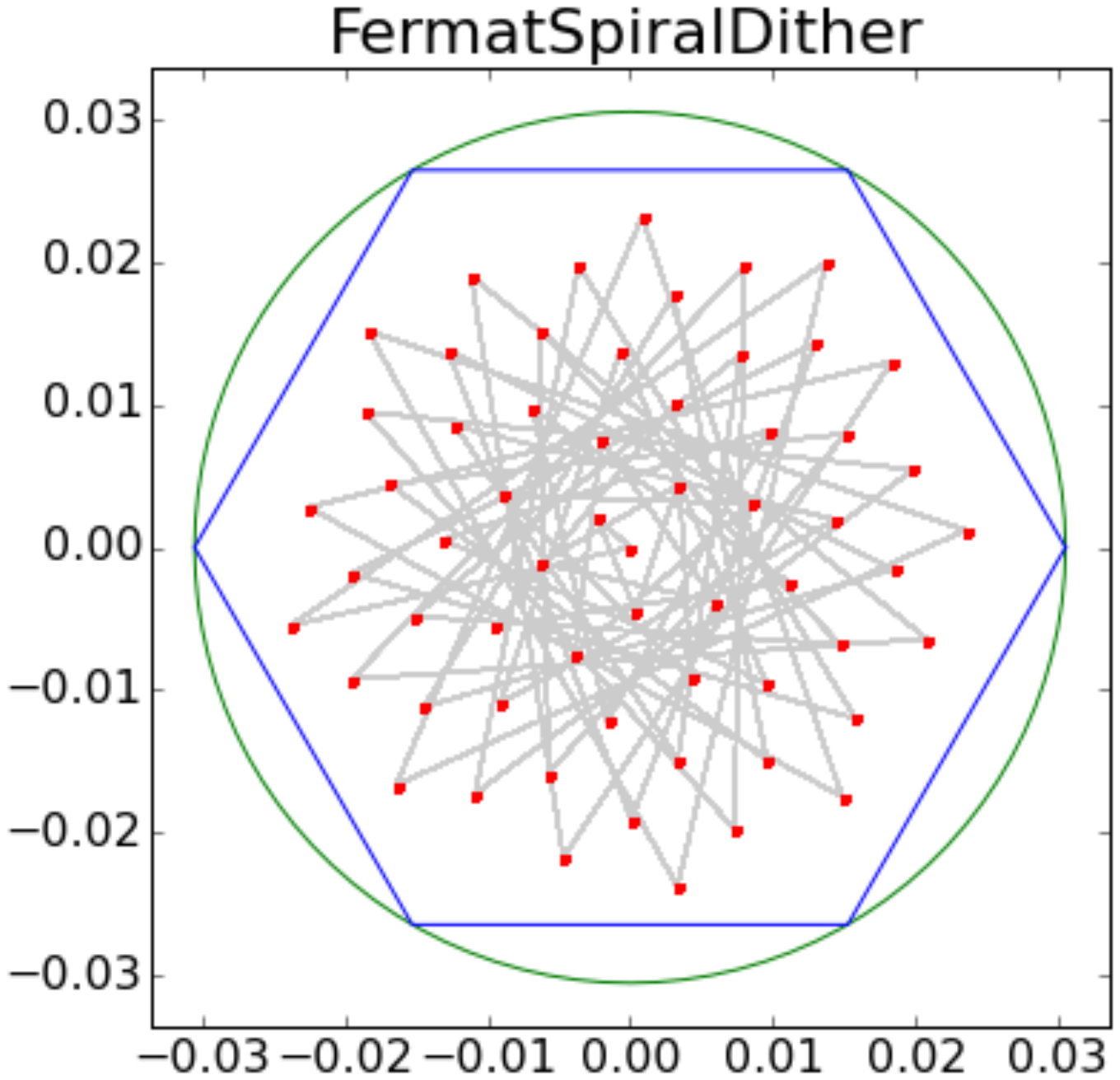}
	\end{minipage}\
	\hspace*{2em}	
	\vspace*{-4em}
	\figcaption[]{Dither geometries: PentagonDither is implemented only for \cl{per-season } timescale, while the rest are implemented for \cl{per-visit, per-night and field-per-night } timescales. The green curve represents the circular FOV with radius of 0.305 radians, the blue hexagon represents the hexagonal tiling of the sky originally adopted for the undithered observations, and the red points are the dither positions, connected with gray lines. The axes are labelled in radians. See Section~\ref{DitheringStrategies} for details. \label{ditherGeometry}}
\end{figure*}

Figure~\ref{ditherGeometry} shows these geometries and the possible dither positions. We also considered some other variants. For by-season timescale, we implemented a PentagonDiamond geometry where the first point is at the center of the FOV, followed by 9 points arranged along a diamond circumscribed by a pentagon. We find that PentagonDiamond leads to results similar to Pentagons, and discuss only the latter here. We also considered Spiral dithers, where equidistant points are chosen along a spiral centered on the FOV and the number of points and coils can be varied, as well as variants of Fermat spiral, in terms of the number of points and $\theta$ as a multiple of 77.508$^{\circ}$ or 177.508$^{\circ}$. Our preliminary analysis shows that these spiral geometries behave similarly as the 60-point, golden-angle Fermat spiral described above. \

To identify the various strategies, we follow a consistent naming scheme: [Geometry]Dither[Field]Per[Timescale], where the absence of \cl{`Field' implies } dither assignment to all fields, while its presence implies that each field is tracked and assigned a dither position independent of other fields. For instance, SequentialHexDitherPerNight assigns the new dither position to all fields every night, while SequentialHexDitherFieldPerNight assigns it to \cl{a } field only when it is observed on a new night.

\section{Analysis $\&$ Results}{\label{Results and Analysis}}
We use OpSim dataset \texttt{enigma\_1189}\footnote{\url{https://confluence.lsstcorp.org/display/SIM/OpSim+Datasets+for+Cadence+Workshop+LSST2015}}, which includes the wide-fast-deep (WFD) survey region as well as five Deep Drilling fields; we focus only on WFD survey for our analysis. We implement various dithers within MAF by building $Stackers$ corresponding to each \cl{dither } strategy and post-processing the OpSim output to find the survey results using the dithered positions. First, we examine the $r$-band coadded depth (i.e. the final depth after the 10-year survey) as a function of sky location, followed by an analysis of the fluctuations in the galaxy counts, in order to probe the effects of \cl{dither } strategies on large-scale structure studies.

\begin{figure*} 	 
	\vspace*{1em}
	\hspace*{1.5em}
	\begin{minipage}{0.37\paperwidth}
		\includegraphics[trim={20 70 5 -5},clip=true,width=.35\paperwidth]{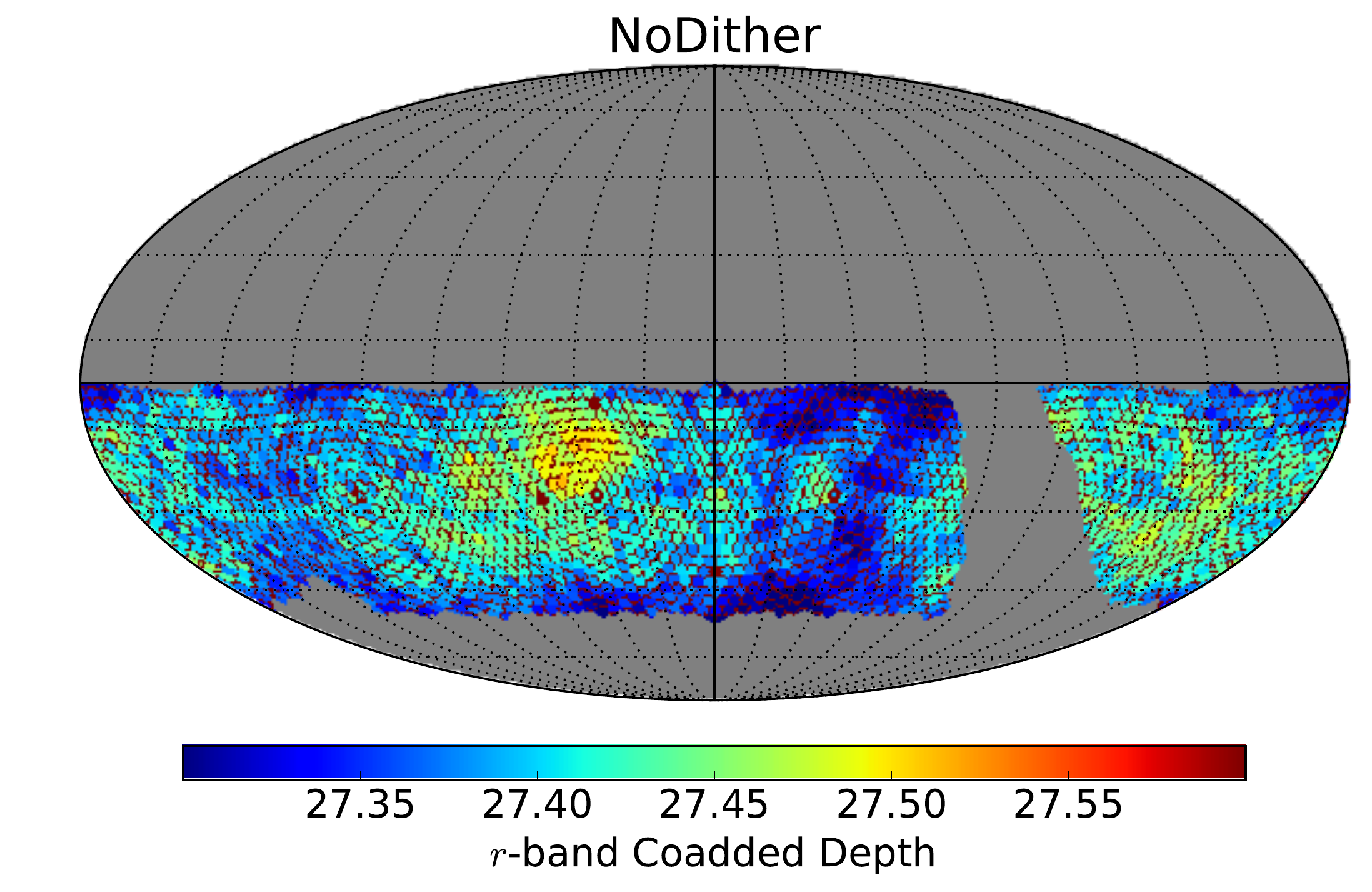}
	\end{minipage}\
	\hspace*{3em}
	\begin{minipage}{0.37\paperwidth}
		\includegraphics[trim={20 70 5 -5},clip=true,width=.35\paperwidth]{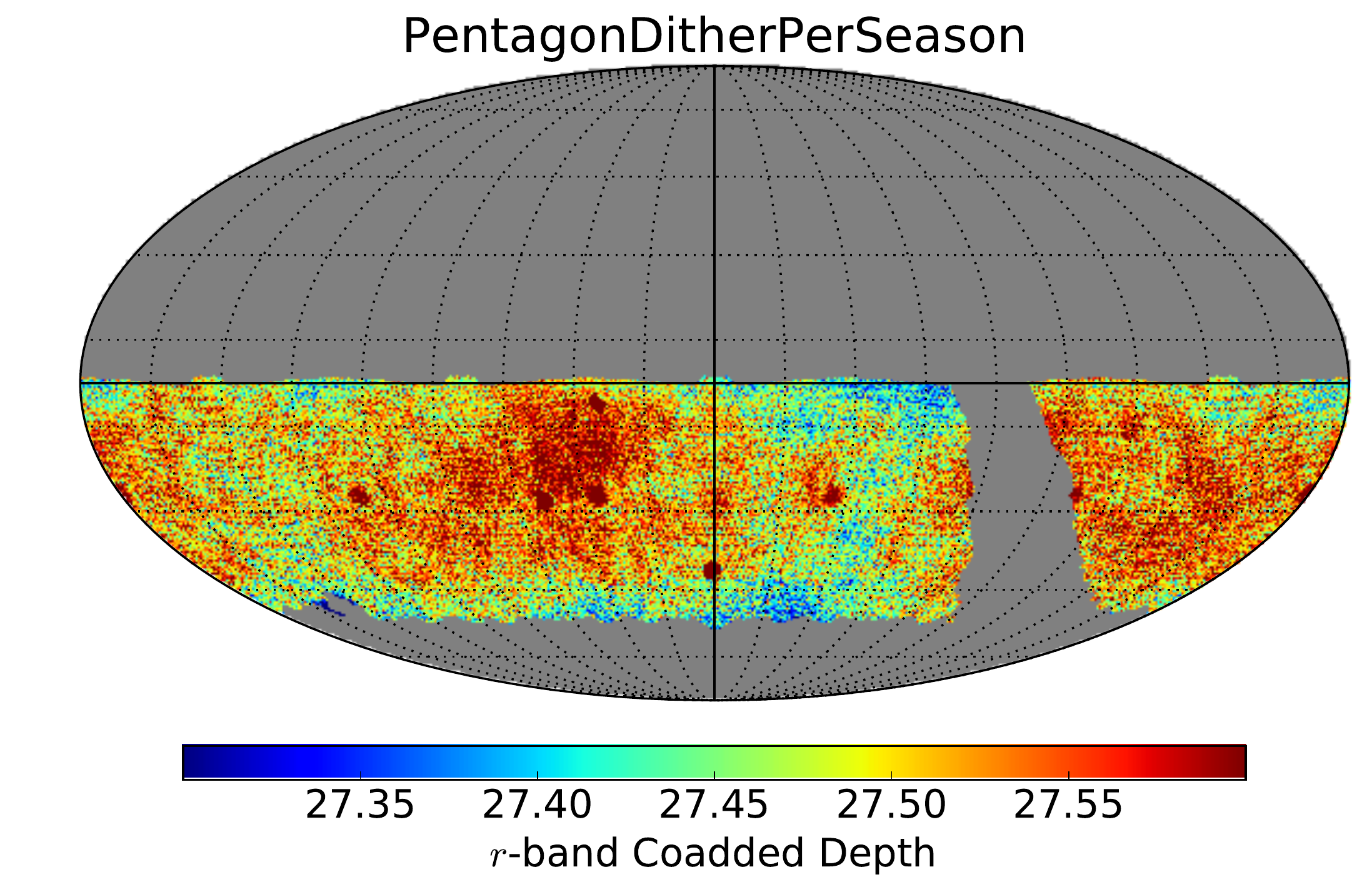}
	\end{minipage}\
	\hfill
	
	\hspace*{-2.5em}
	\vspace*{-0.5em} 
	\begin{minipage}{0.43\paperwidth}
		\includegraphics[trim={-10 75 5 5},clip=true,width=.43\paperwidth]{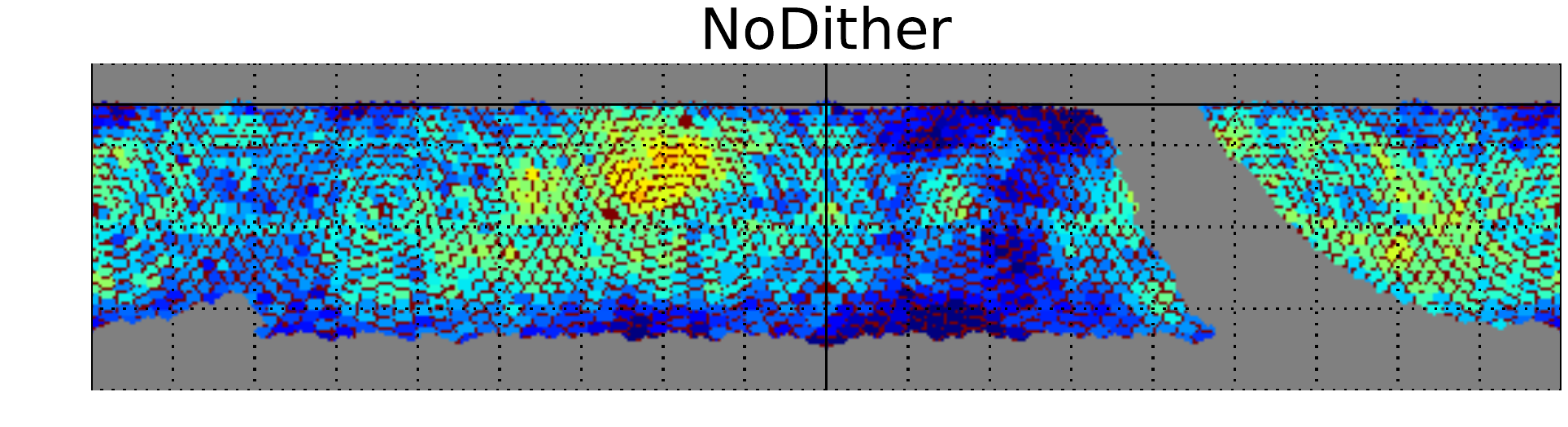}
	\end{minipage}\
	\hspace*{-1em}
	\begin{minipage}{0.43\paperwidth}
		\includegraphics[trim={-10 75 5 5},clip=true,width=.43\paperwidth]{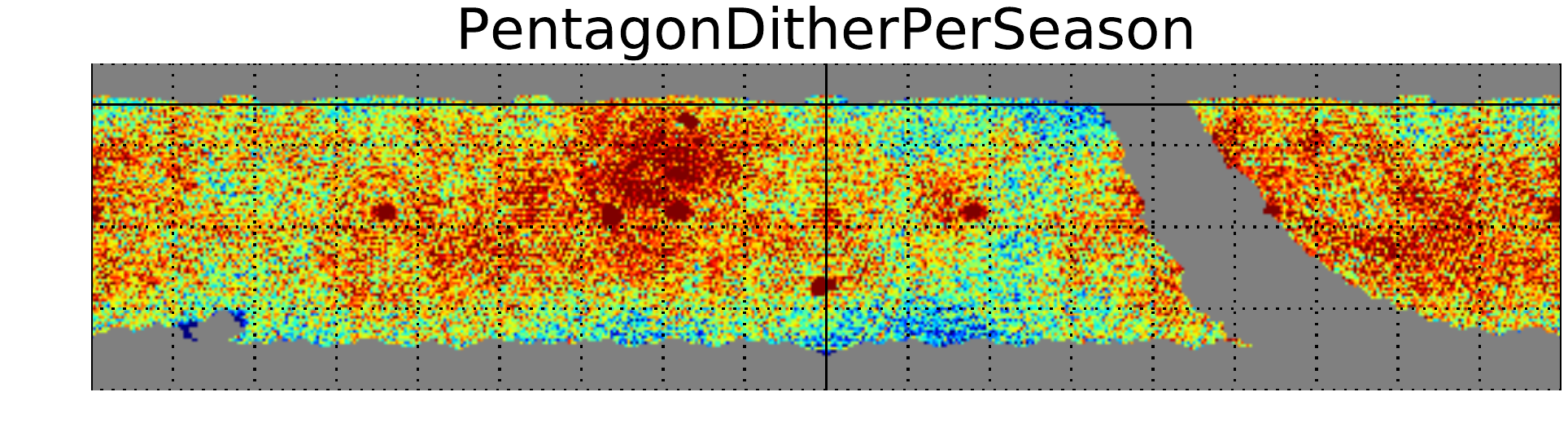}
	\end{minipage}\
	\hfill
	
	\hspace*{-2.5em}
	\vspace*{-0.5em} 
	\begin{minipage}{0.43\paperwidth}
		\includegraphics[trim={-10 75 5 5},clip=true,width=.43\paperwidth]{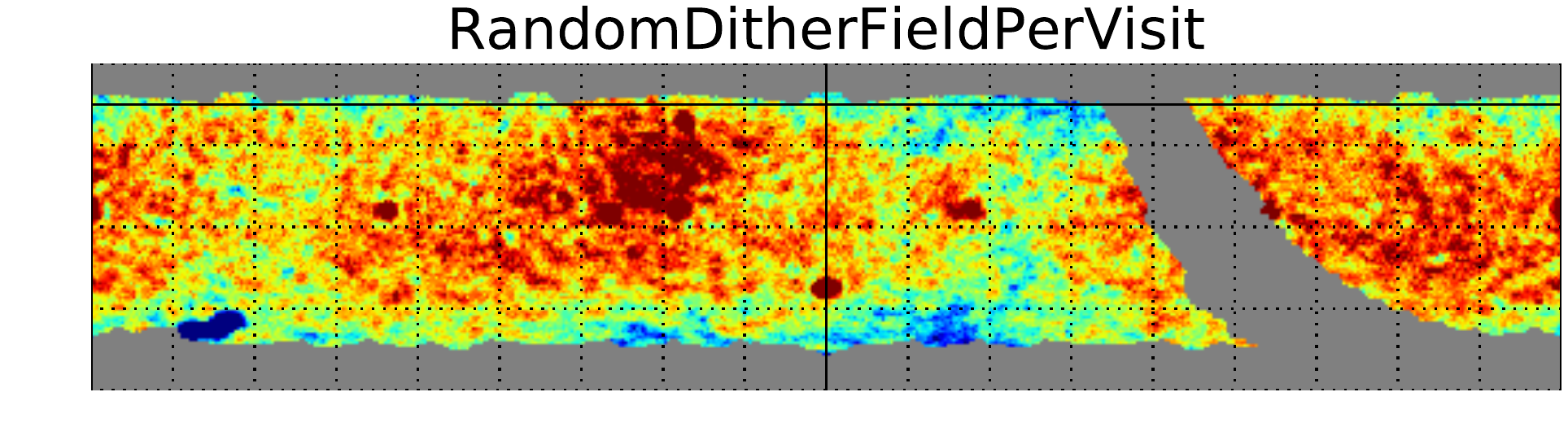}
	\end{minipage}\
	\hspace*{-1em}
	\begin{minipage}{0.43\paperwidth}
		\includegraphics[trim={-10 75 5 5},clip=true,width=.43\paperwidth]{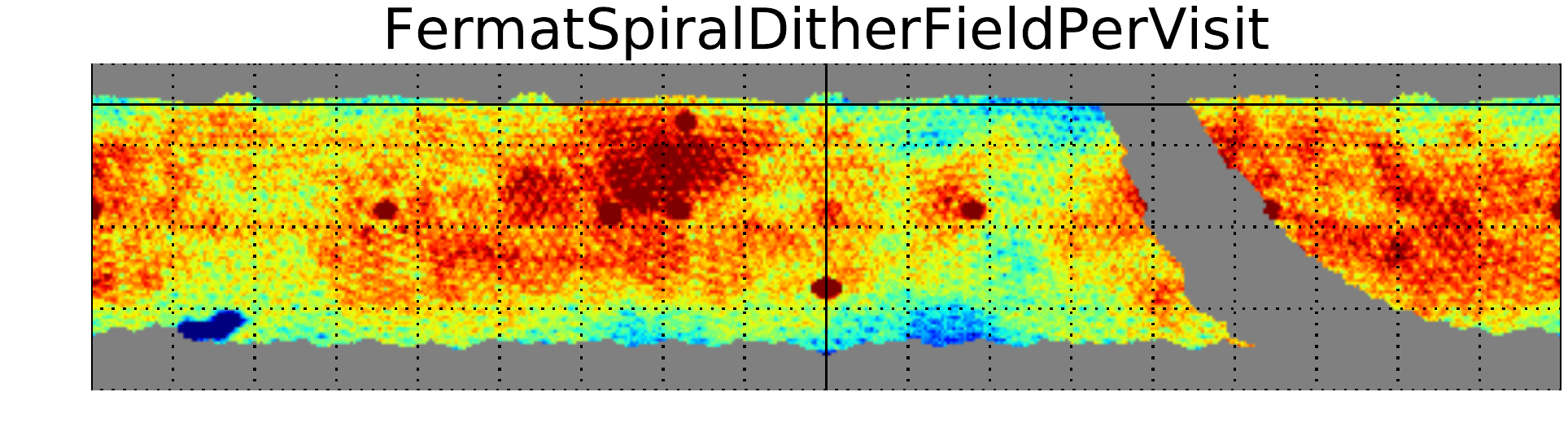}
	\end{minipage}\
	\hfill
	
	\hspace*{-2.5em}
	\vspace*{-0.5em} 
	\begin{minipage}{0.43\paperwidth}
		\includegraphics[trim={-10 75 5 5},clip=true,width=.43\paperwidth]{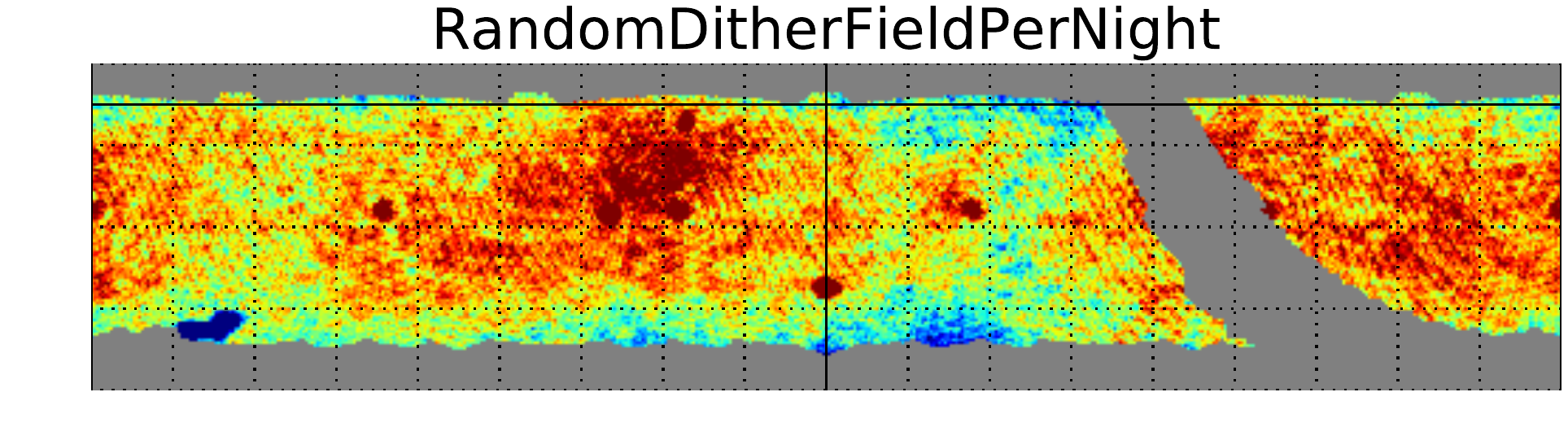}
	\end{minipage}\
	\hspace*{-1em}
	\begin{minipage}{0.43\paperwidth}
		\includegraphics[trim={-10 75 5 5},clip=true,width=.43\paperwidth]{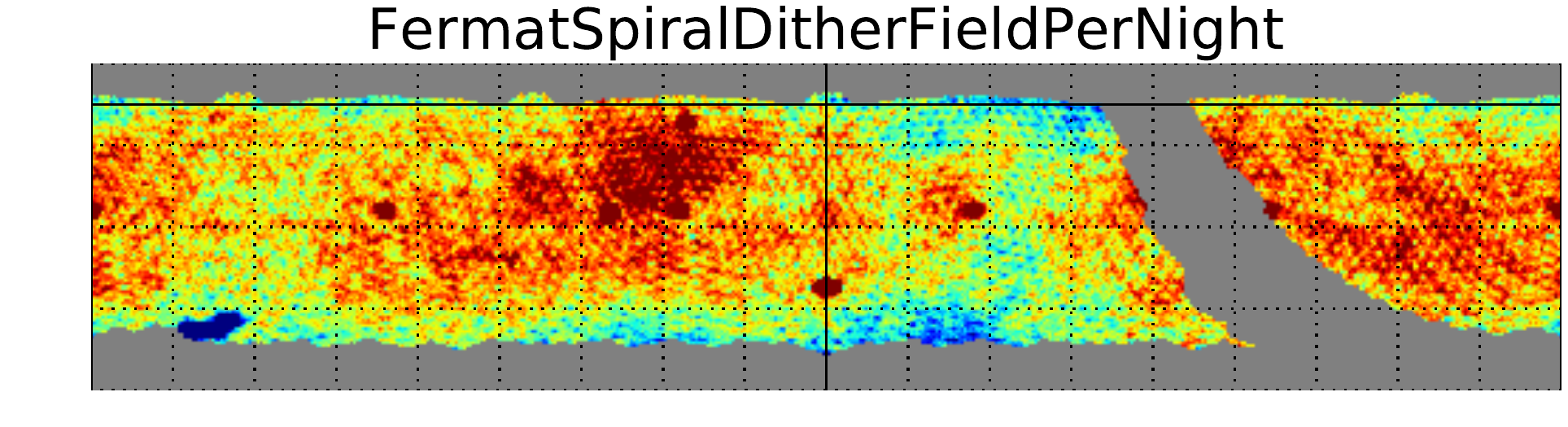}
	\end{minipage}\
	\hfill
		
	\hspace*{-2.5em}
	\vspace*{-0.5em} 
	\begin{minipage}{0.43\paperwidth}
		\includegraphics[trim={-10 75 5 5},clip=true,width=.43\paperwidth]{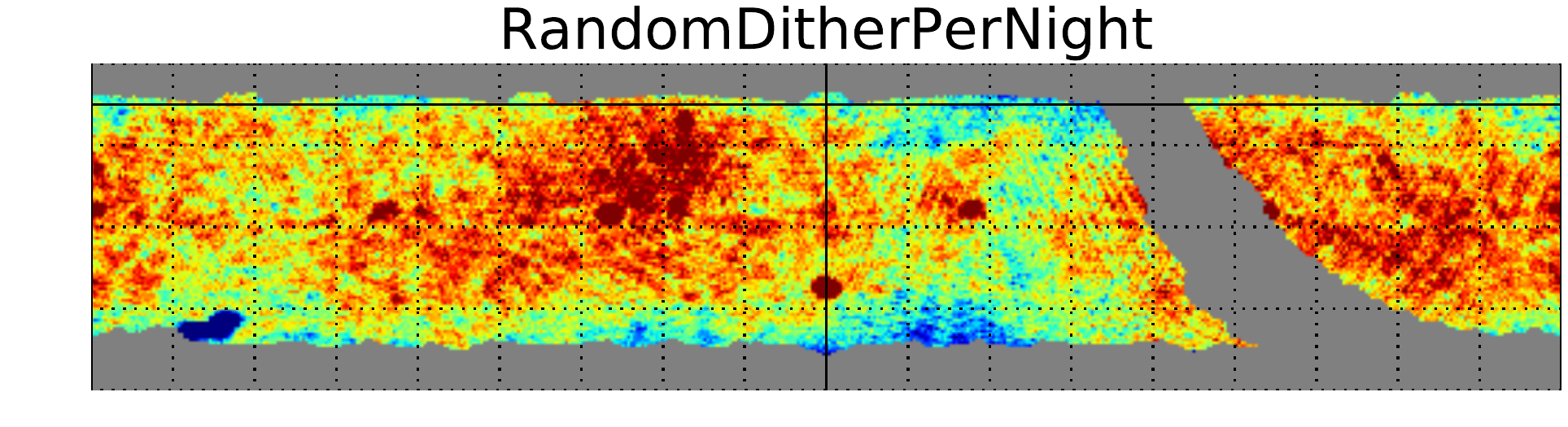}
	\end{minipage}\
	\hspace*{-1em}
	\begin{minipage}{0.43\paperwidth}
		\includegraphics[trim={-10 75 5 5},clip=true,width=.43\paperwidth]{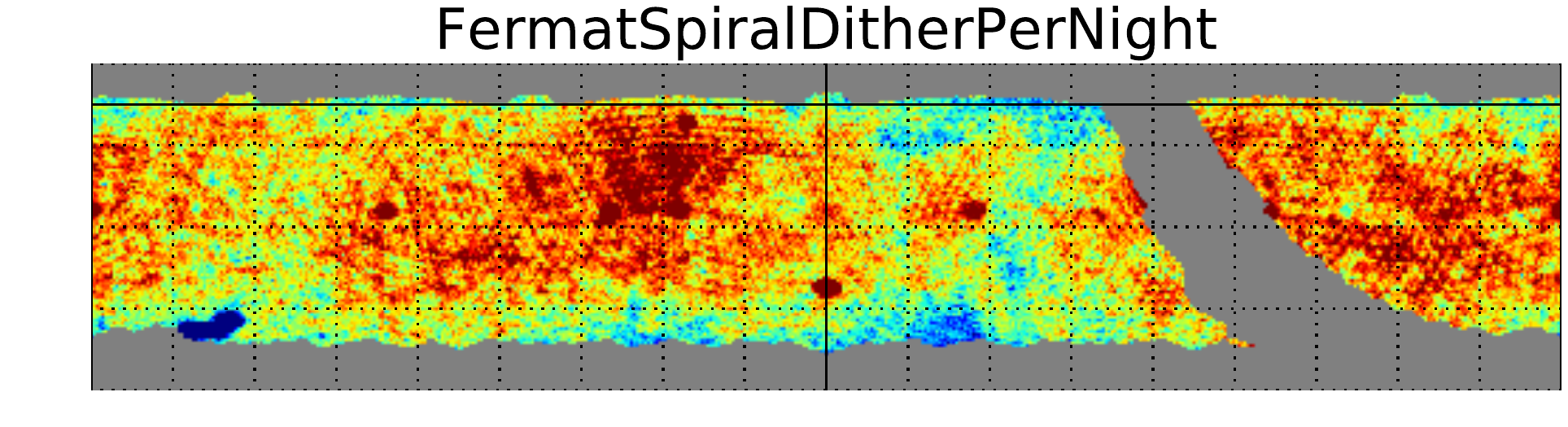}
	\end{minipage}\
	\hfill
	
	\hspace*{-2.5em}
	\vspace*{-0.5em} 
	\begin{minipage}{0.43\paperwidth}
		\includegraphics[trim={-10 75 5 5},clip=true,width=.43\paperwidth]{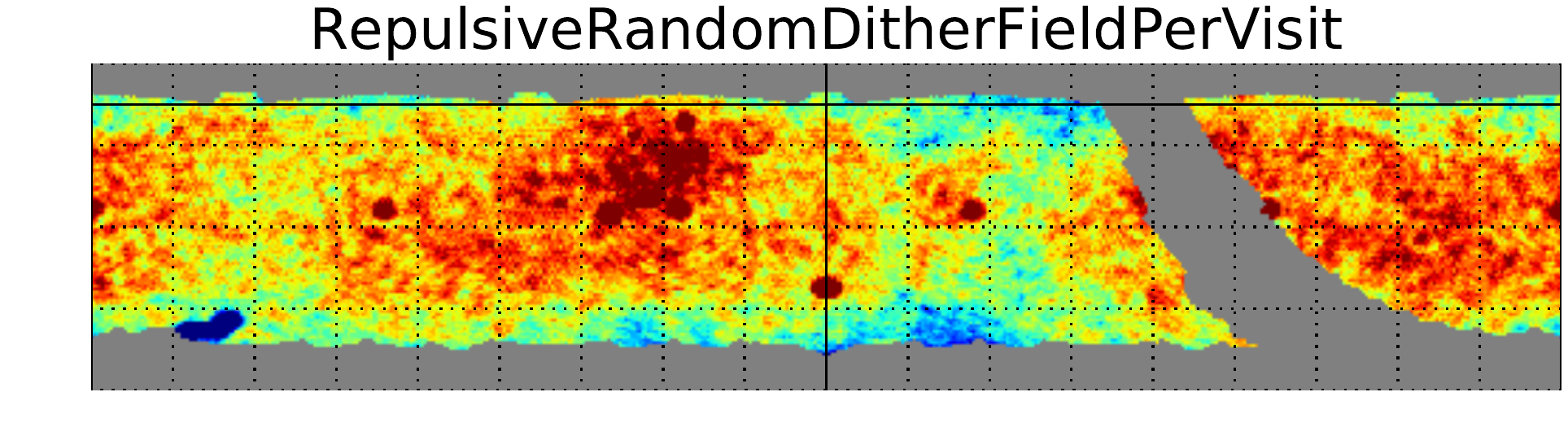}
	\end{minipage}\
	\hspace*{-1em}
	\begin{minipage}{0.43\paperwidth}
		\includegraphics[trim={-10 75 5 5},clip=true,width=.43\paperwidth]{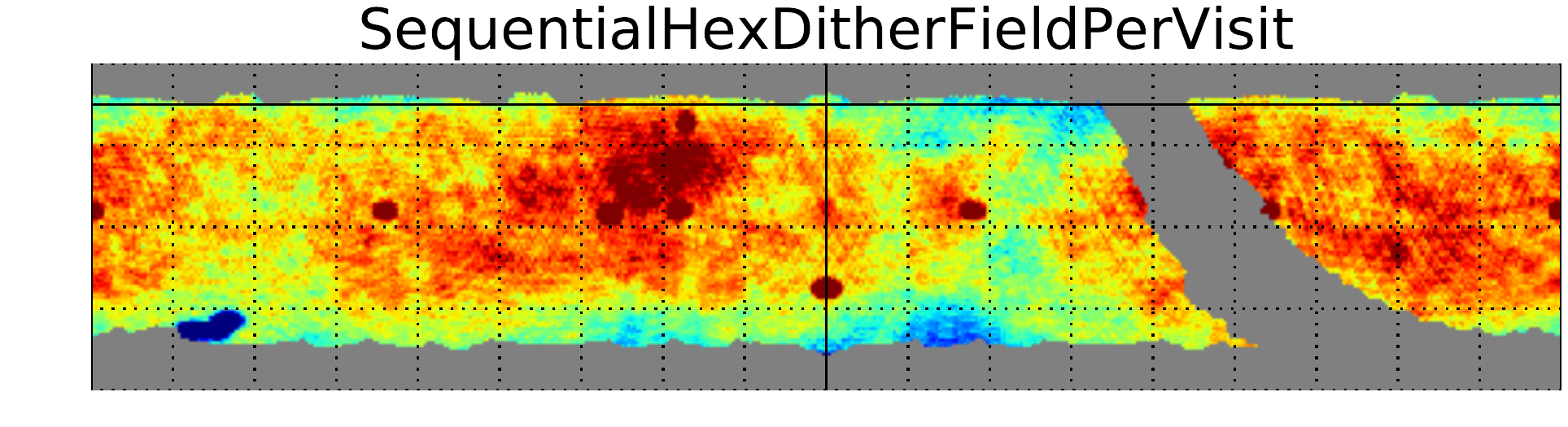}
	\end{minipage}\
	\hfill
	
	\hspace*{-2.5em}
	\vspace*{-0.5em} 
	\begin{minipage}{0.43\paperwidth}
		\includegraphics[trim={-10 75 5 5},clip=true,width=.43\paperwidth]{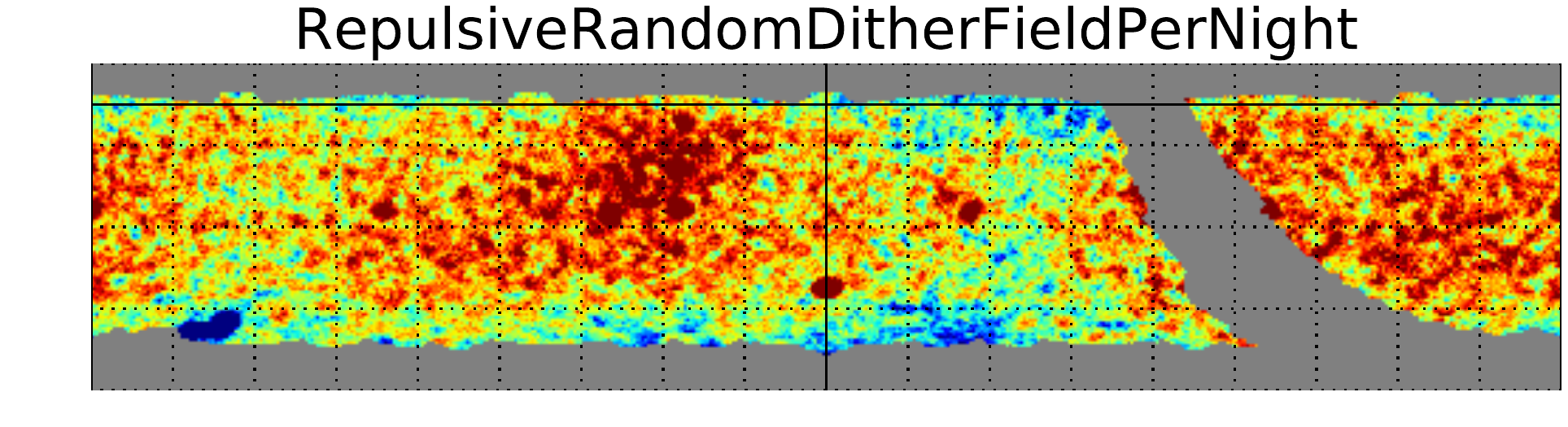}
	\end{minipage}\
	\hspace*{-1em}
	\begin{minipage}{0.43\paperwidth}
		\includegraphics[trim={-10 75 5 5},clip=true,width=.43\paperwidth]{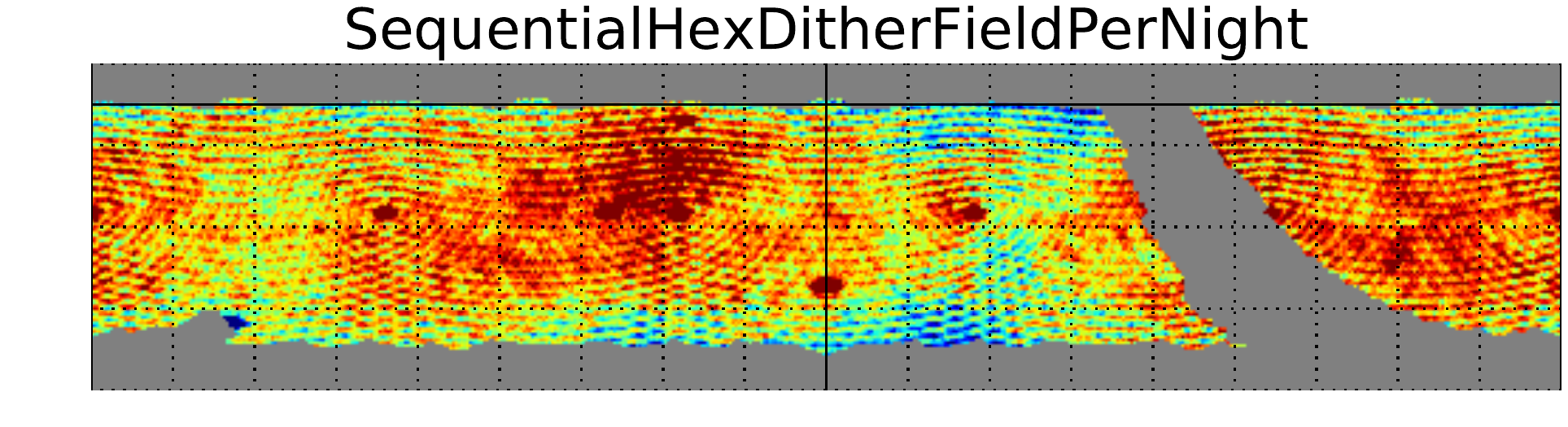}
	\end{minipage}\
	\hfill
	
	\hspace*{-2.5em}
	\vspace*{-0.5em} 
	\begin{minipage}{0.43\paperwidth}
		\includegraphics[trim={-10 75 5 5},clip=true,width=.43\paperwidth]{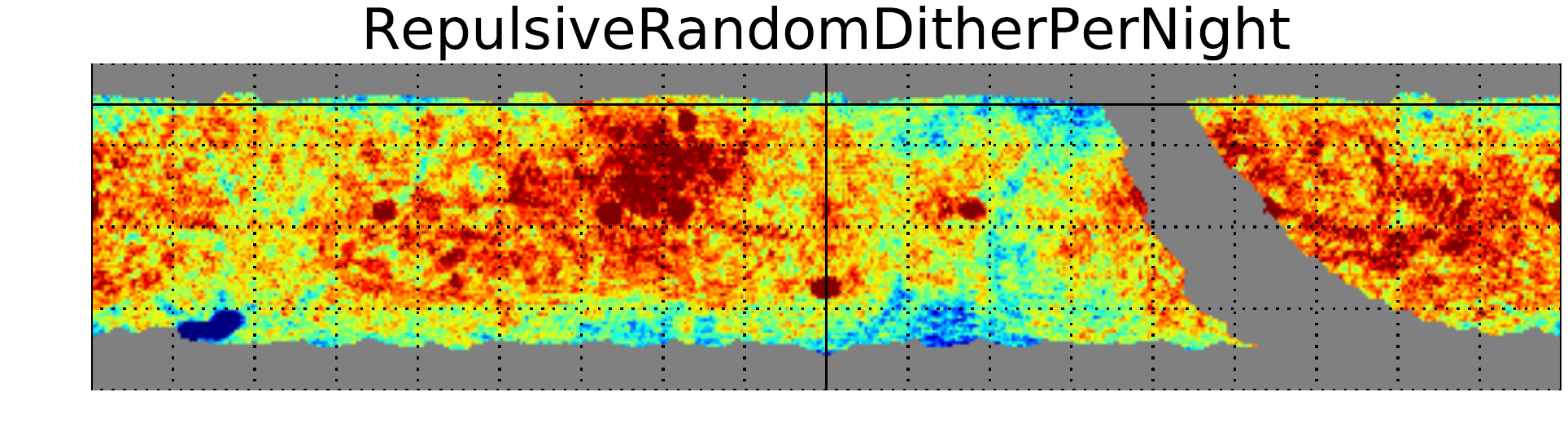}
	\end{minipage}\
	\hspace*{-1em}
	\begin{minipage}{0.43\paperwidth}
		\includegraphics[trim={-10 75 5 5},clip=true,width=.43\paperwidth]{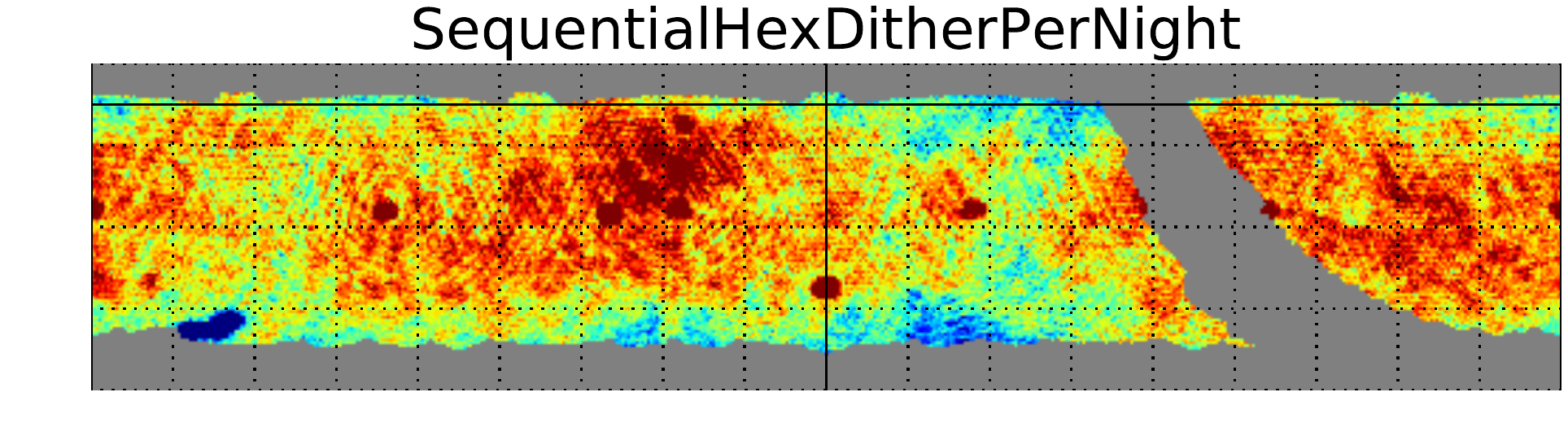}
	\end{minipage}\
	\hfill
		
	\hspace*{8em}
	\vspace*{-0.5em} 
	\begin{minipage}{0.5\paperwidth}
		\includegraphics[trim={-10 5 5 155},clip=true,width=.55\paperwidth]{f2c.pdf}
	\end{minipage}\
	\hspace*{2em}
	\figcaption[]{Plots for \cl{$r$-band coadded 5$\sigma$ depth } from various \cl{dither } strategies, after masking the \cl{shallow-depth } border. \cl{The top row shows the Mollweide projection for two observing strategies while the 2nd row shows the Cartesian projection restricted to $180^\circ$$>$RA$>$$-180^\circ$ (left-right), $-70^\circ$$<$Dec$<$10$^\circ$ (bottom-top); we only show the latter for the rest of the strategies. } We note that the strong honeycomb pattern present in the undithered survey is weaker in the dithered surveys, while for SequentialHexDitherFieldPerNight, we observe strong horizontal striping across the survey region. See Section~\ref{Coadd} for further details.\label{coaddSkymaps_noBorder}}
\end{figure*}
\subsection{Coadded 5$\sigma$ Depth}{\label{Coadd}}
In order to calculate the coadded depth, we use the modified 5$\sigma$ limiting magnitude data from OpSim, where the limiting magnitude is `modified' in order to represent a real point source detection depth \citep{Ivezic2008}. Assuming that the signal-to-noise ratio adds in quadrature, as it should for optimal weighting of individual images \citep[see \cl{e.g.,}][]{Gawiser2006}, we calculate the coadded depth, $\mathrm{5\sigma_{stack}}$, in each HEALPix pixel from the modified 5$\sigma$ limiting magnitude summed over individual observations, $\mathrm{5\sigma_{mod, \mathit{i}}}$:
\begin{equation}
	\mathrm{5\sigma_{stack}= 1.25 \: log_{10} \brac{ \sum_\mathit{i} {10^{0.8 \times 5\sigma_{mod, \mathit{i}}}}}}
	\label{eq: depth}
\end{equation}

We find that dithered surveys lead to shallower depth near the borders of the survey region, adding significant noise to the corresponding angular power spectra. In order to clean the spectra, we develop a border masking algorithm to discount pixels at the edges of the survey region, comprising nearly 15$\%$ of the survey area. See Appendix~\ref{appendix} for details of the masking algorithm. \

Figure~\ref{coaddSkymaps_noBorder} shows two projections for the \cl{$r$-band coadded 5$\sigma$ depth } for the various \cl{dither } strategies after the shallow border has been masked. \cl{The first row shows the Mollweide projection of the coadded depth for NoDither and PentagonDitherPerSeason, while the second row shows the corresponding Cartesian projection, zoomed on the LSST WFD survey area ($-180^\circ$$<$RA$<$$180^\circ$, $-70^\circ$$<$Dec$<$$10^\circ$). To conserve space, we show only the latter projection for the rest of the dither strategies}. We observe that the survey pointings without any dithering lead to deeper overlapping regions between the fields, and consequently a strong honeycomb pattern in the coadded depth. In contrast, the dithered skymaps have comparatively more uniform depth across the survey region, with smaller-scale variations amongst the \cl{dither } strategies.\

Here we note that although dithering in general weakens the honeycomb pattern seen in the undithered survey, we observe horizontal striping from SequentialHexDitherFieldPerNight; in contrast, SequentialHexDitherPerNight and SequentialHexDitherFieldPerVisit show no such behavior. This is an example where a specific \cl{dither } strategy's behavior is highly dependent on the timescale on which it is implemented: for PerNight timescale, a new dither is assigned to all fields every night, implying that the 217-point lattice is traversed multiple times during the $\sim$3650-night survey. Similarly, for PerVisit timescale, although a new dither is assigned to each field every time it is visited, the lattice is traversed multiple times given that every field is visited $\sim$150 times in the $r$-band throughout the survey. In contrast, for FieldPerNight timescale, a new dither point is assigned to each field only when it is observed a new night. Since a given field is only visited on $\sim$50 nights in a given filter, only the lower part of the lattice is traversed (as the lattice is traversed starting from bottom left), leading to horizontal striping. We verified this conclusion by rotating the hexagonal lattice by 90$^\circ$, and observing vertical striping for FieldPerNight timescale. \

\begin{figure}
	\centering
	\includegraphics[trim={25 40 65 50},clip=true,width=.4\paperwidth]{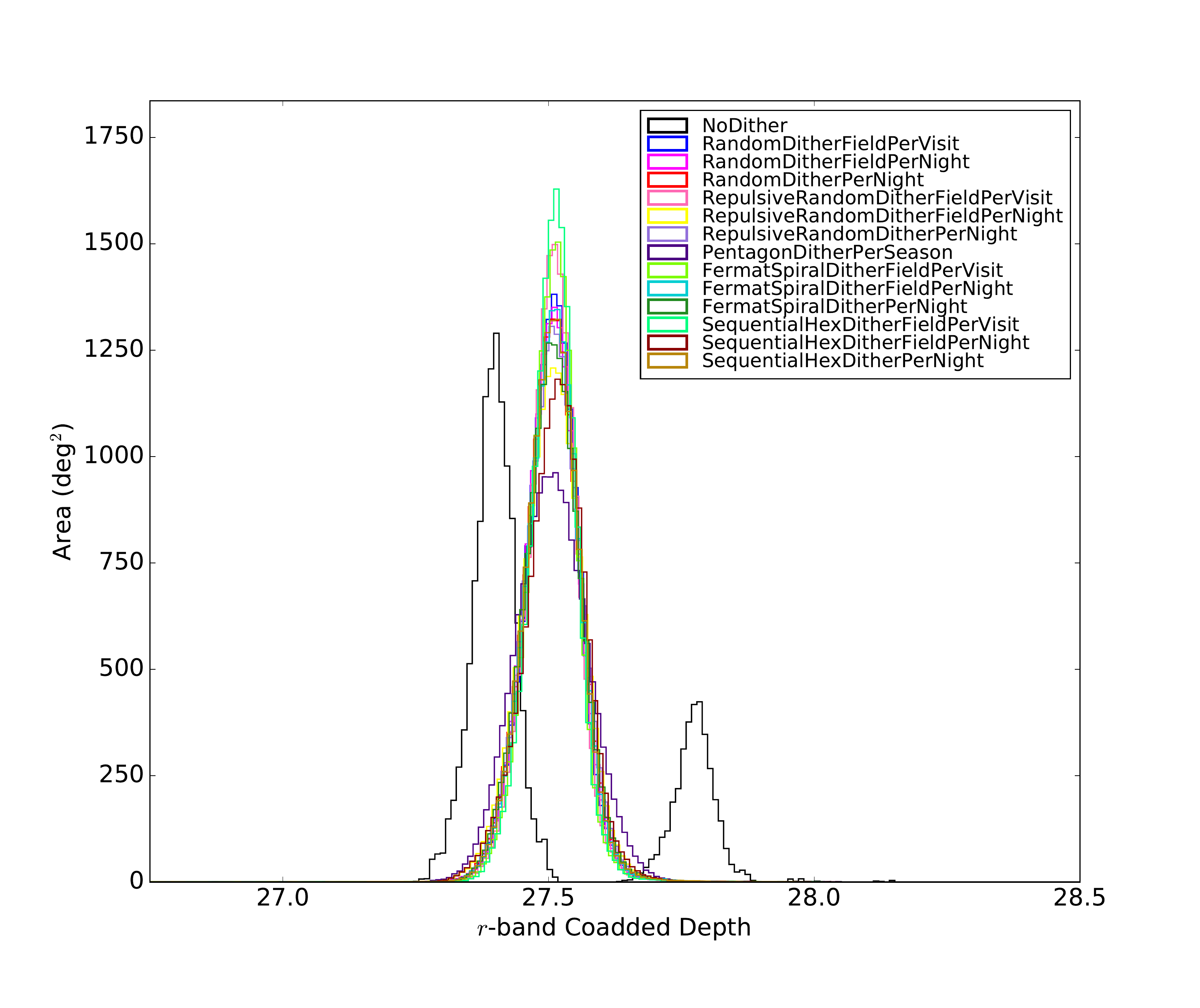}
	\figcaption[]{Histogram for the $r$-band coadded \cl{5$\sigma$ } depth, indicating a bimodal distribution from the undithered survey, and unimodal distributions from the dithered ones. \label{coadd_hist}}
\end{figure}

In Figure~\ref{coadd_hist}, we show a histogram of the $r$-band coadded depth. We see that the undithered survey leads to a bimodal distribution, with the overlapped regions observed much deeper than the rest of the survey. On the other hand, all the dithered surveys lead to unimodal distributions, as dithering leads to observing the data more uniformly, in agreement with \citet{Carroll2014}. \

\begin{figure*}
	\hspace*{4em}
	\includegraphics[trim={90 220 90 205},clip=true,width=.7\paperwidth]{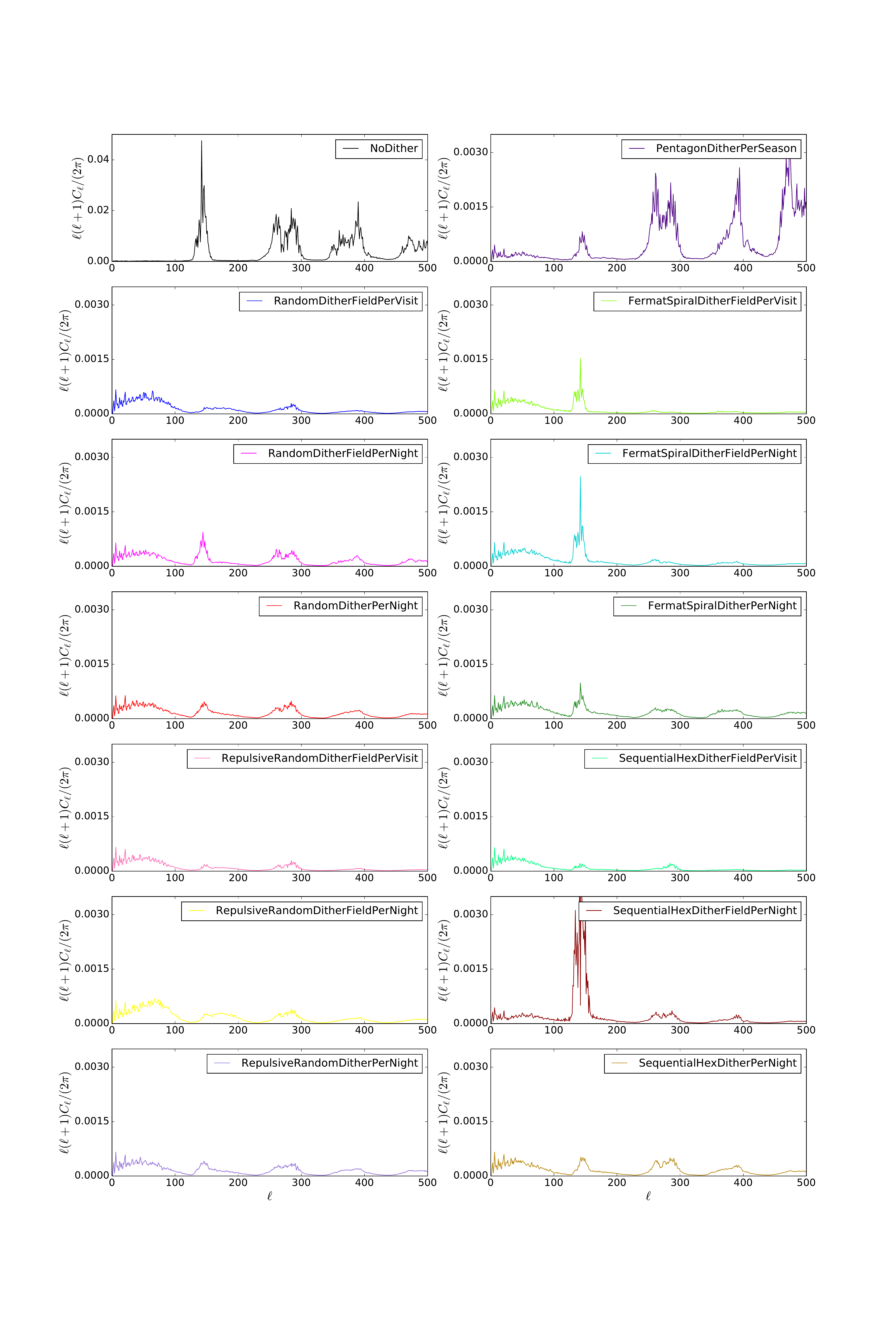}
	\figcaption[]{Angular power spectra for the $r$-band coadded depth for all the \cl{dither } strategies. We note that dithering reduces the angular power by at least a factor of 10 as compared to NoDither. The honeycomb pattern in the undithered survey generates a large peak around $\ell$$\sim$150, while dithering of all kinds decreases the spurious power. The horizontal striping in SequentialHexDitherFieldPerNight also creates a moderate peak around $\ell$$\sim$150. \label{coadd_power}}
\end{figure*}

In order to quantify the angular characteristics reflected in the skymaps, we measure the power spectrum associated with each of the skymaps. Figure~\ref{coadd_power} shows the power spectra for the coadded depth from each of the \cl{dither } strategies considered here\cl{; we have removed the monopole and dipole using HEALPix routine remove\_dipole}. We note that the spectrum corresponding to the undithered survey has a very large peak around $\ell$$\sim$150, resulting from the strong honeycomb pattern. In comparison, we see over 10 times less power  in the dithered surveys; the $\ell$$\sim$150 peak in these surveys is much more comparable to the rest of the spectrum.  \cl{More specifically, we find that the FieldPerVisit timescale is the most effective in reducing the power for a given dither geometry, while Random and RepulsiveRandom dithers perform well on all three timescales. Also, we }confirm the origins of the  $\ell$$\sim$150 peak by creating a pure honeycomb, and observing a power spectrum similar to that from the undithered survey.\

Furthermore, we see that the horizontal striping in the SequentialHexDitherFieldPerNight skymap generates a large peak around $\ell$$\sim$150, while the rest of the dithered spectra do not exhibit such a strong peak. Curiously, the PentagonDitherPerSeason strategy leads to two large peaks around $\ell$$\sim$270 -- a characteristic different from the rest of the \cl{dither } strategies' but similar to NoDither, with much less power. \

\begin{figure*}
	\vspace*{1em}
	\hspace*{-1.em}
	\begin{minipage}{0.27\paperwidth}
		\includegraphics[trim={-2 -5 2 4},clip=true,width=.28\paperwidth]{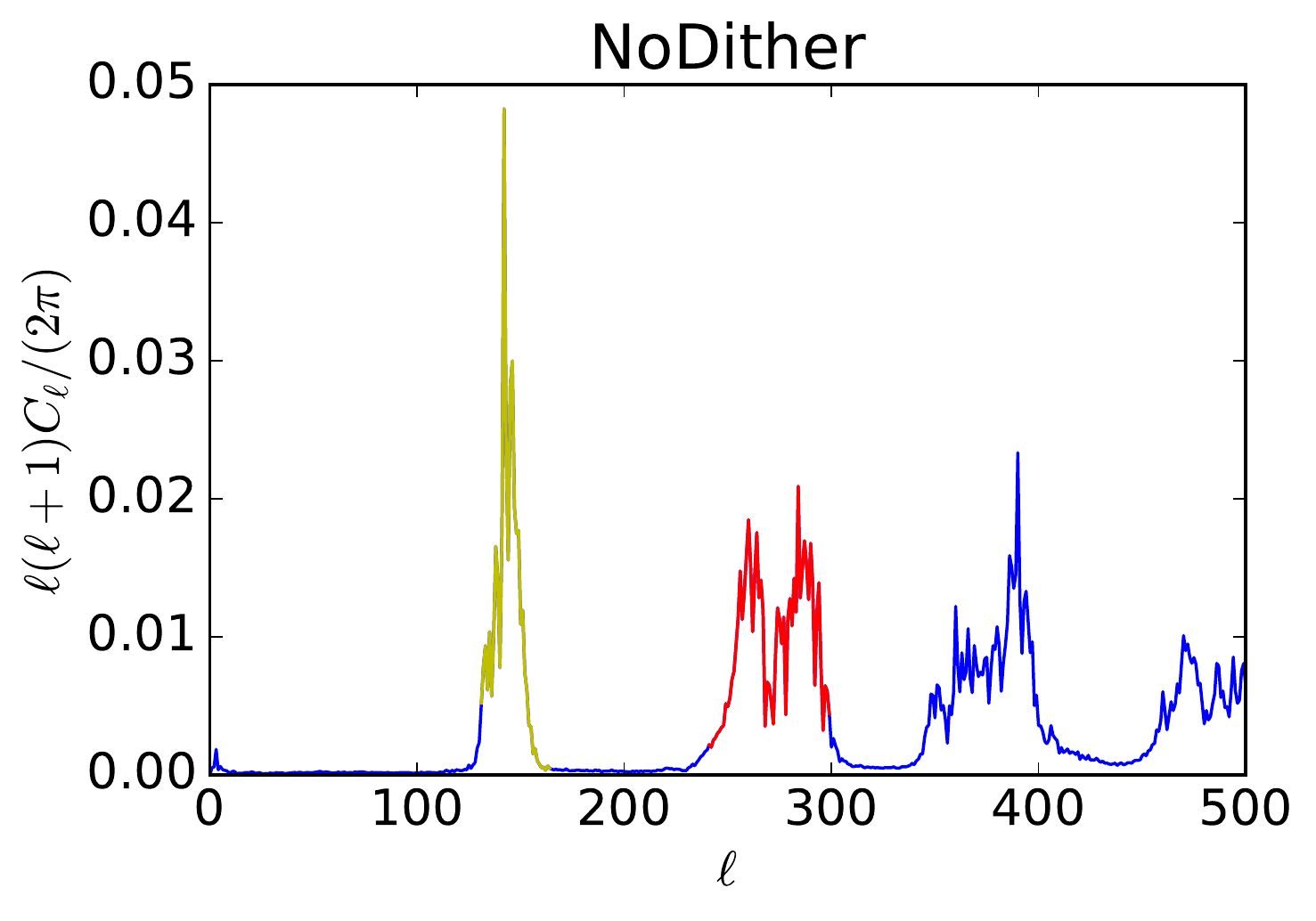}
	\end{minipage}\
	\hspace*{0.25em}
	\begin{minipage}{0.27\paperwidth}
		\includegraphics[trim={27 -5 2 4},clip=true,width=.28\paperwidth]{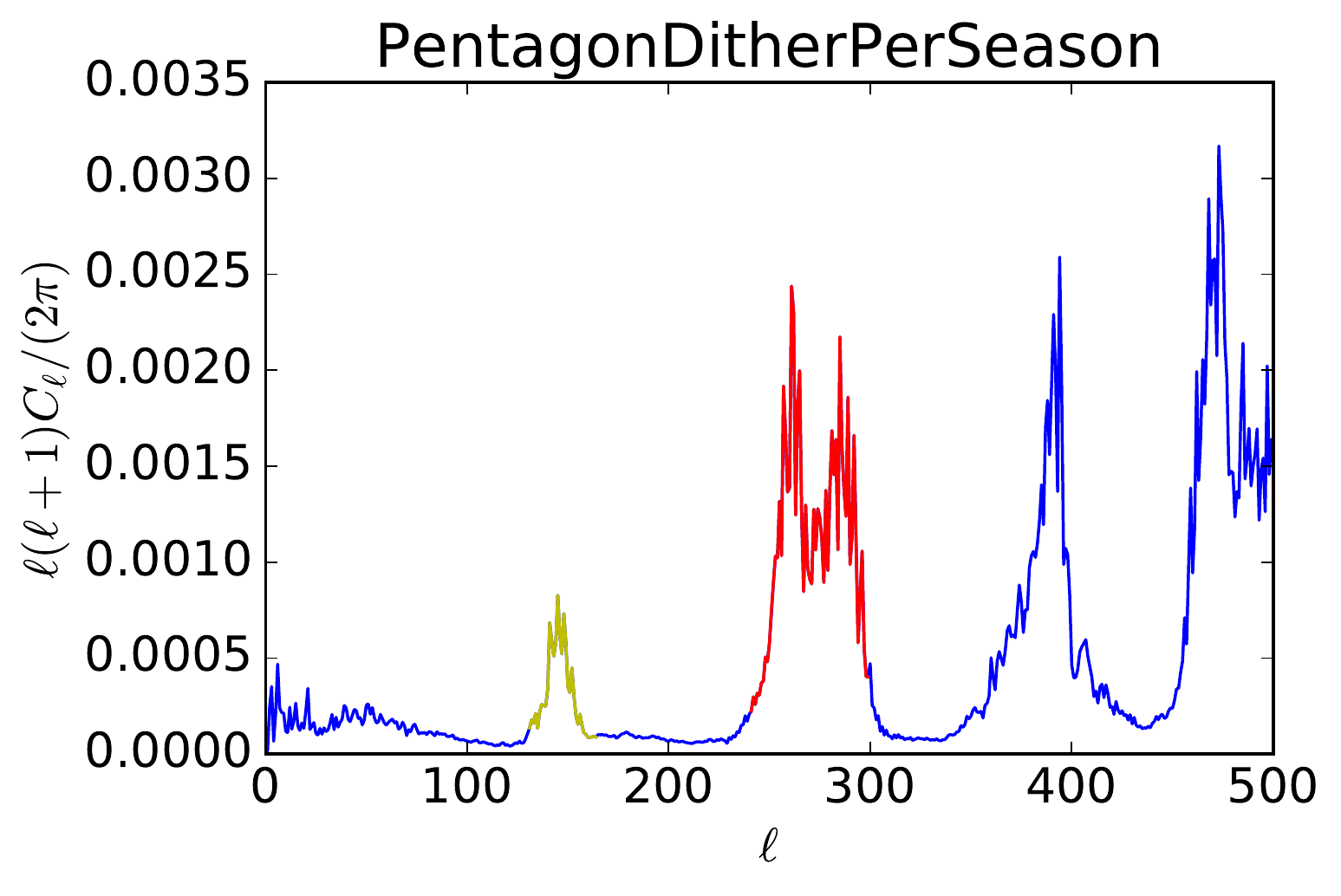}
	\end{minipage}\
	\hspace*{0.25em}
	\begin{minipage}{0.27\paperwidth}
		\includegraphics[trim={27 -5 2 4},clip=true,width=.28\paperwidth]{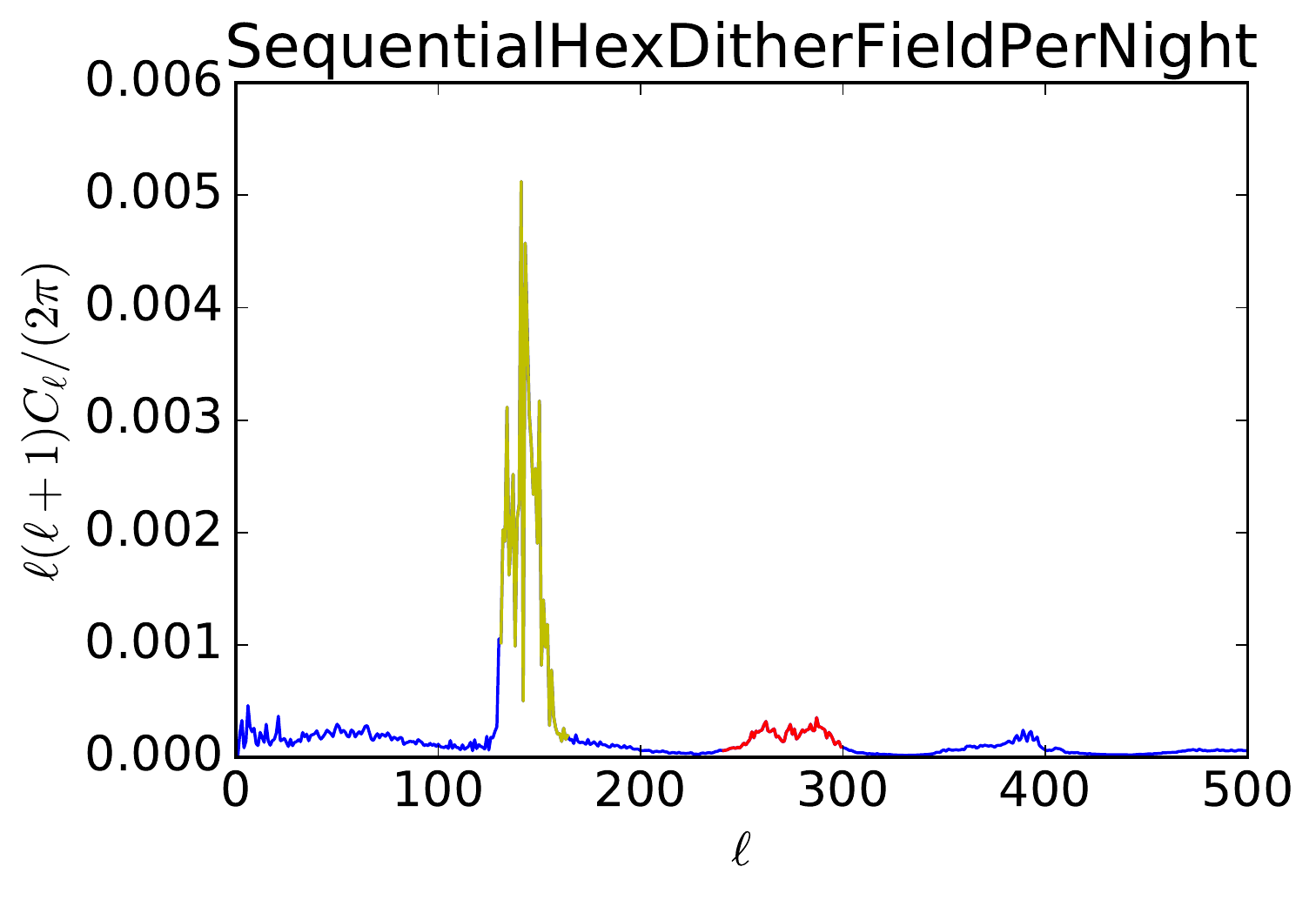}
	\end{minipage}\
	\hspace*{0em}
	\hfill
	
	\vspace*{-.5em}
	\hspace*{0.3em}
	\begin{minipage}{0.27\paperwidth}
		\includegraphics[trim={-2 -5 2 4},clip=true,width=.28\paperwidth]{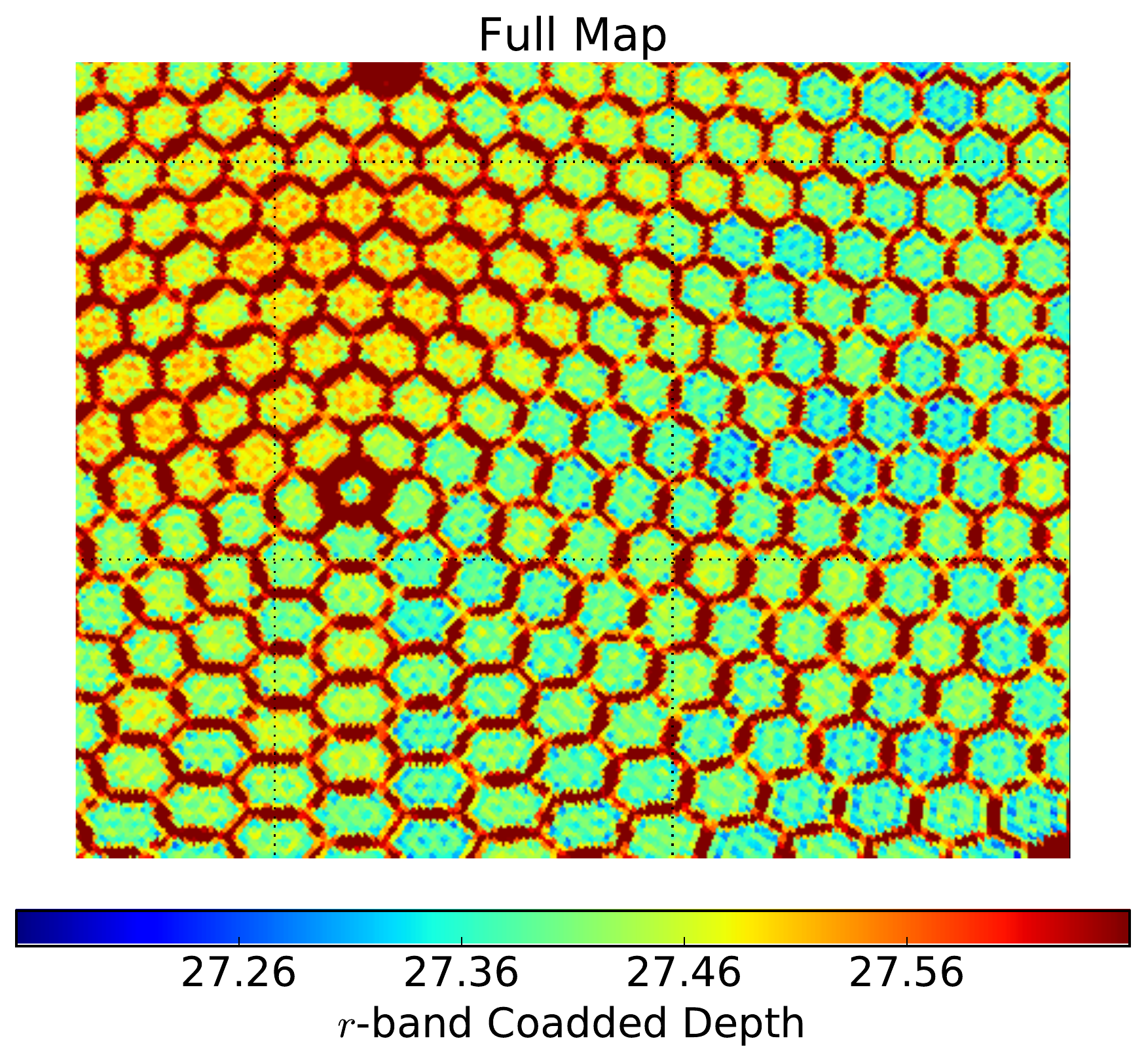}
	\end{minipage}\
	\hspace*{-0.1em}
	\begin{minipage}{0.27\paperwidth}
		\includegraphics[trim={-2 -5 2 4},clip=true,width=.28\paperwidth]{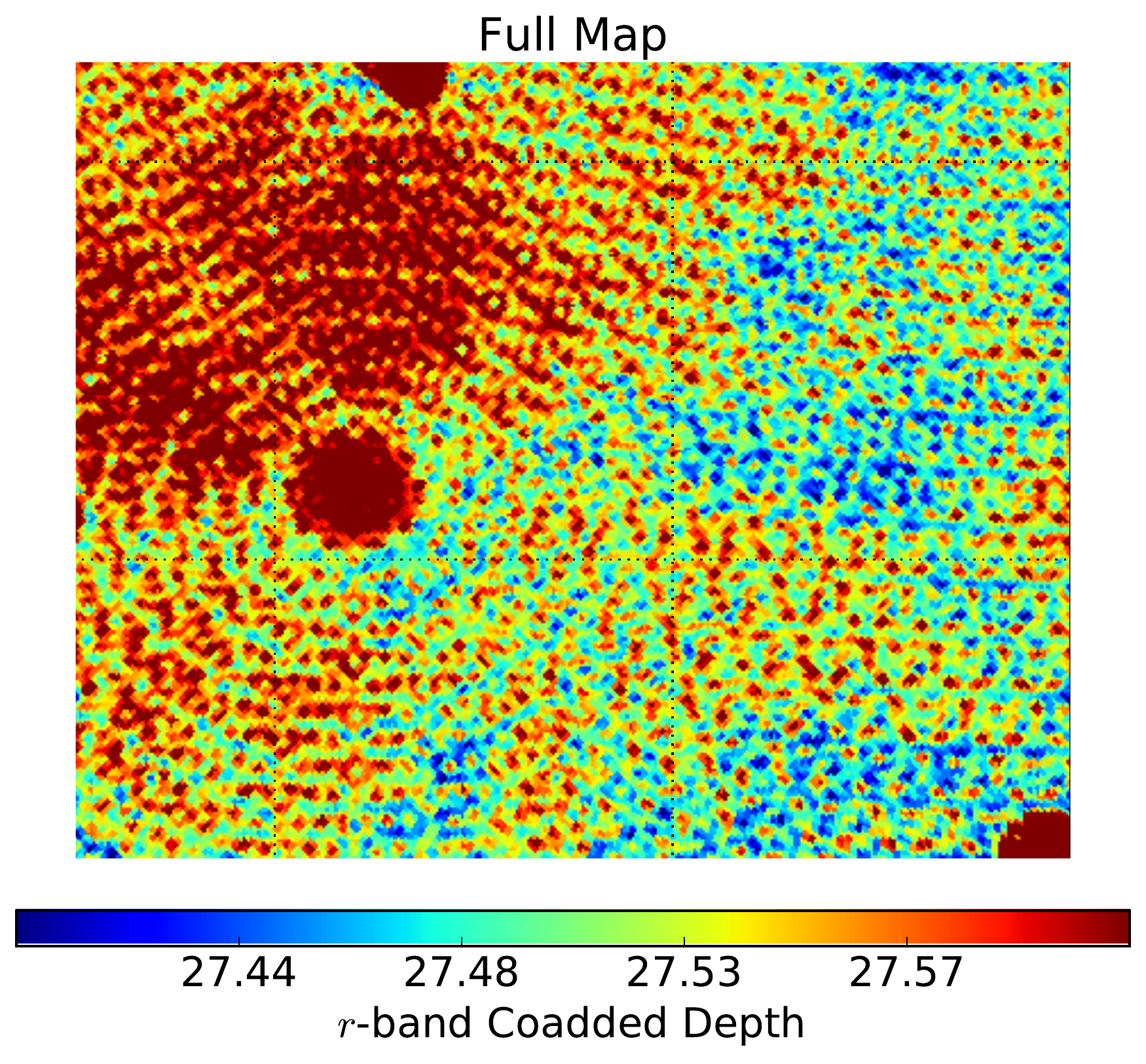}
	\end{minipage}\
	\hspace*{-0.1em}
	\begin{minipage}{0.27\paperwidth}
		\includegraphics[trim={-2 -5 2 4},clip=true,width=.29\paperwidth]{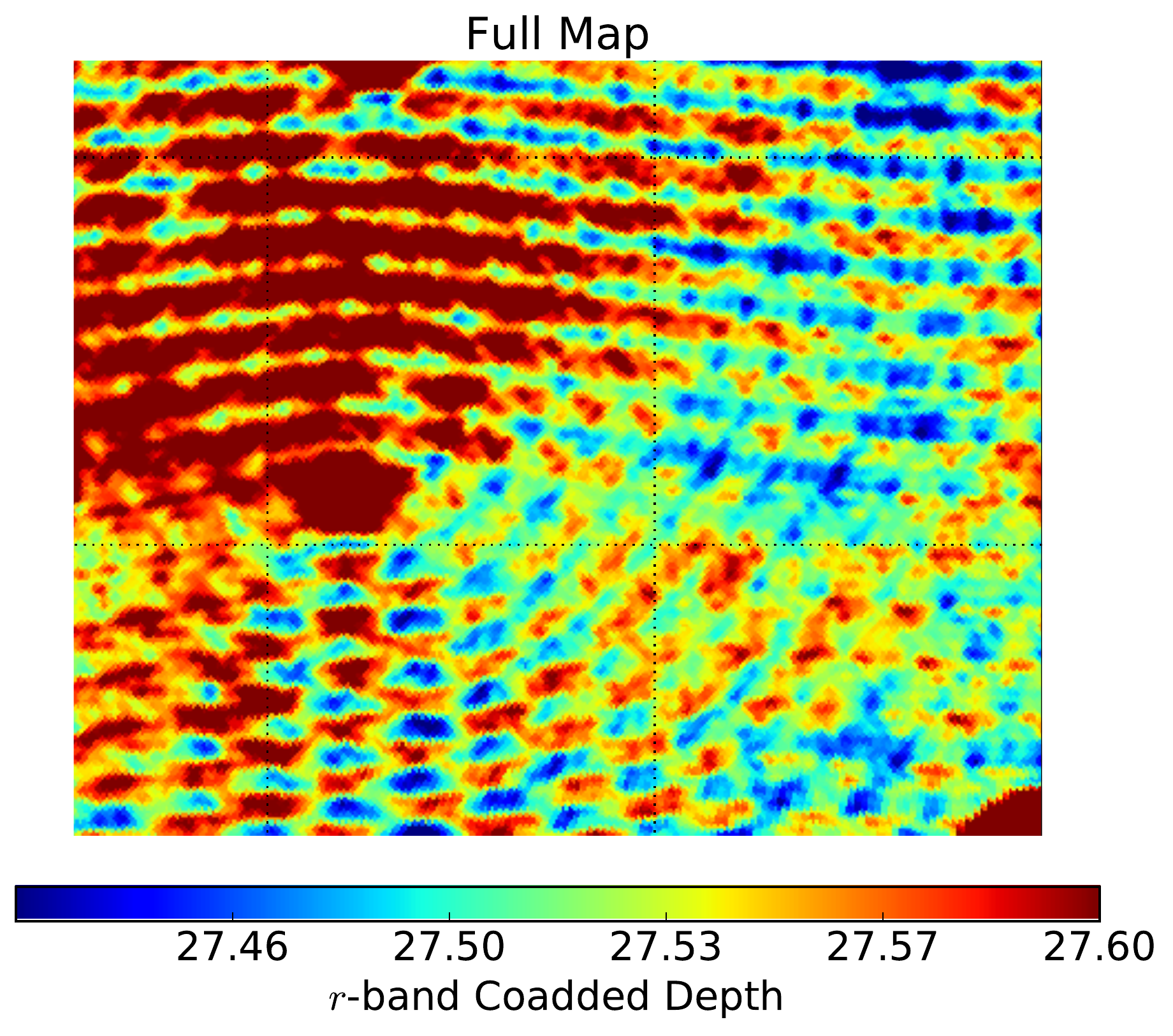}
	\end{minipage}\
	\hspace*{-5em}
	\hfill
		
	\hspace*{0.3em}
	\begin{minipage}{0.27\paperwidth}
		\includegraphics[trim={-2 75 2 4},clip=true,width=.28\paperwidth]{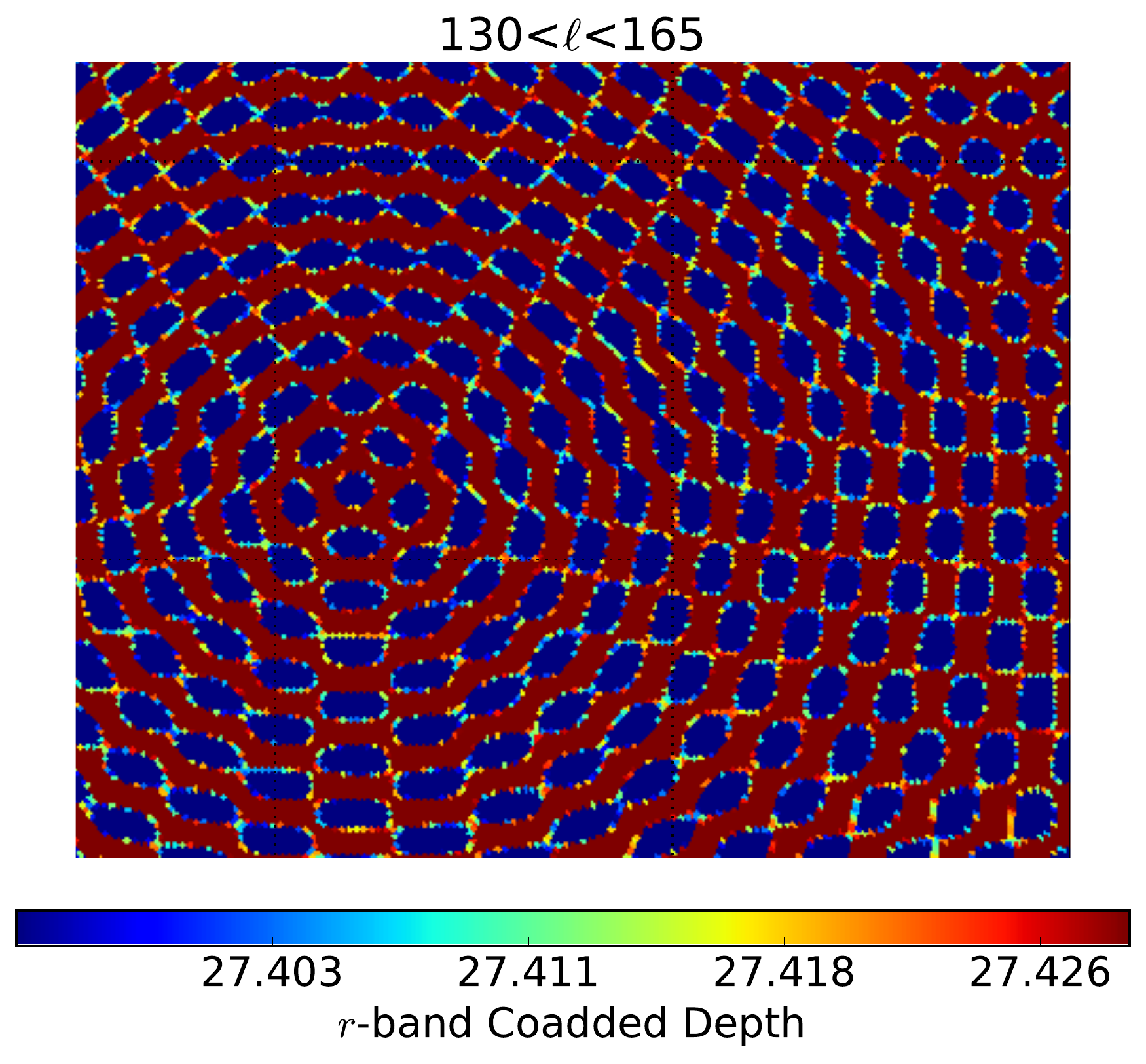}
	\end{minipage}\
	\hspace*{-0.1em}
	\begin{minipage}{0.27\paperwidth}
		\includegraphics[trim={-2 75 2 4},clip=true,width=.28\paperwidth]{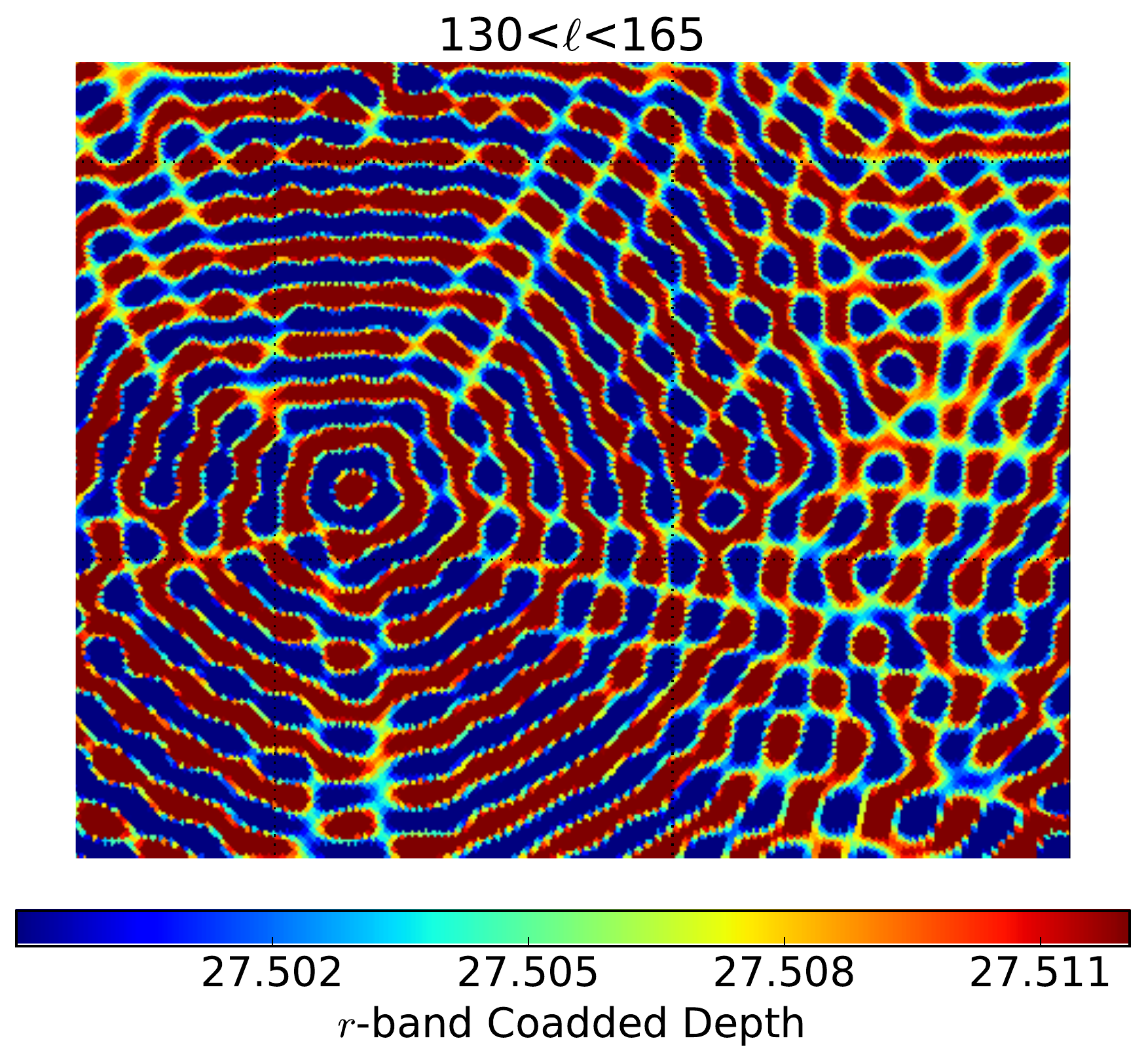}
	\end{minipage}\
	\hspace*{-0.1em}
	\begin{minipage}{0.27\paperwidth}
		\includegraphics[trim={-2 75 2 4},clip=true,width=.28\paperwidth]{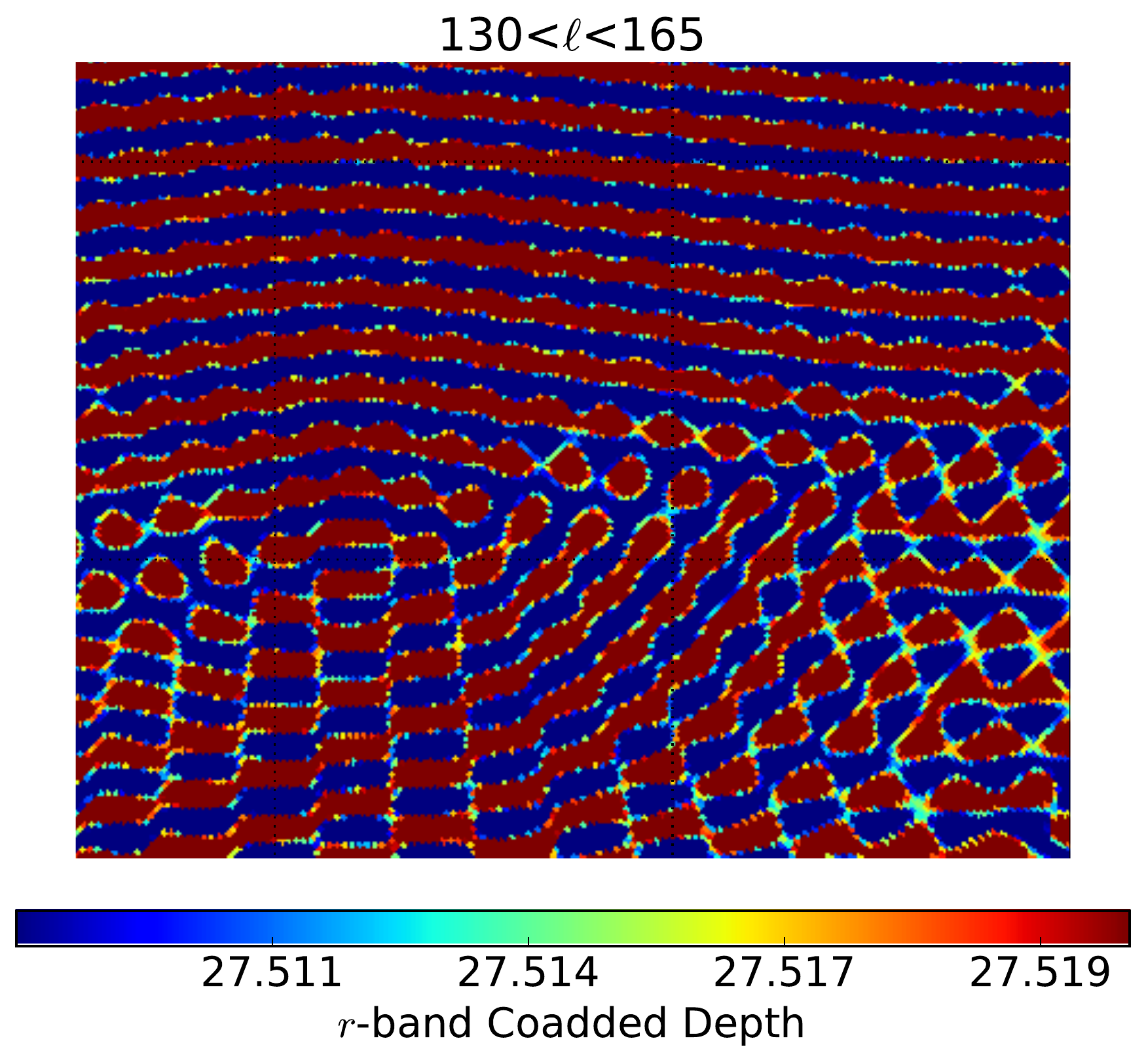}
	\end{minipage}
	\hspace*{2em}
	\hfill

	\vspace*{-0.5em}
	\hspace*{0.3em}
	\begin{minipage}{0.27\paperwidth}
		\includegraphics[trim={-2 -5 2 4},clip=true,width=.28\paperwidth]{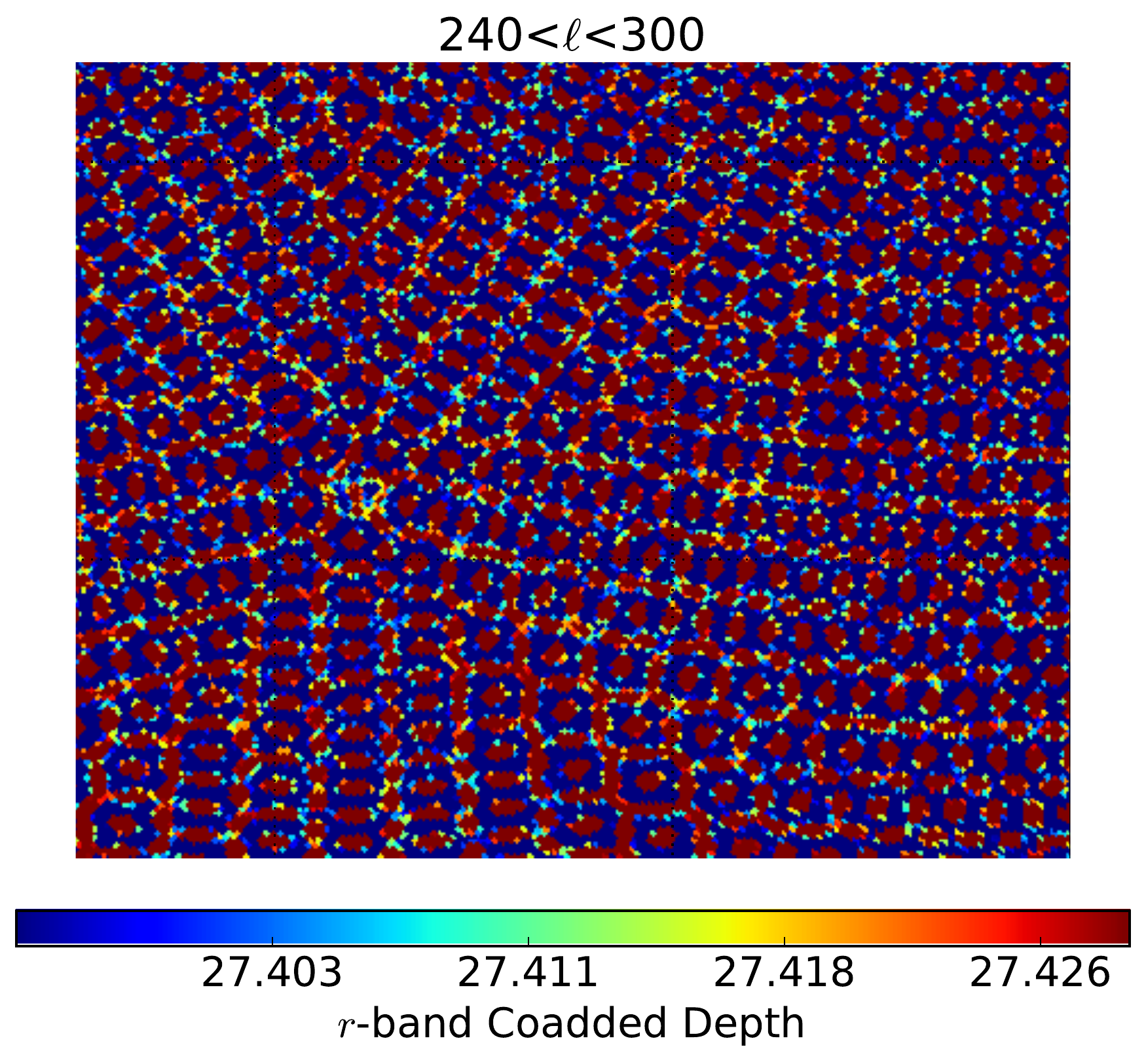}
	\end{minipage}\
	\hspace*{-0.1em}
	\begin{minipage}{0.27\paperwidth}
		\includegraphics[trim={-2 -5 2 4},clip=true,width=.28\paperwidth]{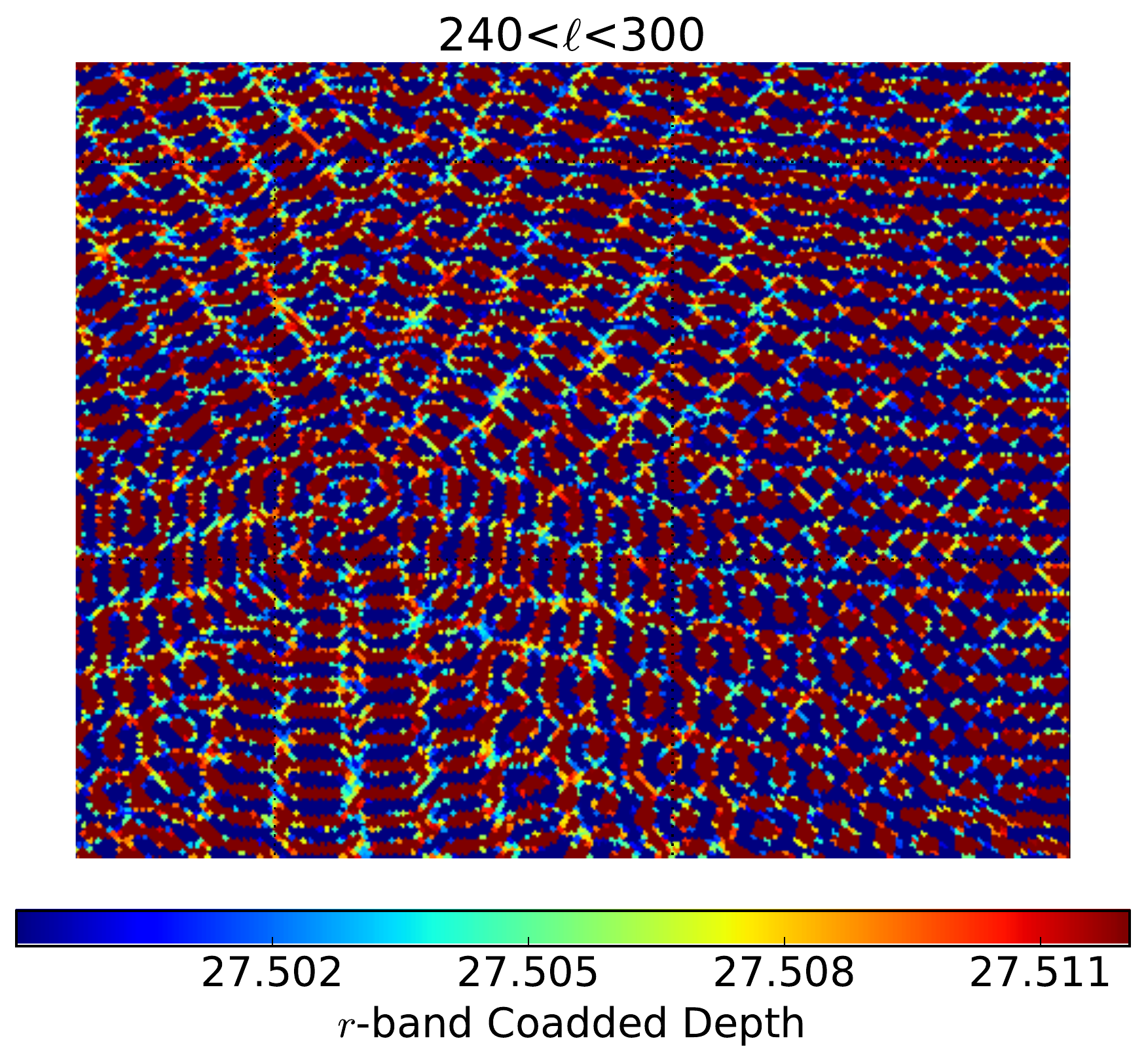}
	\end{minipage}\
	\hspace*{-0.1em}
	\begin{minipage}{0.27\paperwidth}
		\includegraphics[trim={-2 -5 2 4},clip=true,width=.28\paperwidth]{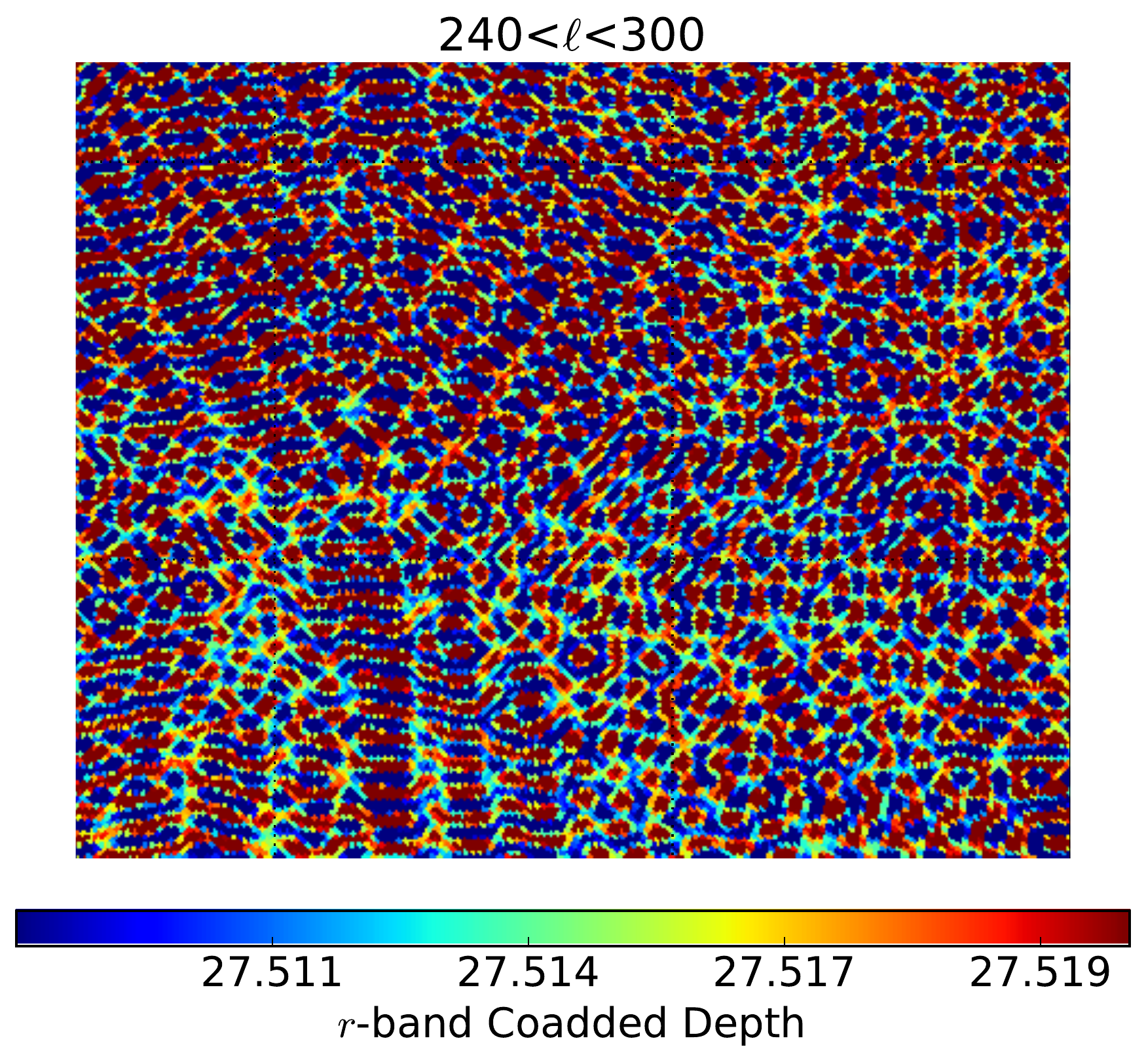}
	\end{minipage}\
	\hspace*{2em}
	\vspace*{1em}
	\figcaption[]{a$_{\ell m}$ analysis plots for two $\ell$-ranges in the $r$-band coadded depth power spectra (colored peaks in the \cl{top } row). The first row shows the full power spectrum for three \cl{observing } strategies; the second row shows the corresponding skymaps for \cl{$50^\circ$$>$RA$>$$0^\circ$ (left-right), $-45^\circ$$<$Dec$<$$-5^\circ$ (bottom-top)}. The third row is for 130$<$$\ell$$<$165 (yellow in the power spectra in the first row),\cl{ and the fourth is for 240$<$$\ell$$<$300 (red in the top row), all in the same RA, Dec range as the second row}. The leftmost column corresponds to NoDither, the middle one to PentagonDitherPerSeason, and the right one to SequentialHexDitherFieldPerNight. We see that the honeycomb pattern in the undithered survey and the horizontal striping in SequentialHex generates the $\ell$$\sim$150 peak. \cl{Also, we see one (partial) Deep Drilling Field at the top as well as a pentagonal tile at Dec$= -30^\circ$ resulting from the tiling of the sphere, both of which are smeared out by dithering. } \label{almAnalysis}}
\end{figure*}

To understand the origins of the characteristic patterns in the skymaps, we consider the a$_{\ell m}$ coefficients  of their spherical harmonic transforms. This allows us to produce the skymaps corresponding to specific ranges of the angular scale $\ell$. We show our results in Figure~\ref{almAnalysis} for NoDither, PentagonDitherPerSeason and SequentialHexDitherFieldPerNight strategies. The top row includes the full power spectrum for each strategy, and the second row shows the corresponding \cl{Cartesian projection for $0^\circ$$<$RA$<$$50^\circ$, $-45^\circ$$<$Dec$<$$-5^\circ$}. The third and fourth rows show the partial skymaps arising from each of the colored peaks shown in the power spectra in the top row. We observe that for the undithered survey, the $\ell$$\sim$150 peak arises from the strong honeycomb pattern, while the second peak arises from structure on the small angular scales. For PentagonDitherPerSeason, we see a milder honeycomb for the $\ell$$\sim$150 peak, while the 240$<$$\ell$$<$300 peak arises from structure similar to the corresponding one in the undithered survey. Finally, for SequentialHexDitherFieldPerNight, we can see the source of the strong $\ell$$\sim$150 peak: the horizontal striping.  For higher-$\ell$ peaks, we note the weaker structure as compared to the other two strategies. We also performed this a$_{\ell m}$ analysis individually for the two peaks in 240$<$$\ell$$<$300 and found the underlying structure to be very similar.

\subsection{Artificial Galaxy Fluctuations}{\label{LSS}}
Given our knowledge of the characteristics induced in the coadded depth due to the observing strategy, we now consider the effects of these artifacts on BAO studies. We model the artificial fluctuations in galaxy counts, accounting for photometric calibration errors, dust extinction, and galaxy catalog magnitude cuts. Since BAO studies are redshift dependent, we consider five redshift bins: 0.15$<$$z$$<$0.37, 0.37$<$$z$$<$0.66, 0.66$<$$z$$<$1.0, 1.0$<$$z$$<$1.5, and 1.5$<$$z$$<$2.0. \

\cl{We first estimate the number of galaxies in specific redshift bins detected in each pixel at a particular depth using a mock LSST catalog, which is constructed using the outputs of the {\small{SAG}} semi-analytic model for galaxy formation \citep{Cora2006, Lagos2008, Tecce2010, Orsi2014, Gargiulo2015, Munoz2015}. The model incorporates differential equations for gas cooling, quiescent star formation, energetic and chemical supernova feedback, the growth of a supermassive black hole, the associated AGN feedback, bursty star formation in mergers and disk instabilities, all coupled to the merger trees extracted from a dark matter simulation run with {\small{GADGET2}} \citep{Springel2005} assuming the standard $\Lambda$CDM model \citep{Jarosik2011}. The subhalo populations of merger trees are found using  {\small{SUBFIND}} \citep{Springel2001}  after the DM haloes were identified using a friends-of-friends algorithm.} \

We normalize the total $r$-band galaxy counts to the empirical \cl{cumulative } galaxy count estimates for LSST \citep[see][Section 3.7.2 for details]{LSST2009} at a \cl{magnitude cut $r$$<$25.9 (corresponding to the CFHTLS Deep survey completeness limit of $i$$<$25.5; see \citealt{Hoekstra2006, Gwyn2008} for details)}.  In contrast with \citet{Carroll2014}, where Fleming's function \citep{Fleming1995} was used to account for the incompleteness near the 5$\sigma$ limit, we use an erfc function. When multiplied by power-law number counts, Fleming's function causes completeness to drop to 20$\%$ of its peak at $r$$\sim$30 before rising again, while the erfc incompleteness function correctly damps down for higher magnitudes. We calculate the number of galaxies, $\mathrm{N_{gal}}$, in each HEALPix pixel in a given redshift bin as
\begin{equation}
	\mathrm{N_{gal}= 0.5\int_{-\infty}^{m_{max}} {erfc[}}a\mathrm{{(m-5\sigma_{stack})]} 10^{c_1m + c_2}dm}
	\label{eq: erfc}
\end{equation}
where $a$ is the rollover speed and is chosen to be 1, $\mathrm{5\sigma_{stack}}$ is the coadded magnitude depth in the given HEALPix pixel,  $\mathrm{m_{max}}$ is the magnitude cut, and c$_1$ and c$_2$ are the power-law constants determined from the mock catalogs for specific redshift bins. Here, we assume galaxies to have average colors, i.e. $u-g= g-r= r-i= 0.4$, and take this into account by modifying $\mathrm{c_2}$ and $\mathrm{m_{max}}$ in equation~\ref{eq: erfc} for $u$, $g$, $i$ vs. $r$. Given the sharp decline of the erfc function at high magnitudes and the consequent decline in the differential galaxy counts, we consider a magnitude limit of $r$= 32.0 as no magnitude limit. \

Using the number of galaxies in each pixel, we calculate the fluctuations in the galaxy counts  $\Delta$N/$\rm{\overline{N}}$ as ($\mathrm{N_{gal}/{N_{avg}})-1}$, where $\mathrm{N_{avg}}$ is the average number of galaxies per pixel across the survey area. Within MAF, this procedure amounts to using a metric to calculate the number of galaxies and then post-processing the galaxy counts to find $\Delta$N/$\rm{\overline{N}}$. \

Here we note that  artificial fluctuations in galaxy counts induced by the observing strategy (OS) scale the fluctuations arising due to actual LSS. In our calculations, we assume that LSS affects the local normalization of the galaxy luminosity function in a given redshift bin, not its shape. This assumption \cl{is } valid as long as LSS does not alter the shape of the faint end of the luminosity function, which dominates the galaxy number counts. More precisely, in the $i$th pixel,
\begin{equation}
	\mathrm{ \left(\frac{N_{gal}}{N_{avg}} \right )_{observed, \mathit{i}} = \left(\frac{N_{gal}}{N_{avg}} \right )_{OS,\mathit{i}}  \left(\frac{N_{gal}}{N_{avg}} \right )_{LSS,\mathit{i}}  }
\end{equation}
Defining $\delta_\mathit{i}$= $\Delta$N$_\mathit{i}$/$\rm{\overline{N}}$= ($\mathrm{N_{gal,\mathit{i}}/{N_{avg}})-1}$, we have
\begin{equation}
	\mathrm{ (1+\delta_{observed,\mathit{i}}) = (1+\delta_{OS,\mathit{i}})(1+\delta_{LSS,\mathit{i}})}
	\label{eq: delta_obs}
\end{equation}
\cl{Since the ensemble average of LSS is zero, we have
\begin{equation}
\begin{split}
	\mathrm{ \ev{\delta_{observed, \mathit{i}}}} &= \mathrm{\ev{\delta_{OS, \mathit{i}}  } + \ev{\delta_{LSS, \mathit{i}} } + \ev{\delta_{OS, \mathit{i}} \delta_{LSS, \mathit{i}}}} \\
	&=\mathrm{{\delta_{OS, \mathit{i}}} + \ev{\delta_{OS,\mathit{i}} \delta_{LSS,\mathit{i}}}}
\end{split}
\label{eq: <delta_obs>}
\end{equation}
where the angular brackets $\ev{..}$ indicate an ensemble average defined as an average over many realizations of the Universe with one LSST survey. Hence, we have $\ev{\delta_{\rm{OS,} \mathit{i}}}= \delta_{\rm{OS,}\mathit{i}}$, as the OS-induced structure represents a fixed pattern on the sky for a given LSST observing strategy and OpSim run}. Since there is generally no correlation between the \cl{OS--induced structure and LSS}, the cross-term  \cl{$\rm{\ev{\delta_{OS,\mathit{i}} \delta_{LSS,\mathit{i}}} }$ }   should be negligible; we check and confirm this for a typical dither pattern. Also, we note that this assumption \cl{about the correlation between OS-induced structure and LSS } breaks down if the survey strategy is correlated with LSS, \cl{e.g., } Deep Drilling Fields focused on galaxy clusters, as then \cl{$\rm{\ev{\delta_{OS,\mathit{i}} \delta_{LSS,\mathit{i}}} } \neq 0$.} \ 

Using equations~\ref{eq: delta_obs}-\ref{eq: <delta_obs>}, \cl{we calculate the power in $\rm{\delta_{observed, \mathit{i}}}$}:
\cl{\begin{equation}
\begin{split}
	\mathrm{\ev{ \delta_{observed, \mathit{i}}^2}} & =  \mathrm{\ev{\delta_{OS, \mathit{i}}^2}  + \ev{\delta_{LSS, \mathit{i}}^2}  + 2\ev{\delta_{OS, \mathit{i}} \delta_{LSS, \mathit{i}}}} \\
	& \hspace*{-2em} \mathrm{ + \ 2\ev{\delta_{OS, \mathit{i}}^2\delta_{LSS, \mathit{i}}} +  2 \ev{\delta_{OS, \mathit{i}}\delta_{LSS, \mathit{i}}^2} + \ev{\delta_{OS, \mathit{i}}^2\delta_{LSS, \mathit{i}}^2}}
\end{split}
\label{eq: deltaErr}
\end{equation}}
As mentioned earlier, \cl{$\rm{\ev{\delta_{OS, \mathit{i}} \delta_{LSS, \mathit{i}}} }$ } is negligible since \cl{OS--induced structure and LSS are generally not correlated}. To check how the higher order terms like \cl{$\rm{\ev{\delta_{OS, \mathit{i}}^2\delta_{LSS, \mathit{i}}^2}}$ } compare with \cl{$\rm{\ev{\delta_{OS, \mathit{i}} \delta_{LSS, \mathit{i}}} }$}, we calculate the cross-spectra for a typical dither pattern. We find that \cl{$\rm{\ev{\delta_{OS, \mathit{i}} \delta_{LSS, \mathit{i}}} }$ } is dominant over \cl{$\rm{\ev{\delta_{OS, \mathit{i}}^2\delta_{LSS, \mathit{i}}^2}}$ } and therefore these higher order terms are also negligible. \cl{Therefore, 
\begin{equation}
\begin{split}
	\mathrm{\ev{\delta_{observed, \mathit{i}}^2} \approx  \ev{\delta_{OS, \mathit{i}}^2}  + \ev{\delta_{LSS, \mathit{i}}^2}   =  {\delta_{OS, \mathit{i}}^2}  + \ev{\delta_{LSS, \mathit{i}}^2}  }
\end{split}
\label{eq: deltaErr2}
\end{equation}
implying that the OS and LSS contribute independently to the observed power. $\rm{\delta_{OS, \mathit{i}}^2}$ thus represents a bias in our measurement of LSS. }\

\begin{figure*}
	\hspace*{-0.3em}
	\begin{minipage}{0.38\paperwidth}
		\includegraphics[trim={20 70 5 -5},clip=true,width=.4\paperwidth]{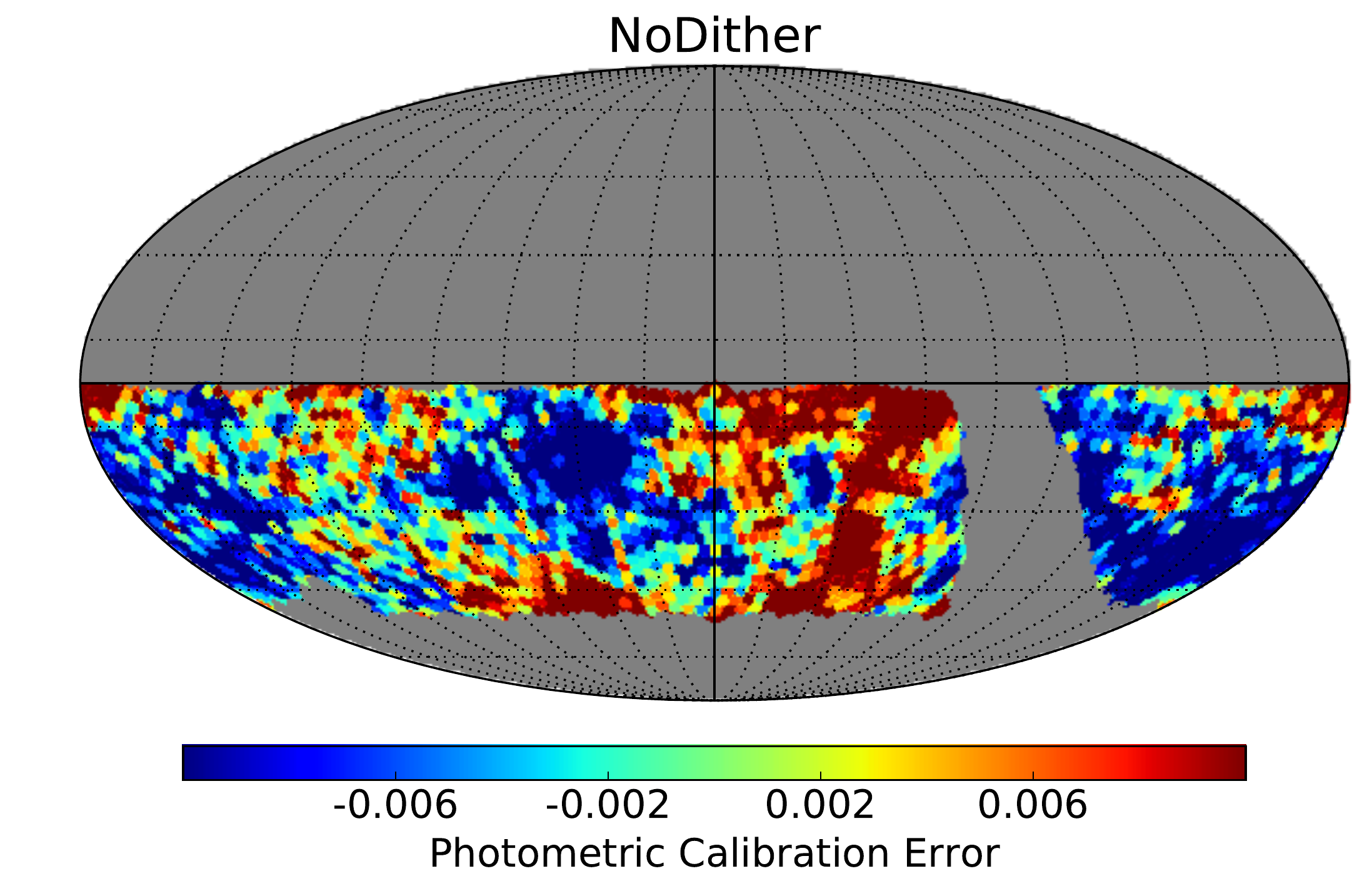}
	\end{minipage}\
	\hspace*{1em}
	\begin{minipage}{0.38\paperwidth}
		\includegraphics[trim={20 70 5 -5},clip=true,width=.4\paperwidth]{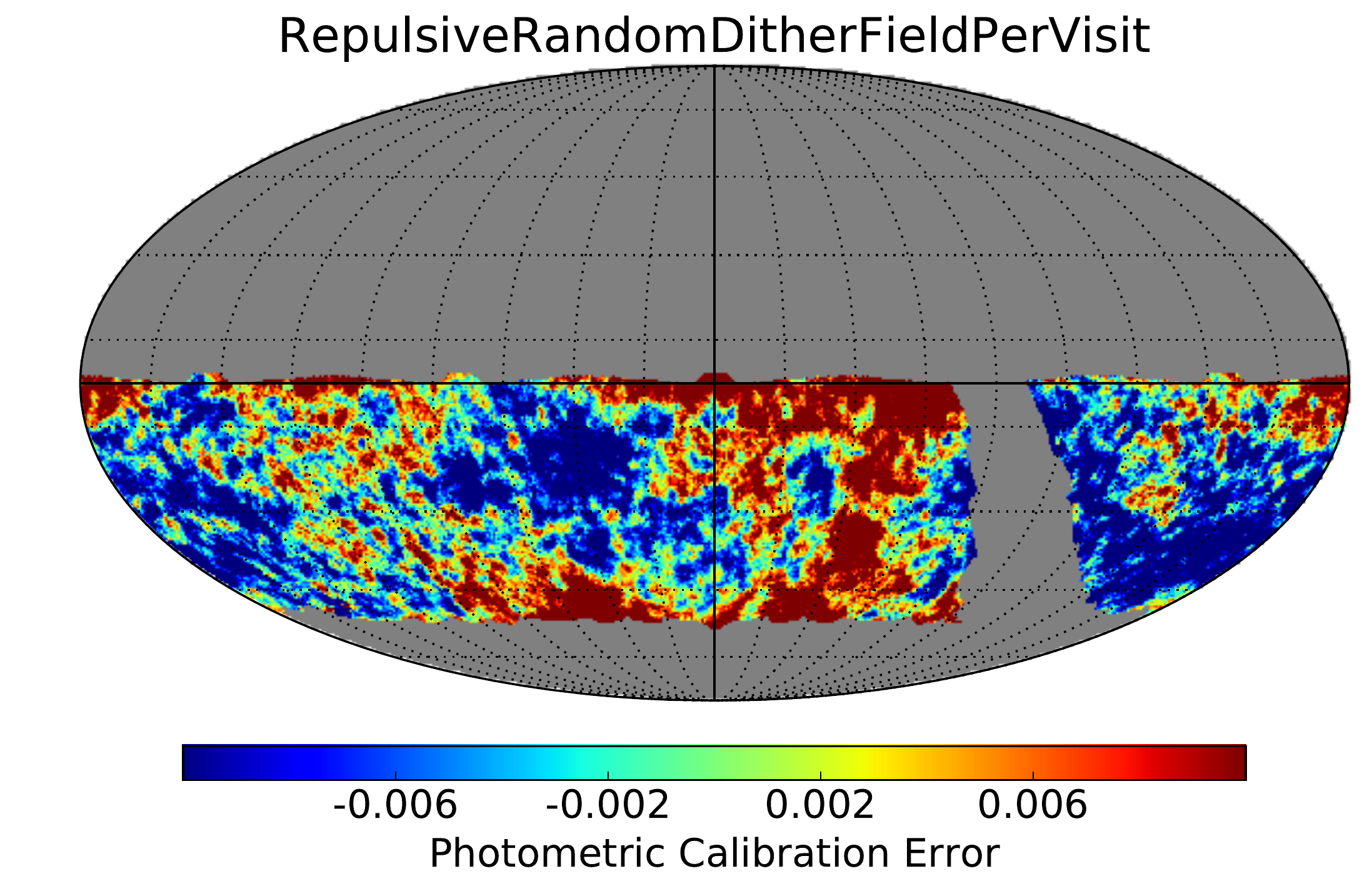}
	\end{minipage}\
	\hspace*{2em}	
	\hfill
	
	\hspace*{10em}
	\begin{minipage}{0.5\paperwidth}
		\includegraphics[trim={20 0 5 330},clip=true,width=.5\paperwidth]{f6a.pdf}
	\end{minipage}\
	\figcaption[]{Skymaps of simulated photometric calibration uncertainties for example \cl{dither } strategies.\label{0ptSkymaps}}
\end{figure*}
\begin{figure*}
	\centering{\normalsize{No Photometric Calibration Uncertainties, Dust Extinction or Poisson Noise} \\ }
	\vspace*{1em}
	\hspace*{-0.5em}
	\begin{minipage}{0.38\paperwidth}
		\includegraphics[trim={20 80 5 -5},clip=true,width=.4\paperwidth]{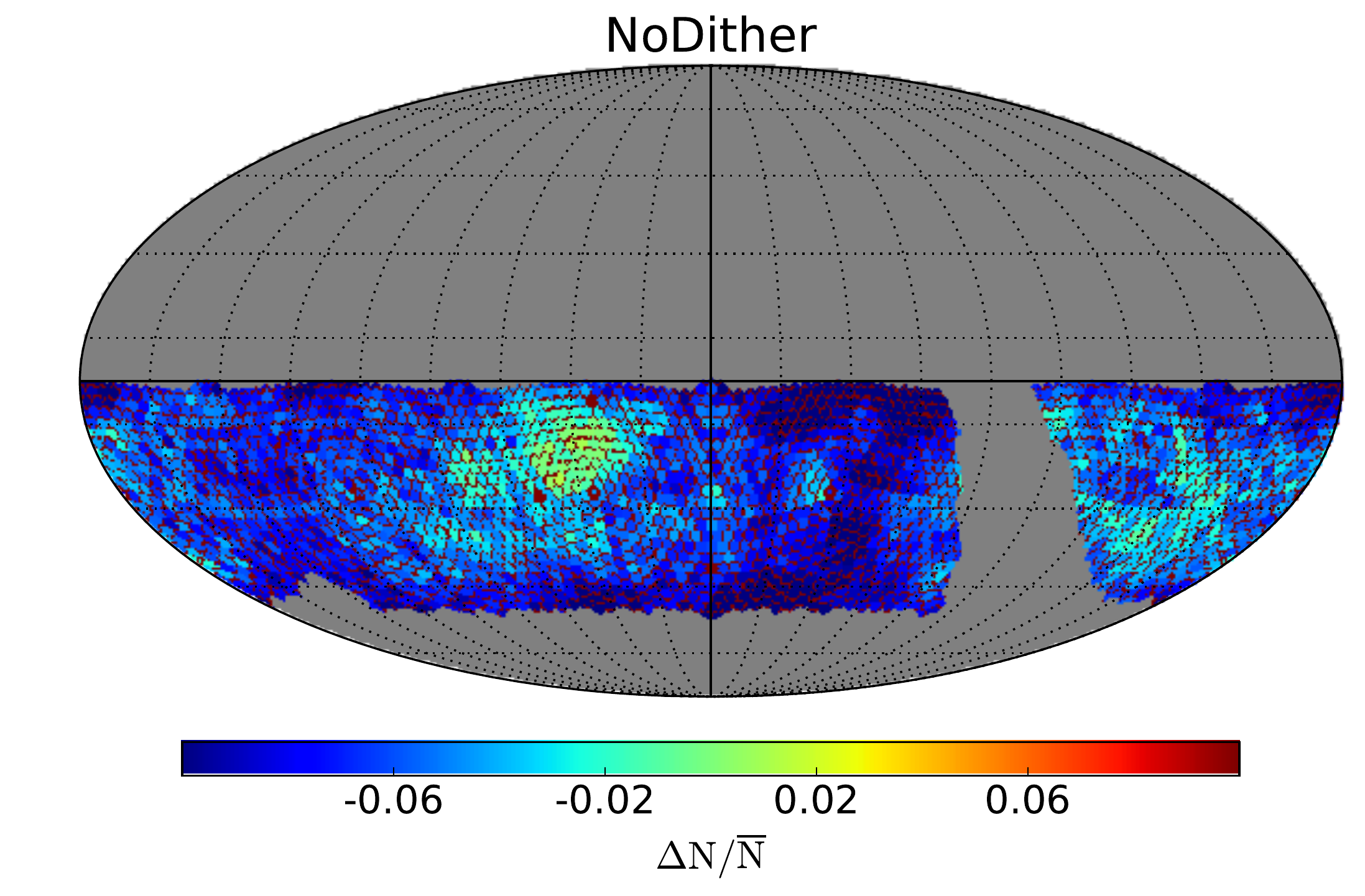}
	\end{minipage}\
	\hspace*{1em}
	\begin{minipage}{0.38\paperwidth}
		\includegraphics[trim={20 80 5 -5},clip=true,width=.4\paperwidth]{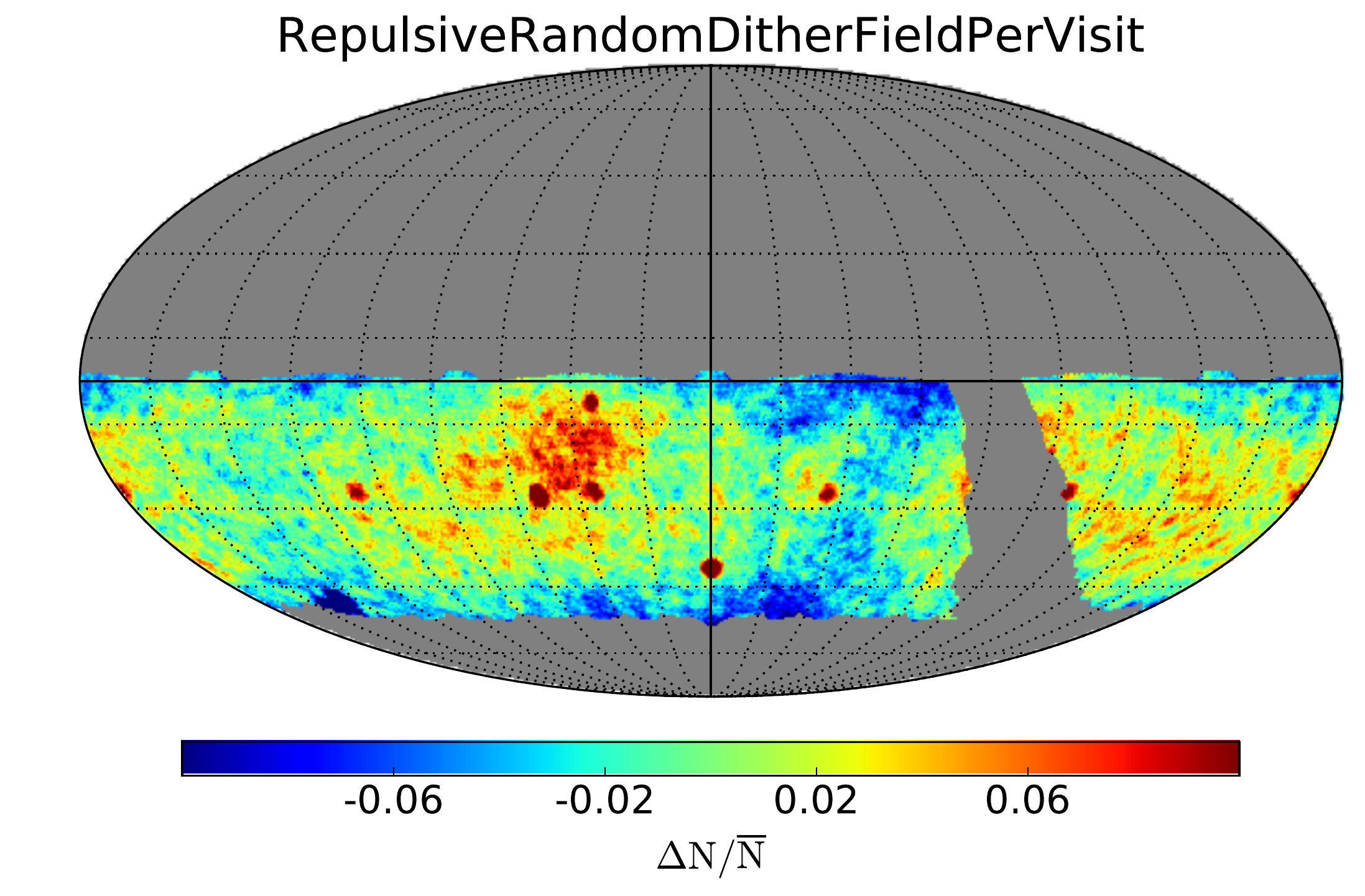}
	\end{minipage}\
	\hspace*{2em}
	\hfill
	
	\vspace*{2em}
	\centering{\normalsize{With Photometric Calibration Uncertainties, Dust Extinction and Poisson Noise} \\}
	\vspace*{1em}
	\hspace*{-0.5em}
	\begin{minipage}{0.38\paperwidth}
		\includegraphics[trim={20 80 5 -5},clip=true,width=.4\paperwidth]{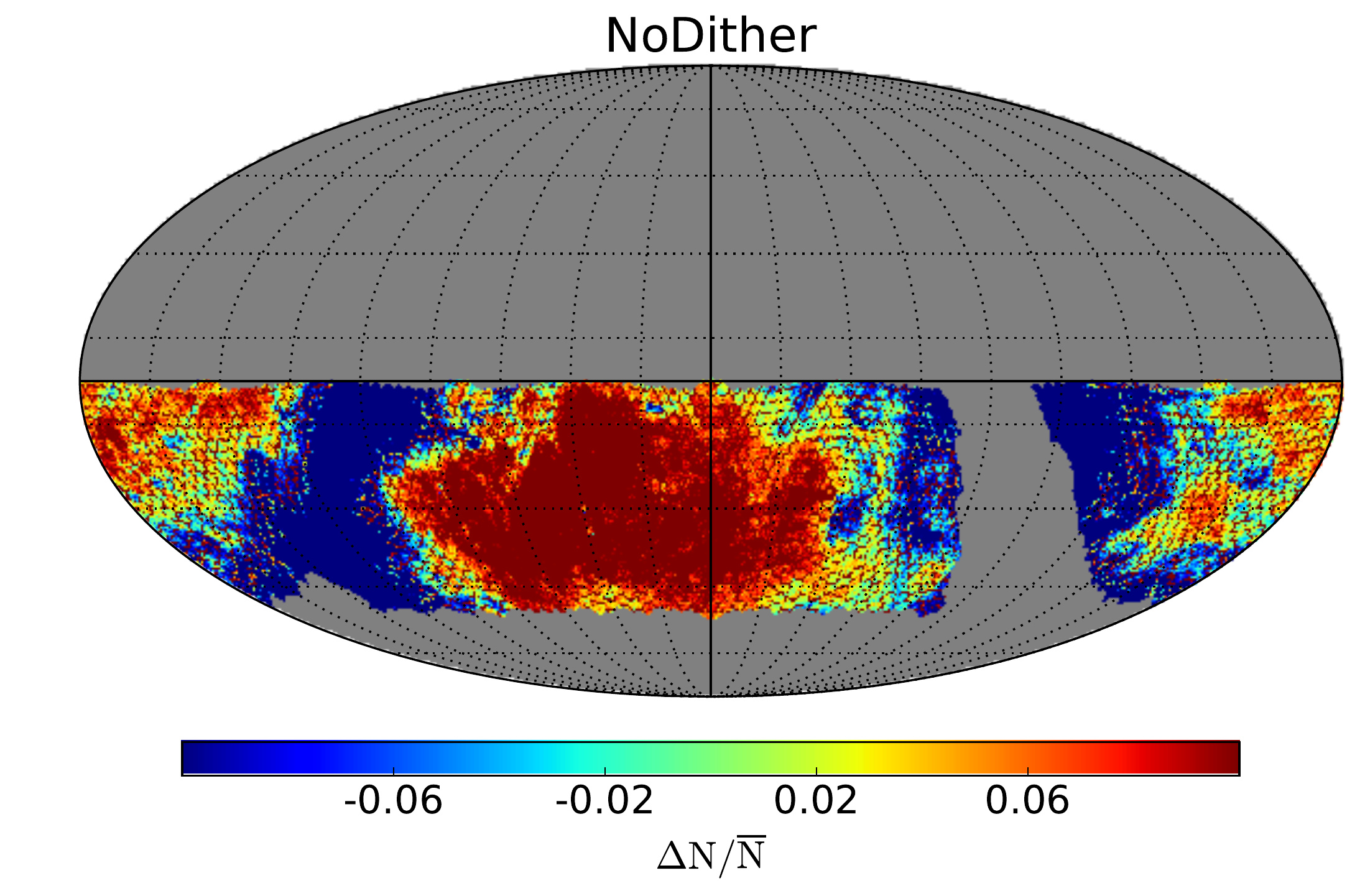}
	\end{minipage}\
	\hspace*{1em}
	\begin{minipage}{0.38\paperwidth}
		\includegraphics[trim={20 80 5 -5},clip=true,width=.4\paperwidth]{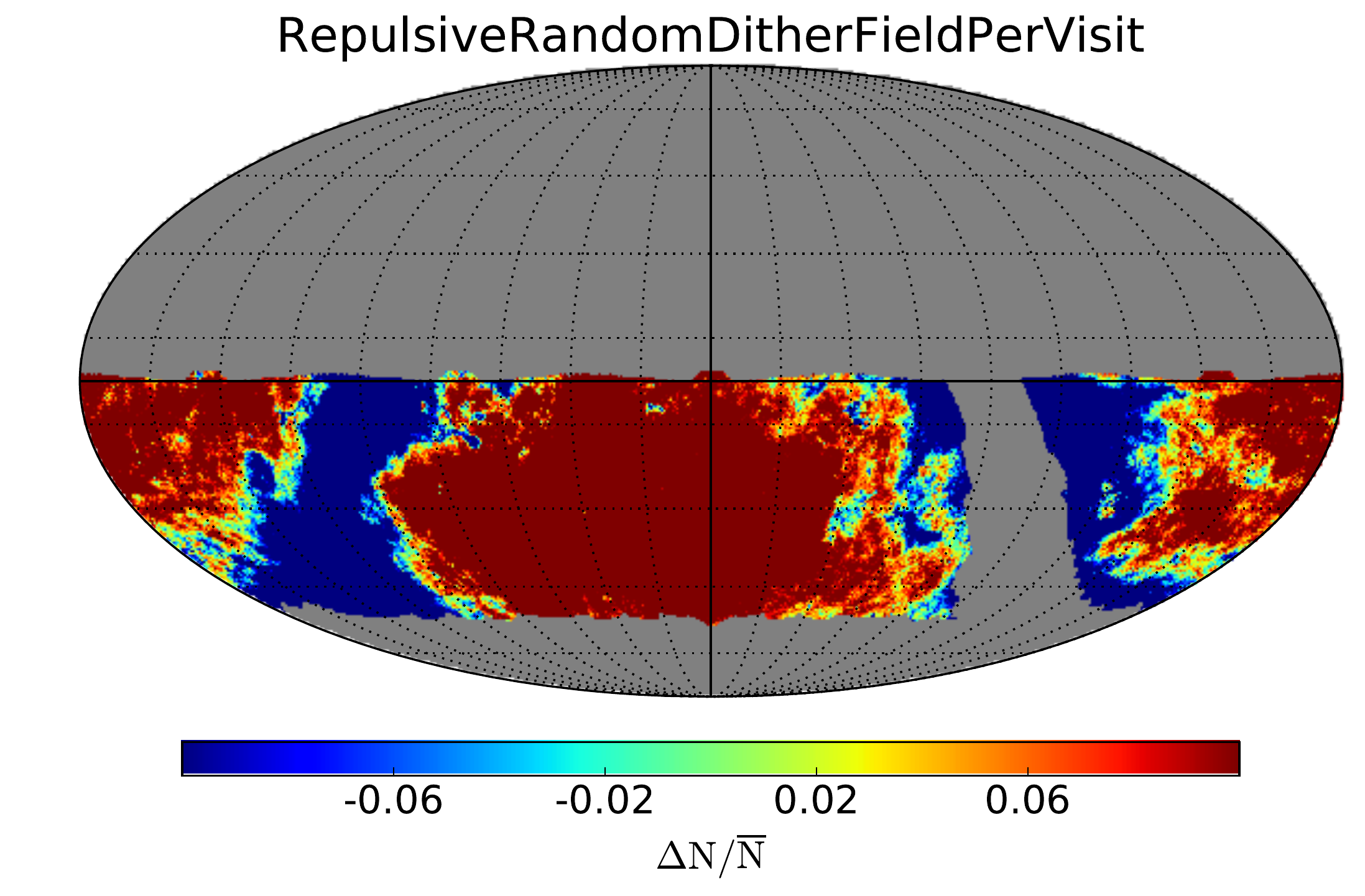}
	\end{minipage}\
	\hspace*{2em}
	\hfill
	
	\hspace*{0.2em}
	\begin{minipage}{0.5\paperwidth}
		\includegraphics[trim={20 0 5 330},clip=true,width=.5\paperwidth]{f7d.pdf}
	\end{minipage}\
	\hspace*{2em}	
	\figcaption[]{Skymaps for artificial galaxy fluctuations for example \cl{dither } strategies for 0.66$<$$z$$<$1.0. \textit{Top row}: \cl{without calibration errors, dust extinction or poisson noise. \textit{Bottom row}: After including calibration uncertainties, dust extinction and poisson noise. We do not see significant differences in the fluctuations after including the photometric calibration uncertainties or poisson noise; } the skymaps match those in the top row. However, we see that dust extinction dominates the structure on large angular scales. These trends remain consistent across all five $z$-bins. \label{galDev_skymaps_after0pt&Dust}}
\end{figure*}

To consider realistic behavior of the \cl{observing } strategies, we account for the uncertainties arising from photometric calibrations. Given that related systematic errors correlate with seeing \citep{Leistedt2015} and are expected to decrease with the number of observations, we model the calibration uncertainty $\mathrm{\Delta_{\mathit{i}}}$ in the \textit{i}th HEALPix pixel as
\begin{equation}
	\mathrm{\Delta_{\mathit{i}}= \frac{\mathit{k} \Delta s_{\mathit{i}}}{\sqrt{N_{obs, \mathit{i}}}}}
\end{equation}
where $\mathrm{\Delta s_{\mathit{i}}}$ is the difference between the average seeing in the $i$th HEALPix pixel and the average seeing across the map,  N$\mathrm{_{obs, \mathit{i}}}$ is the number of observations in the \textit{i}th pixel, and $k$ is a constant such that the variance $\mathrm{\sigma_{\Delta_\mathit{i}}^2=0.01^2}$, ensuring the expected 1$\%$ errors in photometric calibration \citep{LSST2009}. Figure~\ref{0ptSkymaps} shows skymaps for these simulated uncertainties for example \cl{dither } strategies. We note that while dithering does not alter the amplitudes of the photometric calibration uncertainties in our model, it helps mitigate the sharp hexagonal pattern seen in the uncertainties in the undithered survey. \

In order to account for the fluctuations in the galaxy counts arising due to the photometric calibration uncertainties, we modify the upper limit on the magnitude in equation~\ref{eq: erfc} to be $\mathrm{m_{max}+\Delta_{\mathit{i}}}$ for the \textit{i}th pixel. Since the calibration uncertainties are small, the skymaps for the fluctuations in the galaxy counts after accounting for the calibration uncertainties are indistinguishable from those without. These are shown in the top row in Figure~\ref{galDev_skymaps_after0pt&Dust}. \

Furthermore, we include dust extinction by using the Schlegel-Finkbeiner-Davis dust map \citep{Schlegel1998} when calculating the coadded depth \cl{as well as poisson noise in the galaxy counts after accounting for both dust extinction and photometric calibration}. The bottom row in Figure~\ref{galDev_skymaps_after0pt&Dust} shows the skymaps for the artificial fluctuations for 0.66$<$$z$$<$1.0 after accounting for photometric calibration uncertainties\c{, dust extinction and the poisson noise.} We find that dust extinction \cl{dominates both photometric calibration uncertainties and poisson noise; it } induces power on large angular scales, \cl{but } it does not wash out the honeycomb pattern in the undithered survey or its low-level residual in the dithered surveys. These trends remain consistent across the five redshift bins. \

Finally, in order to account for the spurious power introduced by the depth variations, we consider the relationship between the measured power spectrum and the true one, for a perfectly uniform survey:
\begin{equation}
	\mathrm{\ev{P_{measured}(\mathbf{k})}= \int d\mathbf{k'} \ P_{true} (k') \ |W(\mathbf{k}-\mathbf{k'})|^2}
	\label{eq: windowFunction}
\end{equation}
where $\mathrm{W(\mathbf{k}-\mathbf{k'})}$ is the survey window function, accounting for the effective survey geometry \citep{Feldman1994, Sato2013}. Projecting the 3D \textbf{k}-space onto the 2D $\ell$-space, we have
\begin{equation}
	\mathrm{C_{\ell, measured}= \sum_{\ell'}|W_{\ell-\ell'}|^2 \ev{C_{\ell'}}+ \delta C_\ell}
\end{equation}
where $\rm{\ev{C_\ell}}$ is the expected power spectrum on the full sky, and $\delta \rm{C_\ell}$ is an error term whose minimum variance is given by \citep[see][Chapter 8 for details]{Dodelson}
\begin{equation}
	\mathrm{ \left ( \Delta C_\ell \right )^2= \frac{2}{f_{sky} (2\ell + 1)}\ev{C_\ell}^2}
	\label{eq: statFloor}
\end{equation}
where f$\rm{_{sky}}$ is the fraction of the sky observed, accounting for the reduction in observed power due to incomplete sky coverage. Since we consider only the WFD survey with masked shallow borders, \cl{f$\rm{_{sky}} \approx 37\%-39\%$ for the dithered surveys while f$\rm{_{sky}} \approx 36\%$ for the undithered survey}. The expected power spectrum can be defined as
\begin{equation}
	\mathrm{\ev{C_\ell}= C_{\ell,LSS} + \frac{1}{\bar{\eta}}}
	\label{eq: expectedC}
\end{equation}
where $\rm{\bar{\eta}}$ is the surface number density in steradians$^{-1}$; see \citet{Fall1978}, \citet{Huterer2001}, \citet{Jing2005} for details. The first term in equation~\ref{eq: expectedC} is the LSS contribution to the expected power spectrum  while the second is the shot noise contribution arising from discrete signal sampling. \

With no LSS and negligible shot noise, $\rm{\ev{C_\ell} \rightarrow 0}$. However, \cl{as shown in equation~\ref{eq: deltaErr2}}, the \cl{observing } strategy induces a bias in the measured power spectrum, leading to non-zero power even when $\rm{\ev{C_\ell} \rightarrow 0}$. The uncertainty in this bias \cl{caused by imperfect knowledge of the survey performance } limits our ability to correct for the OS-induced artificial structure. More quantitatively, we have
\begin{equation}
	\mathrm{\left ( \sigma_{C_{\ell, measured}} \right)^2 =  \left ( \Delta C_\ell \right )^2 + \left(\sigma_{C_{\ell, OS}}  \right)^2}
\end{equation}
where the first term on the right is the minimum statistical uncertainty defined in equation~\ref{eq: statFloor}, while the second term corresponds to the contribution from the uncertainty in the bias induced by the OS. Since the ``statistical floor" $\rm{\Delta C_\ell}$  assumes no bias in $\rm{C_\ell}$ measurements caused by the observing strategy, the OS-induced uncertainty $\rm{\sigma_{C_{\ell, OS}}}$ must be subdominant to the statistical floor for an optimal measurement of BAO at a given redshift, i.e.
\begin{equation}
	\mathrm{\sigma_{C_{\ell, OS}}   <<   \Delta C_\ell = \sqrt{\frac{2}{f_{sky} (2\ell + 1)}} \left(  C_{\ell, LSS}+ \frac{1}{\bar{\eta}} \right) }
	\label{eq: sigmas}
\end{equation}
\cl{Here we note that the right-hand side in equation~\ref{eq: sigmas} is formally derived in \citet{Shafer2005}; also see \citet{Huterer2003}. These papers offer a detailed theoretical treatment of artificial structure induced by calibration errors, and while our approach is similar to theirs, we incorporate the additional effects of dust extinction, variations in survey depth, and incompleteness in galaxy detection.}

Considering the case where $\rm{C_{\ell, LSS}= 0}$, we find $\rm{C_{\ell, measured}}$, giving us $\rm{C_{\ell, OS}}$ for each band and magnitude cut. Since $ugri$ bands are the deepest and appear to have the greatest influence on photometric redshifts (Prakash, priv. comm.), we model the overall bias as the mean $\rm{C_{\ell, OS}}$ across the four bands. We calculate $\rm{\sigma_{C_{\ell, OS}}}$ as the standard deviation of $\rm{C_{\ell, OS}}$ across the $ugri$ bands, modeling uncertainties due to detecting galaxy catalogs in different bands. Therefore,  $\rm{\sigma_{C_{\ell, OS}}}$ should provide a conservative upper limit on the true uncertainty in $\rm{C_{\ell, OS}}$. \ 

\begin{figure*}
	\hspace*{-0.5em}
	\begin{minipage}{0.4\paperwidth}
		\includegraphics[trim={40 25 45 30},clip=true,width=.43\paperwidth]{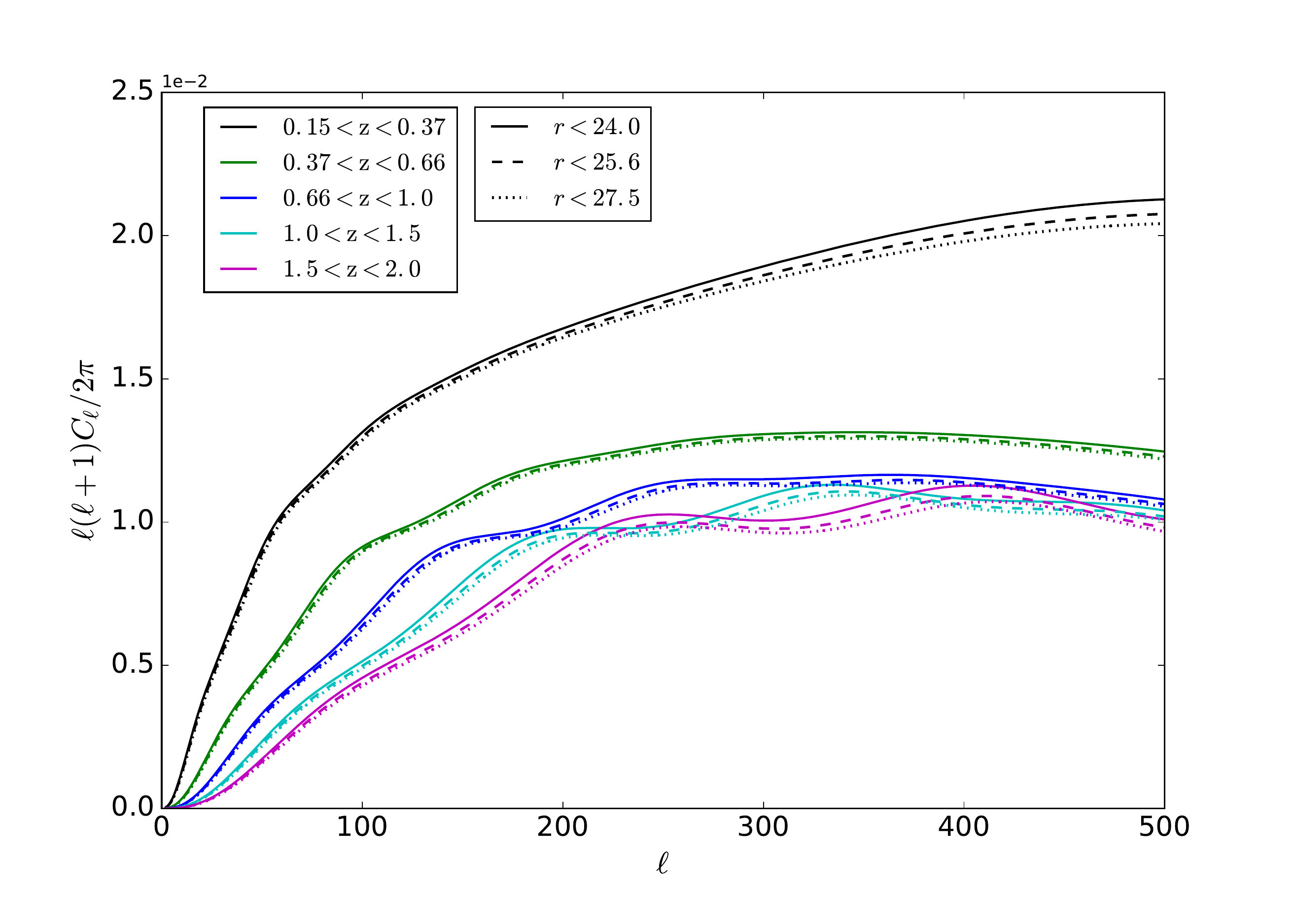}
	\end{minipage}\
	\hspace*{1em}
	\begin{minipage}{0.4\paperwidth}
		\includegraphics[trim={40 25 45 30},clip=true,width=.43\paperwidth]{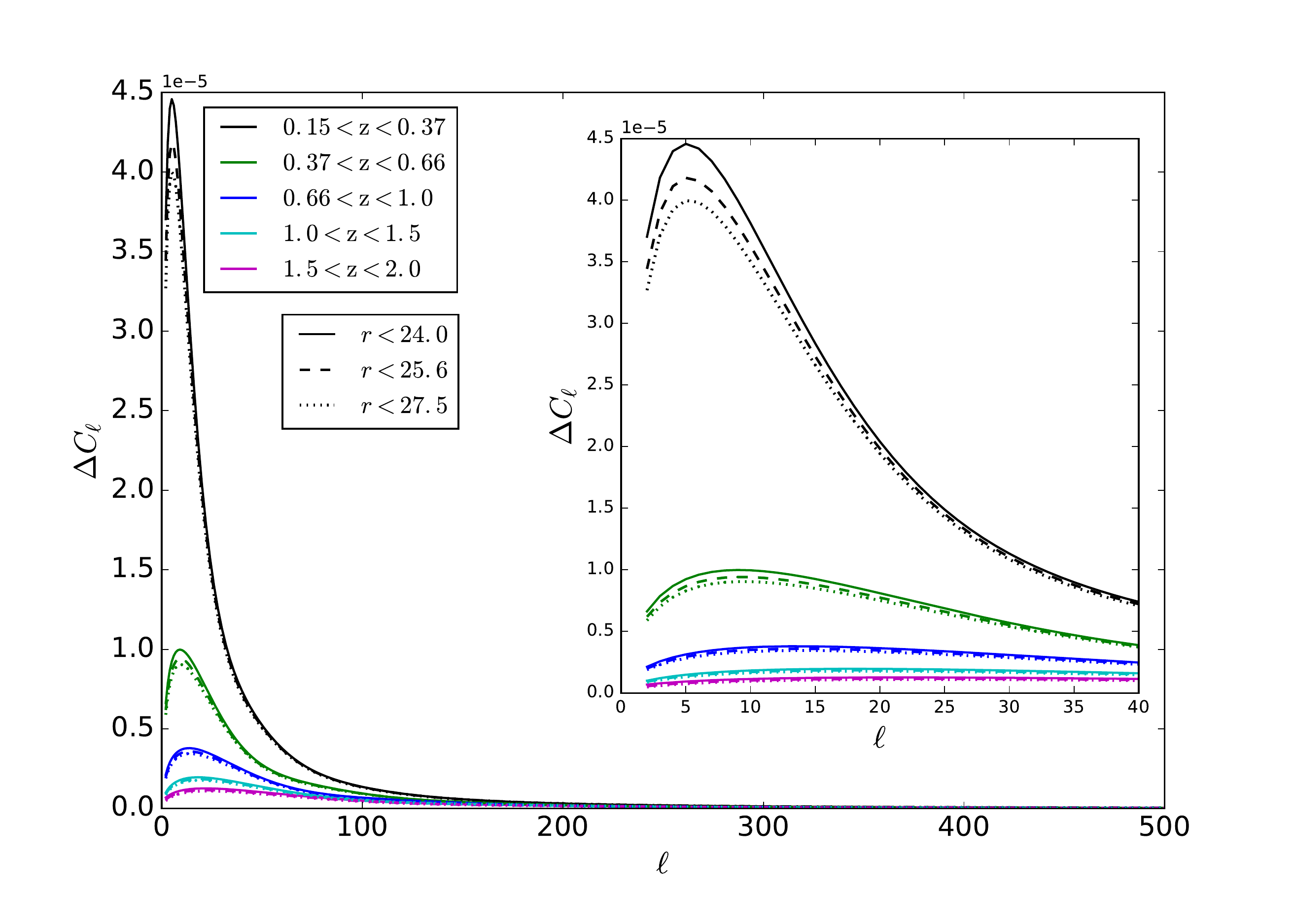}
	\end{minipage}\
	\hspace*{0em}	
	\figcaption[]{{\textit{Left}: Simulated full-sky, \cl{pixelized } galaxy power spectra with BAO signal from five different redshift bins for \cl{three galaxy catalogs: $r$$<$24.0, 25.6, 27.5. \textit{Right}: Minimum statistical error associated with measuring the signal in the left panel, with lower--$\ell$ range shown in the inset. We observe that neither curve changes significantly with magnitude cuts considered in Section~\ref{LSS}.}} \label{BAO}}
\end{figure*}
The left panel in Figure~\ref{BAO} shows the full-sky galaxy power spectrum with the BAO signal for \cl{three galaxy catalogs, $r$$<$24.0, 25.6, 27.5, } for the five redshift bins: 0.15$<$$z$$<$0.37, 0.37$<$$z$$<$0.66, 0.66$<$$z$$<$1.0, 1.0$<$$z$$<$1.5, and 1.5$<$$z$$<$2.0. \cl{These spectra are pixelized in order to account for the finite angular resolution of our survey simulations, especially when comparing the uncertainties in $\rm{C_{\ell, OS}}$ with the minimum statistical error in measuring BAO.  Assuming that all the HEALPix pixels are identical, the pixelized power spectra can be approximated by multiplying the galaxy power spectra with the pixel window function\footnote {See Appendix B in the HEALPix primer: \url{http://healpix.sourceforge.net/pdf/intro.pdf}}. \

The galaxy power spectra are calculated using the code from \citet{Zhan2006}, with modifications to account for BAO signal damping due to non-linear evolution \citep{Eisenstein2007}. Using the galaxy redshift distribution from \citet{LSST2009}, galaxies are assigned to the five redshift bins according to their photometric redshifts, with a time-varying but  scale-independent galaxy bias of $b(z) = 1 + 0.84z$ over scales of  interest and a simple photometric redshift error model, $\sigma_z = 0.05(1+z)$. Here we assume the cosmology with $w_0 = -1$, $w_a = 0$, $\Omega_m = 0.127$, $\Omega_b = 0.0223$,  $\Omega_k = 0$,  spectral index of the primordial scalar perturbation power spectrum $n_s= 0.951$ and primordial curvature power spectrum at $k$ = 0.05/Mpc, $\Delta_R^2 = 2\times10^{-9}$. \

The right panel in Figure~\ref{BAO} shows the minimum statistical uncertainty for the five redshift bins \cl{for all three galaxy catalogs; the uncertainties are calculated using $\rm{f_{sky}}$ from the undithered survey. We observe that while shallower galaxy catalogs lead } to larger C$_\ell$ and $\rm{\Delta{C_{\ell}}}$, the difference is small and decreases with increasing redshift. For the lowest $z$-bin, 0.15$<$$z$$<$0.37, there is only about 8$\%$ increase in C$_\ell$ and $\rm{\Delta{C_{\ell}}}$ when comparing the  $r$$<$25.6 catalog with $r$$<$24.0}. \

\begin{figure*}
	\vspace*{3em}
	\hspace*{8em}
	\includegraphics[trim={74 85 90 70},clip=false,width=.6\paperwidth]{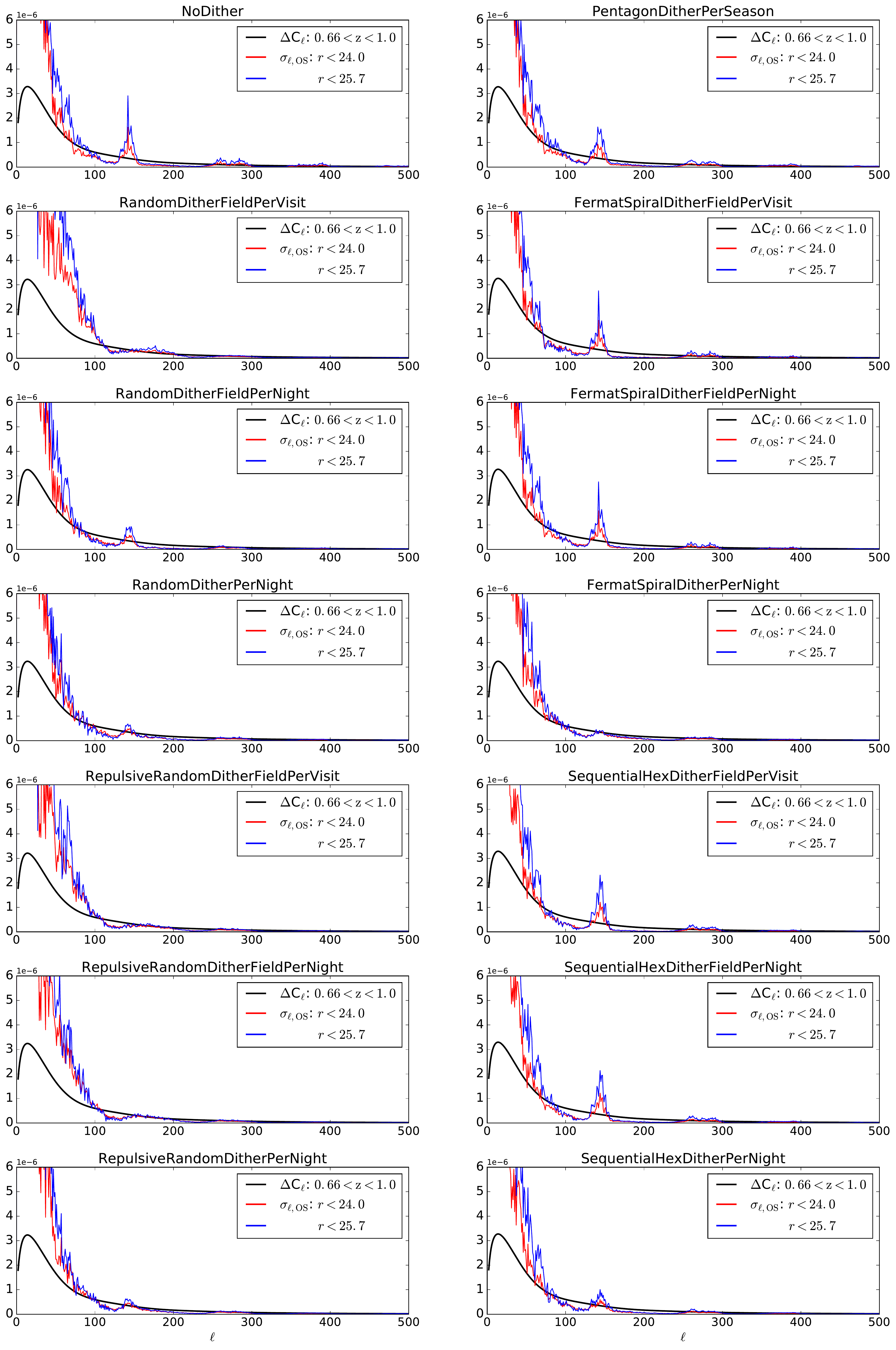}
	\vspace*{3em}
	\figcaption[]{ $\sigma \rm{_{\ell, OS}}$ comparison with the minimum statistical uncertainty $\Delta$C$_\ell$ for 0.66$<$$z$$<$1.0 for different magnitude cuts after only one year of survey. \label{bin3_1yr}}
\end{figure*}

First we calculate $\rm{C_{\ell, OS}}$ and its uncertainties for 0.66$<$$z$$<$1.0 after only one year of survey in order to explore the quality of BAO study the first data release will allow.  Figure~\ref{bin3_1yr} shows the $\rm{C_{\ell, OS}}$ uncertainties as well as the minimum statistical error for 0.66$<$$z$$<$1.0 for various \cl{observing } strategies, for $r$$<$24.0 and $r$$<$25.7 (corresponding to the gold sample, $i$$<$25.3). We observe that  the undithered survey leads to $\rm{C_{\ell, OS}}$ uncertainties \cl{1-3}$\times$ the minimum statistical uncertainty for the gold sample at  $\ell > $100, and only a few \cl{dither } strategies are effective in reducing the difference. In particular, Random and RepulsiveRandom dithers are the most effective, reducing $\rm{\sigma_{C_{\ell, OS}}}$ to nearly 1-2$\times$ the statistical floor. \cl{We note that FermatSpiral and SequentialHex dithers perform nearly as poorly as NoDither when implemented on FieldPerVisit and FieldPerNight timescales, while the PerNight timescale is more effective. On the other hand, we see that FieldPerVisit and FieldPerNight lead to smaller uncertainties for Random and RepulsiveRandom geometries. }As expected, we see that a shallower sample $r$$<$24.0 reduces the $\rm{C_{\ell, OS}}$  uncertainties; the undithered survey still leads to $\rm{\sigma_{C_{\ell, OS}}}$ about 3$\times$ the statistical floor\cl{, while Random and RepulsiveRandom dithers lead the uncertainties comparable to the statistical floor on some timescales. Here we note that since we do not mask borders when considering the one-year data, f$\rm{_{sky}} \approx 42\%-45\%$ for the one year survey, depending on the dither strategy. }\

\begin{figure*}
	\vspace*{3em}
	\hspace*{8em}
	\includegraphics[trim={74 85 90 70},clip=false,width=.6\paperwidth]{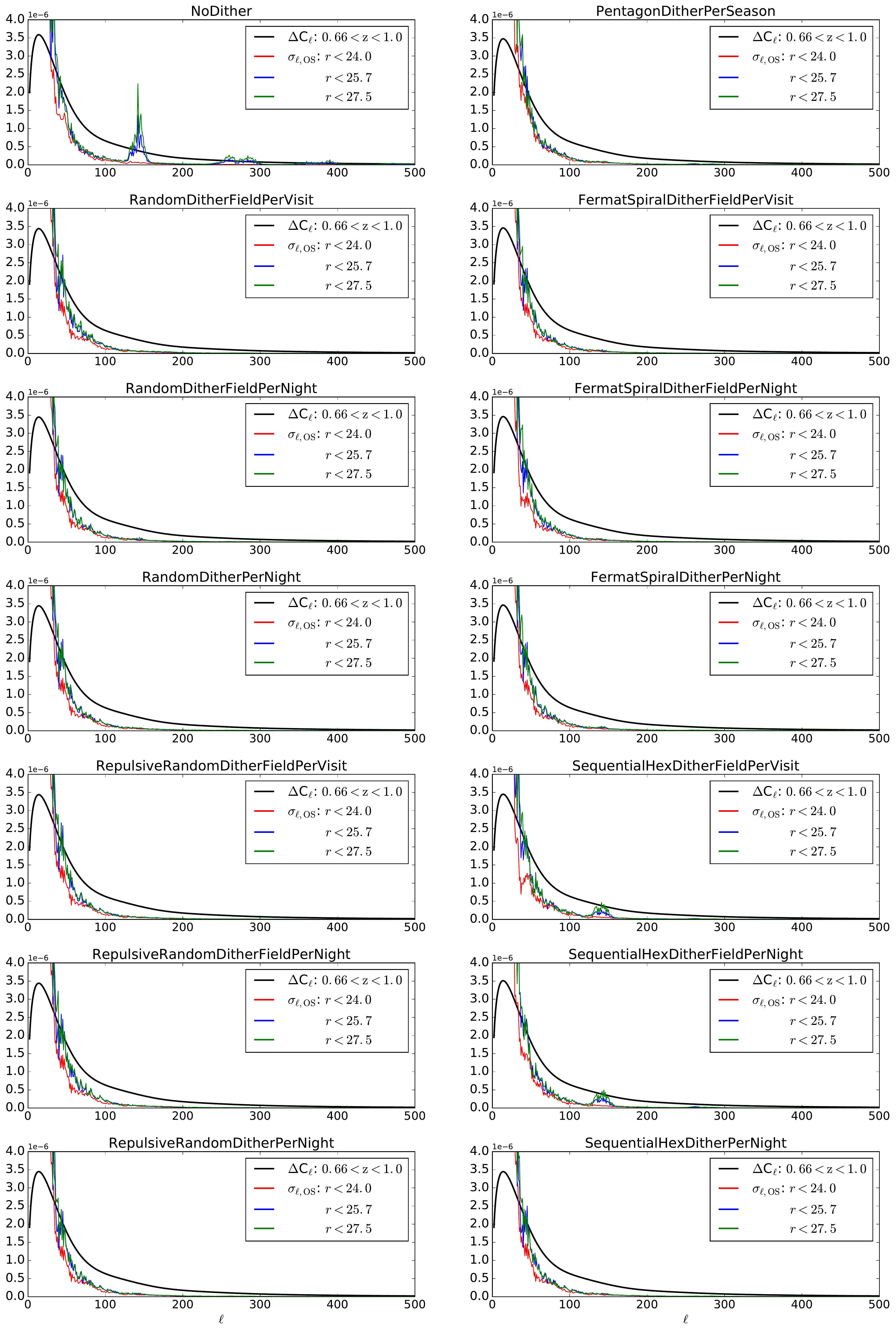}
	\vspace*{3em}
	\figcaption[]{ $\sigma \rm{_{\ell, OS}}$ comparison with the minimum statistical uncertainty $\Delta$C$_\ell$ for 0.66$<$$z$$<$1.0 for different magnitude cuts after the full 10-year survey. \label{bin3_10yr}}
\end{figure*}
We then extend the calculation of the OS-induced power to the full 10-year survey. Figure~\ref{bin3_10yr} shows $\rm{\sigma_{C_{\ell, OS}}}$ as well as $\rm{\Delta C_\ell}$ for 0.66$<$$z$$<$1.0 for three different magnitude cuts: $r$$<$24.0, $r$$<$25.7 and $r$$<$27.5. We find that the undithered survey leads to $\rm{\sigma_{C_{\ell, OS}}}$ \cl{0.2-4 } times the minimum statistical floor for $r$$<$25.7 and $r$$<$27.5; at $\ell > $100, only a very strict cut of $r$$<$24.0 brings $\rm{\sigma_{C_{\ell, OS}}}$ below $\rm{\Delta C_\ell}$. However, most dither strategies reduce the uncertainties below the statistical floor for galaxy catalogs as deep as $r$$<$27.5, \cl{with exceptions of SequentialHex dithers on FieldPerVisit and FieldPerNight timescales}. We note here that systematics correction methods such as template subtraction and mode projection can be applied to further reduce the contribution of  $\rm{C_{\ell, OS}}$ to the total  $\rm{C_{\ell}}$ uncertainties; \cl{e.g., } see \citet{Elsner2016}, \citet{Hogg2012}. Such application appears necessary for the 1-year survey as optimizing the observing strategy alone does not reduce the uncertainties in $\rm{C_{\ell, OS}}$ below $\rm{\Delta C_\ell}$. However, the correction methods may not lead to significant improvements for a dithered 10-year survey, as optimizing the \cl{observing } strategy is effective in reducing $\rm{C_{\ell, OS}}$ well below the statistical floor. \ 

To further our understanding, we repeat the 1-year and 10-year analysis for 1.5$<$$z$$<$2.0. We find similar qualitative results as those from 0.66$<$$z$$<$1.0 analysis: for the 1-year survey,  Random and RepulsiveRandom perform well alongside FermatSpiral and SequentialHex on PerNight timescale, while most \cl{dither } strategies are effective for the ten-year survey, with the exception of SequentialHex on FieldPerVisit and FieldPerNight timescales. \

\begin{deluxetable*}{lcccc} 
\centering
\tablecolumns{4} 
\tablewidth{0pc} 
\tablecaption{Estimated number of galaxies from $r$-band coadded depth after the 10-year survey for 0.15$<$$z$$<$2.0, after accounting for photometric calibration errors\cl{, dust extinction and poisson noise}.} 
\tablehead{ \colhead{ } & \colhead{$r$$<$27.5} & \colhead{$r$$<$25.7} &  \colhead{$r$$<$24.0}}
\startdata
Number of galaxies from NoDither	                         &	\cl{$1.0 \times 10^{10}$}	  &	\cl{$4.3 \times 10^{9}$}	  &	\cl{$1.6  \times 10^{9}$}\\ \hline \hline
Percent improvements in comparison with NoDither	&						  &		                           &				\\ \hline

PentagonDitherPerSeason	                                 & 		\cl{7.0}				  &			 \cl{6.6} 	  &  			\cl{6.6} \\

SequentialHexDitherFieldPerVisit	                        &  		\cl{8.1}				  &			\cl{7.8}	 	  &  			\cl{7.9} \\
SequentialHexDitherFieldPerNight	                        & 		\cl{4.9}				  &			\cl{4.3}	 	  &  			\cl{4.4} \\
SequentialHexDitherPerNight	                                & 		\cl{8.3}				  &			\cl{8.0}	 	  &  			\cl{8.1} \\

FermatSpiralDitherFieldPerVisit	                                 & 		\cl{7.6}				  &			\cl{7.2}	 	  &  			\cl{7.3} \\
FermatSpiralDitherFieldPerNight	                        & 		\cl{7.6}				  &			\cl{7.2}	 	  &  			\cl{7.3} \\
FermatSpiralDitherPerNight	                                 & 		\cl{7.4}				  &			\cl{7.0}	 	  &  			\cl{7.1} \\

RandomDitherFieldPerVisit	                                 &  		\cl{8.7}				  &			\cl{8.4}	 	  &  			\cl{8.5} \\
RandomDitherFieldPerNight	                                 & 		\cl{8.3}				  &			\cl{8.0}	 	  &  			\cl{8.1} \\
RandomDitherPerNight	                        		        & 		\cl{8.5}				  &			 \cl{8.2}		  &  			\cl{8.3} \\

RepulsiveRandomDitherFieldPerVisit                        & 		\cl{8.9}				  &			\cl{8.5}	 	  &  			\cl{8.7} \\
RepulsiveRandomDitherFieldPerNight	                & 		\cl{8.6}				  &			\cl{8.4}	 	  &  			\cl{8.5} \\
RepulsiveRandomDitherPerNight	                        &  		\cl{8.3}				  &			\cl{7.9}	 	  &  			\cl{8.0}
\enddata
\tablecomments{ We observe 6.5-9$\%$ improvement in the estimated number of galaxies from dithered surveys in comparison with undithered \cl{survey}, across the three magnitude cuts. The exception \cl{is SequentialHexDitherFieldPerNight } where the improvement is only 4-5$\%$.  \vspace{-0.5em} \label{tab: numGal}}
\end{deluxetable*} 

The effect of magnitude cuts is further illustrated in Table~\ref{tab: numGal}, which includes the estimated number of galaxies for 0.15$<$$z$$<$2.0 from the $r$-band coadded depth for the 10-year survey after accounting for photometric calibration errors\cl{, dust extinction and poisson noise}.  We see that each magnitude cut eliminates a substantial number of galaxies. Also, as in \citet{Carroll2014}, we see that dithering increases the estimated number of galaxies when compared to the undithered survey; the fractional difference in the number of galaxies from dithered to undithered surveys increases with shallower surveys. 

\section{Conclusions}{\label{Conclusion}}
It is critical to develop an LSST observing strategy that will maximize the data quality for its science goals. In this work, we analyzed the effects of \cl{dither } strategies on $r$-band coadded 5$\sigma$ depth to study the feasibility of increasing the uniformity across the survey region. We investigated different \cl{dither } geometries on different timescales, and illustrated how a specific geometrical pattern (\cl{e.g., } hexagonal lattice) can perform quite differently when implemented on different timescales. We find that per-visit and  per-night implementations outperform field-per-night and per-season timescales, while some dither geometries (like repulsive random dithers) consistently lead to less spurious power for all the timescales on which the dither positions are assigned. We also performed an a$_{\ell m}$ analysis to probe the origins of some of the characteristic patterns induced by the \cl{observing } strategies. Our work illustrates \cl{the sensitivity of depth uniformity to the dither strategy}. \

We then considered how the artifacts in coadded depth produce fluctuations in galaxy counts; we calculate the uncertainties in the bias induced by the \cl{observing } strategy, which limits our ability to correct for the spurious structure. We find that after accounting for photometric calibration uncertainties, dust extinction, \cl{poisson noise } and reasonable magnitude cuts, dithers of most kinds are effective in reducing the uncertainties in the observing-strategy-induced bias below the minimum statistical uncertainty in the measured galaxy power spectrum. Specifically, we find that RepulsiveRandom dithers implemented on per-visit and field-per-night timescales are the most effective for the 0.66$<$$z$$<$1.0 sample after only one year of survey, although they do not bring down the uncertainties in the induced bias below the minimum statistical floor for $r$$<$25.7. As for the \cl{full 10-year survey}, we find that all \cl{dither } strategies (\cl{except per-visit and field-per-night  SequentialHex dithers}) bring down the uncertainties below the statistical floor for a galaxy catalog as deep as $r$$<$27.5. We find similar results for all redshift bins.\

\cl{To precisely determine the limiting uncertainties in the bias induced by the observing strategy, more detailed LSST simulations are needed, including photometric redshifts, input large-scale structure and further systematics reduction methods, e.g., mode projection accounting for imperfect detectors and the consequent instrumental effects}. Also, while our work illustrates the impact of dithers on large-scale structure studies, the differences between some dither geometries are small and therefore need more detailed investigation to determine a conclusively-best \cl{dither } strategy\cl{, alongside an analysis of the impacts of various dither strategies on other science goals. }Such analyses will facilitate a more definitive measure of the precision with which LSST data will allow high redshift studies of large-scale structure.\\

This research was supported by the Department of Energy (grant DE-SC0011636) and National Science Foundation (REU grant PHY-1263280). Hu Zhan was partially supported by the Chinese Academy of Sciences (grant XDB09000000), and Alejandra M. Mu\~{n}oz Arancibia by FONDECYT (grant 3160776) \cl{and BASAL CATA PFB-06.  Nelson D. Padilla also acknowledges support from BASAL CATA PFB-06 and FONDECYT (grant 1150300); the lightcones were run using the Geryon cluster hosted at the Centro de Astro-Ingenieria UC. Sof\'ia A. Cora acknowledges support from Consejo Nacional de Investigaciones Cient\'ificas y T\'ecnicas, Agencia Nacional de Promoci\'on Cient\'ifica y Tecnol\'ogica, and Universidad Nacional de La Plata, Argentina. We thank K. Simon Krughoff for suggesting the hexagonal dither pattern for LSST, and Abhishek Prakash, Jeff Newman, Andy Connolly, Rachel Mandelbaum, Dragan Huterer, Terry Matilsky and Seth Digel for helpful comments}, conversations and insights.  Finally, we thank the LSST Dark Energy Science Collaboration for feedback on design and conduct of this research. 

\newpage
\bibliography{Awan_et_al_paper}

\appendix
\section{Border Masking Algorithm}{\label{appendix}}
In Figure~\ref{coadd_withBorder}, we show skymaps (left column) and the corresponding power spectrum (right column) for the \cl{$r$-band coadded 5$\sigma$ depth } from the undithered survey and an example dithered survey. While the dithered survey does not have the strong honeycomb seen in the undithered case, we notice that the border of the dithered survey area is much shallower than the rest of the survey. This variation in depth carries over to the power spectrum as strong oscillations, especially at small $\ell$. In order to minimize this effect, we develop a border masking algorithm to mask the pixels within a specific `pixel radius' from the edge of the survey area. For this purpose, we utilize the distinction between out-of-survey and in-survey area in MAF: the former is masked, and the analysis only accounts for the data in the unmasked portion of the \cl{data } array. Using this distinction and the HEALpix routine \texttt{get\_all\_neighbours}, we find the unmasked pixels with masked neighbors, effectively finding the edge of the survey. We parametrize the number of iterations for this neighbor finding algorithm, and choose the number of iterations (determined by what we call the pixel radius) that removes the shallow border. The masking algorithm can be found on GitHub\footnote {\url{https://github.com/LSST-nonproject/sims\_maf\_contrib/blob/master/mafContrib/maskingAlgorithmGeneralized.py}}.\

Working at N$\rm{_{side}}$= 256 resolution, we masked all the pixels within a 14-pixel radius from the edge of survey, effectively masking $\sim$15$\%$ of the survey area. The bottom row in Figure~\ref{coadd_withBorder} shows the dithered skymap and the corresponding power spectrum after the shallow border has been removed.   We notice a stark difference between the power spectrum before and after the border masking, as removing the shallow border allows the in-survey variations to be seen much more clearly.

\setcounter{figure}{0} \renewcommand{\thefigure}{A.\arabic{figure}}
\begin{figure*}
	\vspace*{3em}
	\hspace*{-0.5em}
	\begin{minipage}{0.4\paperwidth}
		\includegraphics[trim={20 70 5 -5},clip=true,width=.4\paperwidth]{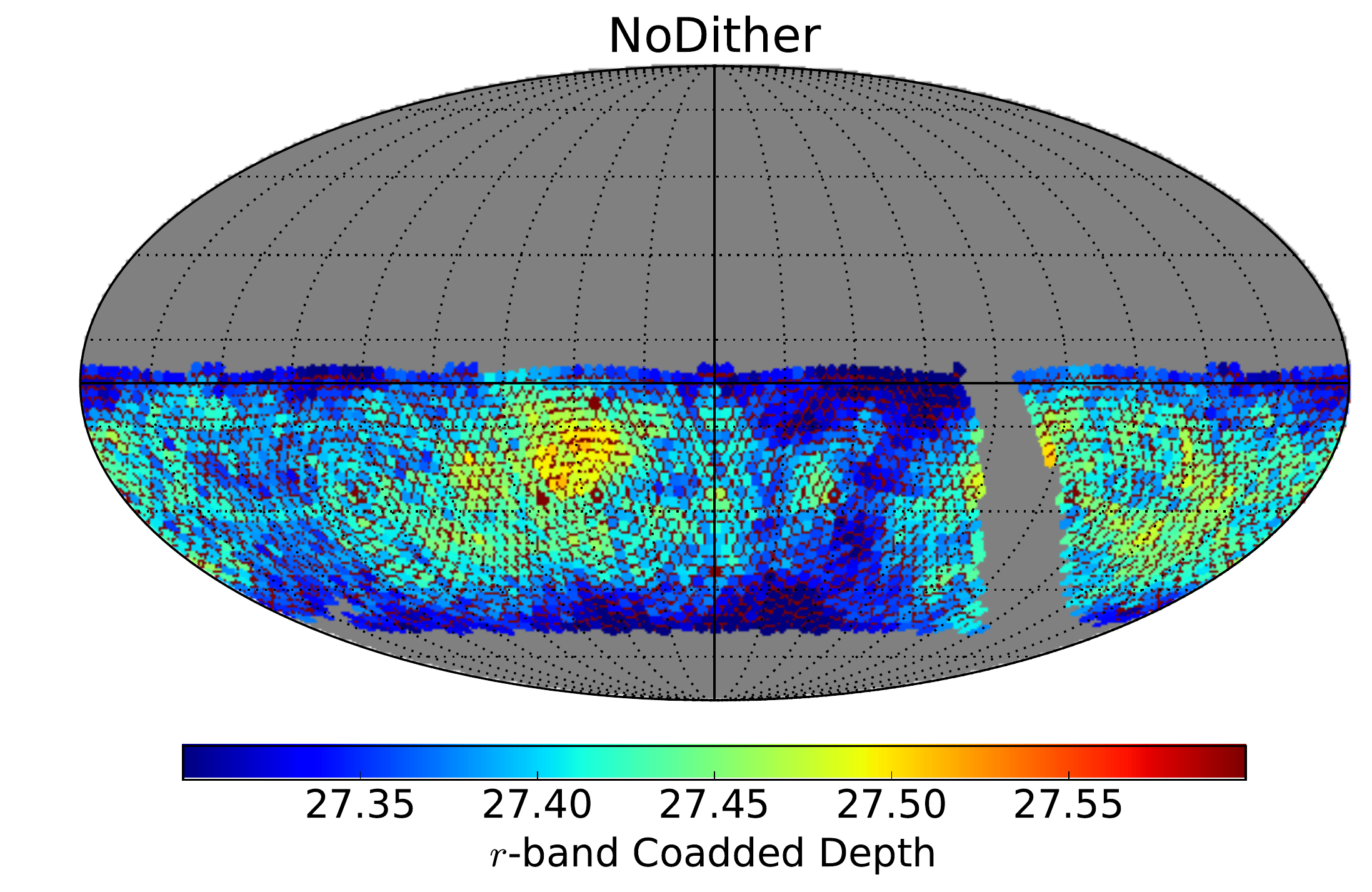}
	\end{minipage}\
	\hspace*{0em}
	\begin{minipage}{0.4\paperwidth}
		\includegraphics[trim={5 3 5 5},clip=true,width=.4\paperwidth]{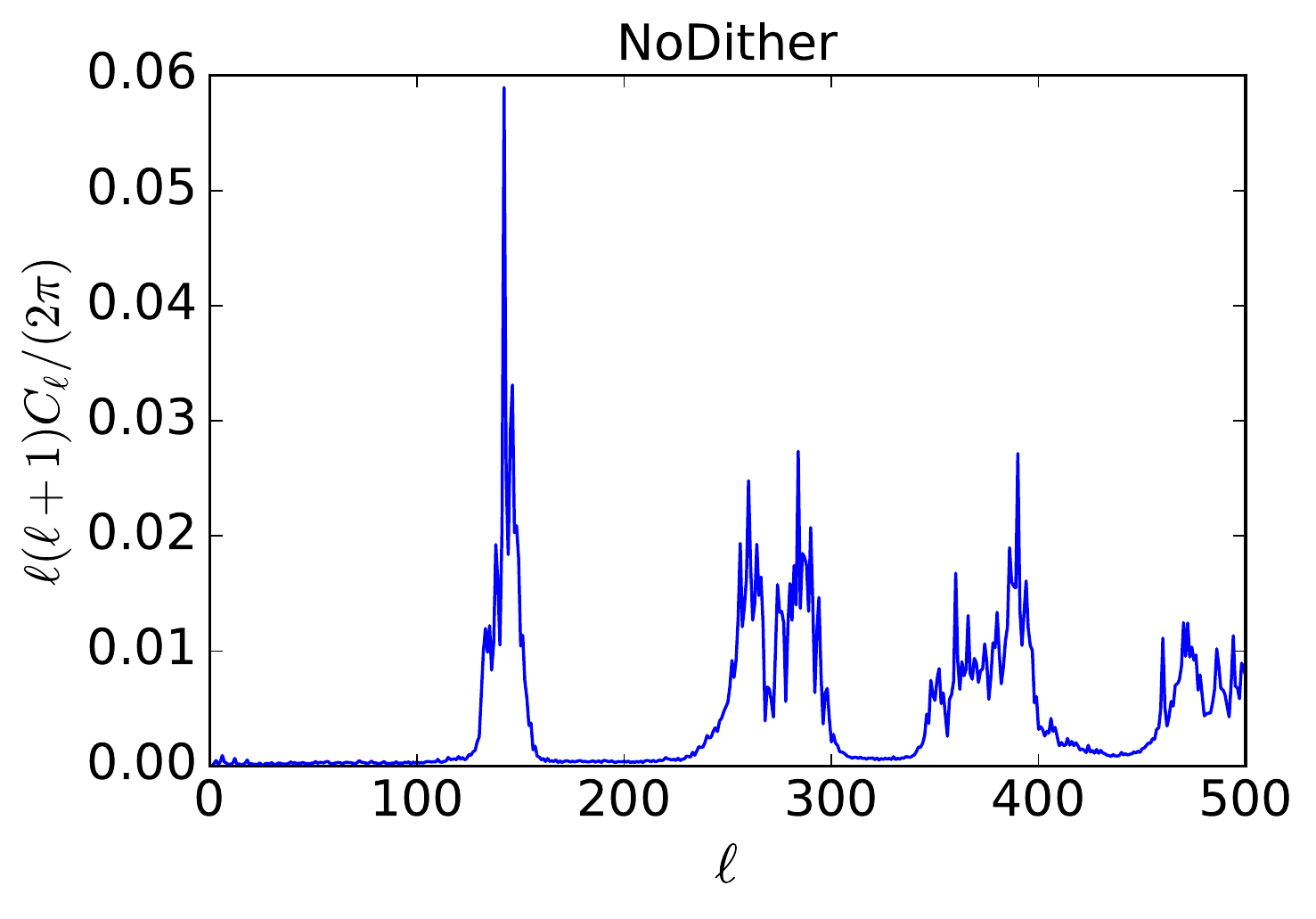}
	\end{minipage}\
	\hspace*{2em}
	\vspace*{.5em}
	\hfill
	
	\hspace*{-0.5em}
	\begin{minipage}{0.4\paperwidth}
		\includegraphics[trim={20 70 5 -5},clip=true,width=.4\paperwidth]{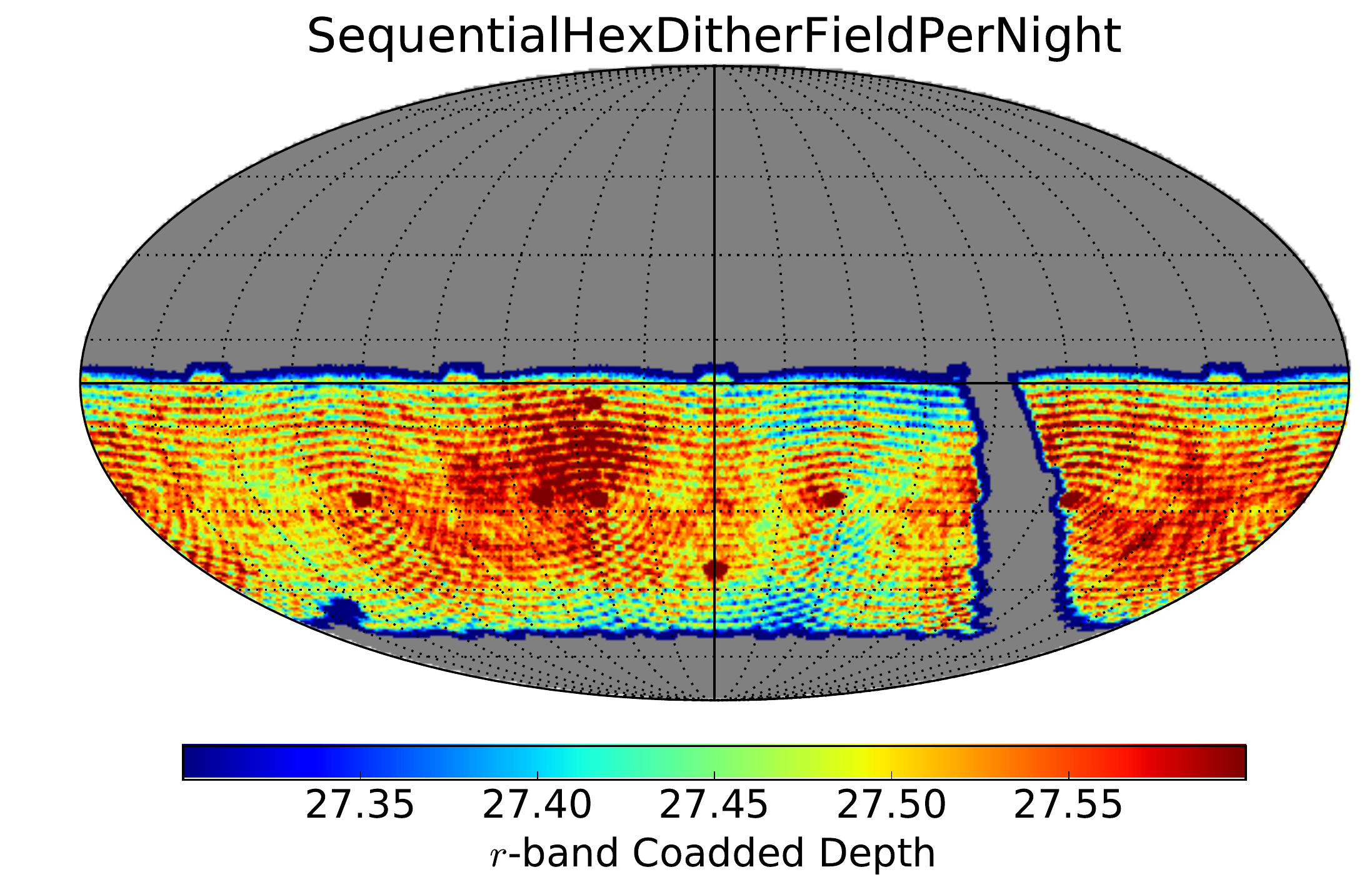}
	\end{minipage}\
	\hspace*{0em}
	\begin{minipage}{0.4\paperwidth}
		\includegraphics[trim={5 3 5 5},clip=true,width=.4\paperwidth]{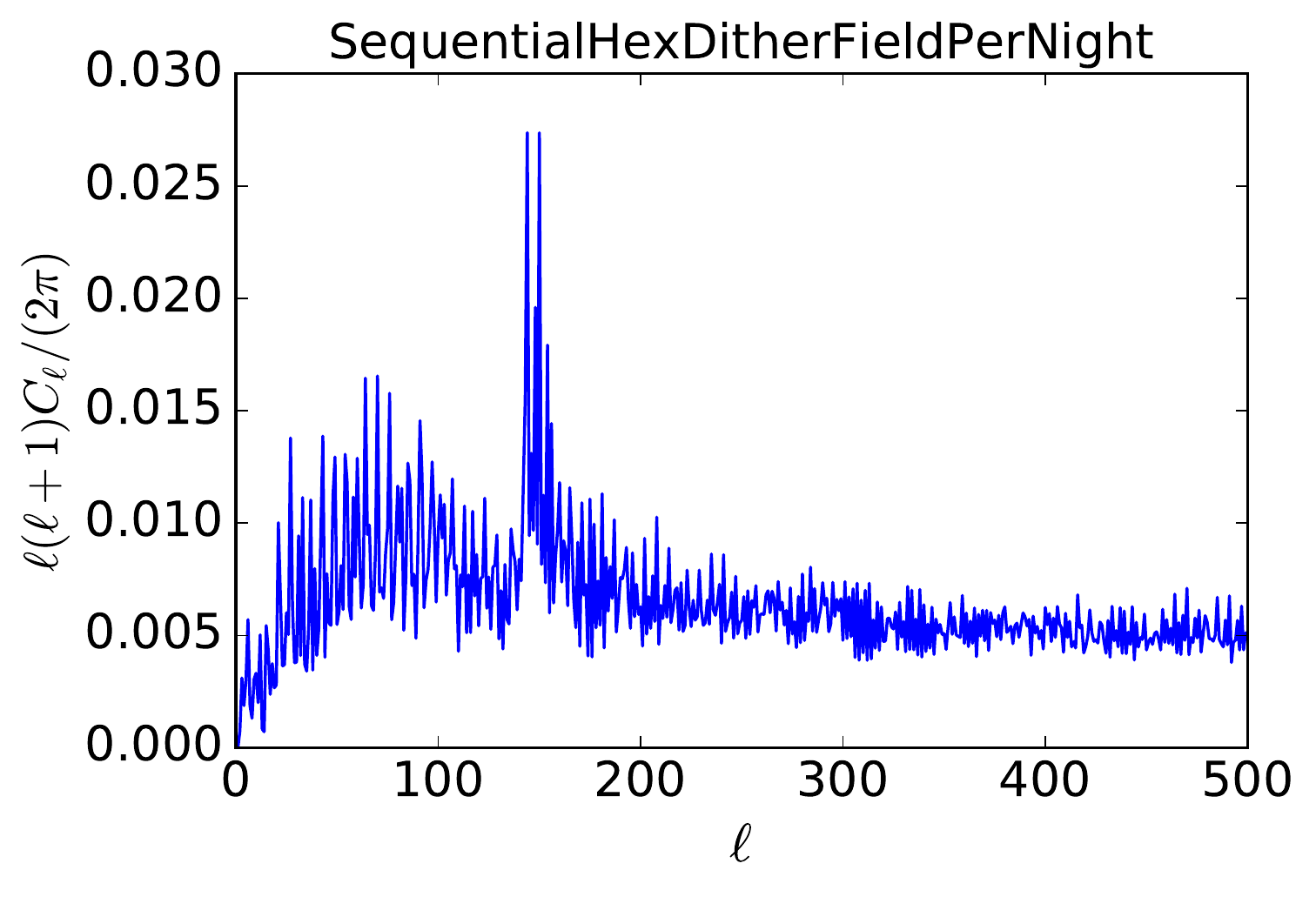}
	\end{minipage}\
	\hspace*{2em}
	\vspace*{.5em}
	\hfill
	
	\hspace*{-0.5em}
	\begin{minipage}{0.4\paperwidth}
		\includegraphics[trim={20 70 5 -5},clip=true,width=.4\paperwidth]{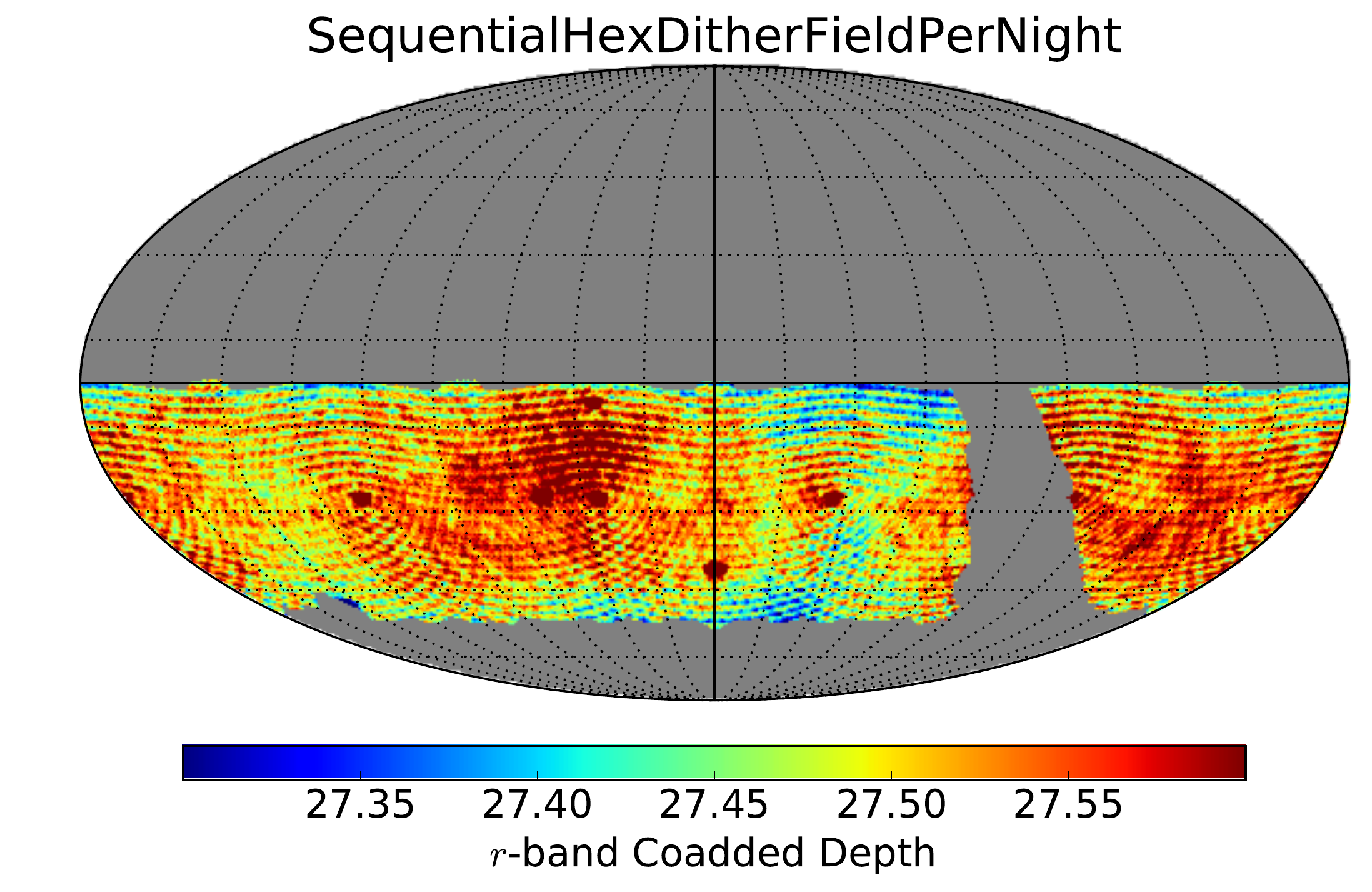}
	\end{minipage}\
	\hspace*{0em}
	\begin{minipage}{0.4\paperwidth}
		\includegraphics[trim={5 3 5 5},clip=true,width=.4\paperwidth]{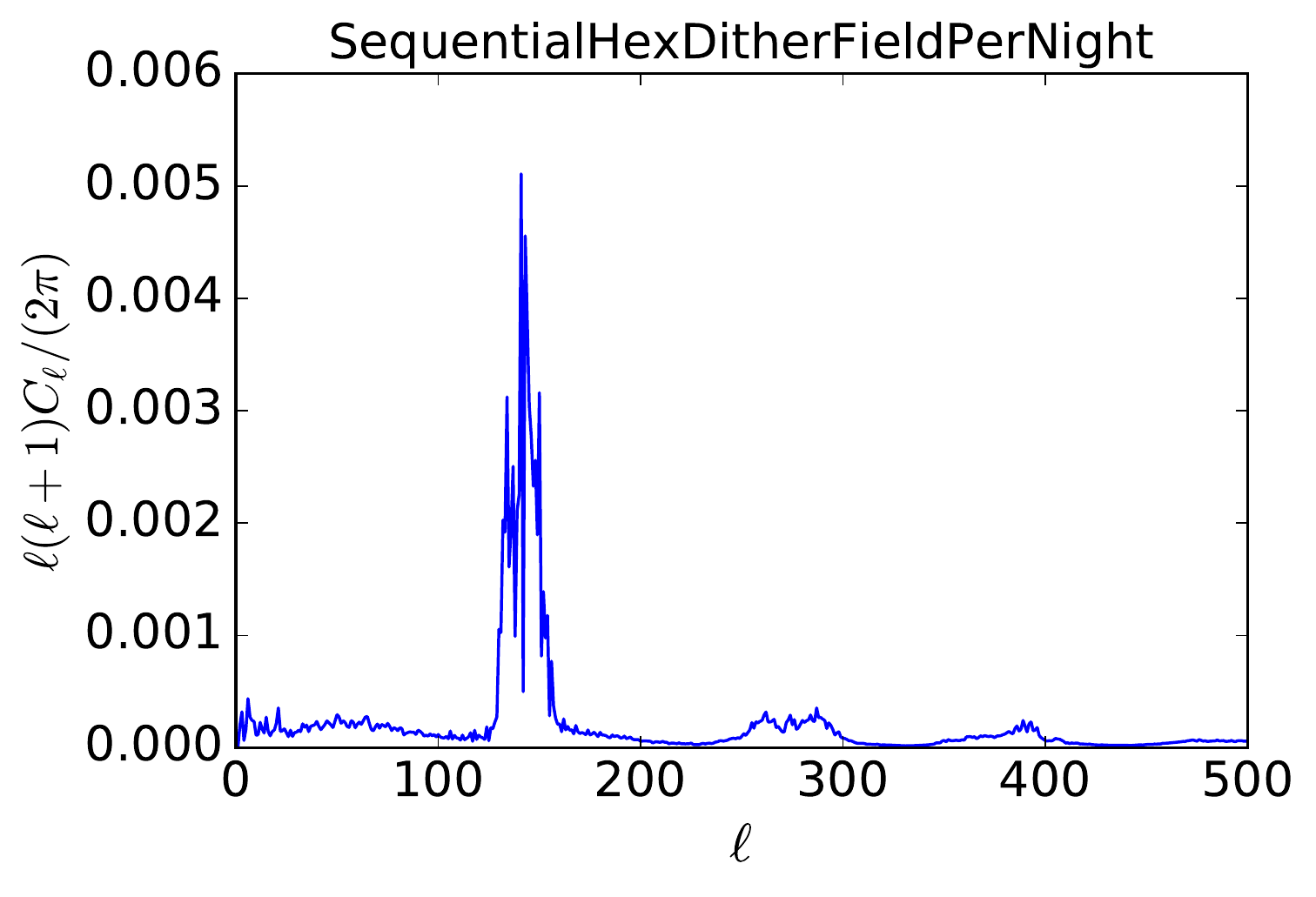}
	\end{minipage}\
	\hspace*{2em}
	\hfill
	
	\hspace*{-1em}
	\begin{minipage}{0.4\paperwidth}
		\includegraphics[trim={20 7 5 340},clip=true,width=.45\paperwidth]{fA1e.pdf}
	\end{minipage}\
	\hspace*{2em}
	\vspace*{.5em}
	\figcaption[]{\textit{Left column}: Skymaps for $r$-band \cl{coadded 5$\sigma$ } depth for example \cl{dither } strategies. \textit{Right column}: Angular power spectra corresponding to the skymaps in the first column. Top and middle rows show the data without any border masking. We note that the undithered survey does not lead to any shallow edges, while dithered survey does. The shallow-depth edge leads to a noisy power spectrum, shown in the middle right panel. After removing the shallow-border by implementing 14 pixel-radius masking, we see a reduction in the low-$\ell$ power, and therefore a cleaner spectrum. \label{coadd_withBorder}}
\end{figure*}

\end{document}